\documentclass[12pt]{article}

\usepackage{setspace,graphicx,epstopdf,amsmath,amsfonts,amssymb,amsthm, versionPO, texintro}
\usepackage{marginnote,datetime,enumitem,rotating,fancyvrb}
\usepackage{hyperref,float}
\usepackage{booktabs}
\usepackage{placeins}
\usepackage{bbm}
\usepackage{csvsimple}
\usdate
\usepackage{booktabs,tabularx,siunitx,threeparttable}
\newcolumntype{Y}{>{\raggedright\arraybackslash}X}
\usepackage{adjustbox}

\usepackage[justification=centering]{caption}
	\usepackage[margin=0.75in]{geometry}%
	
	\usepackage{lscape}

\makeatletter\let\chapter\@undefined\makeatother 

\usepackage{indentfirst} 
\usepackage{endnotes}    
\usepackage{jf}          
\usepackage[labelfont=bf,labelsep=period]{caption}   
\usepackage{bm}
\usepackage{subcaption}

\usepackage{caption}
\usepackage{booktabs}
 \usepackage{threeparttable}
  \usepackage{longtable}
  \usepackage{array}
\newcolumntype{C}[1]{>{\centering\arraybackslash}p{#1}}
\usepackage[table]{xcolor}
\usepackage{makecell}
\usepackage{fouriernc}
\usepackage[T1]{fontenc}

\usepackage{chapterbib} 
\usepackage[round,authoryear]{natbib}
\usepackage{hyperref}
\hypersetup{
    colorlinks=true,
    linkcolor=blue,
    citecolor=blue,
    urlcolor=blue
}

\makeatletter
\AddToHook{env/tabular/begin}{\let\input\@@input}
\makeatother

\graphicspath{{"/Users/gschuber/Gregor_Dropbox Dropbox/Gregor Schubert/Research/Geography of AI/output/figures"}}

\author{Michael Blank\thanks{
Stanford Graduate School of Business, Email: \href{mailto:blankm@stanford.edu}%
{blankm@stanford.edu}.}  \hspace{-.3cm} \and Gregor Schubert\thanks{%
UCLA Anderson School of Management, Email: \href{mailto:gregor.schubert@anderson.ucla.edu}{gregor.schubert@anderson.ucla.edu.}}  
\hspace{-.3cm}  \and Miao Ben Zhang\thanks{USC Marshall School of Business, Email: \href{mailto:Miao.Zhang@marshall.usc.edu}{%
miao.zhang@marshall.usc.edu}.}}

\title{\bf The  Household Impact of Generative AI: \\ Evidence from Internet Browsing Behavior\thanks{We thank participants at the OpenAI economic seminar, the UCLA Finance Brownbag, and conference participants at the 2026 AFA special session and the 2026 AEA meeting.}} 

\date{  \today  } 

\begin{document}

\begin{titlepage}
\maketitle
\thispagestyle{empty}
\begin{abstract}

\onehalfspacing
\noindent This paper studies the impact of generative AI on U.S. households' task allocation at home, using detailed Internet browsing data from a large sample of home devices between 2021 and 2024. Leveraging pre-ChatGPT browsing patterns, we measure households' exposure to ChatGPT and use it as an instrument for ChatGPT adoption during the post-release period. Our IV estimates show that adopting generative AI substantially increases leisure browsing on home devices while leaving time spent on productive digital tasks unchanged. To examine mechanisms, we infer the purpose of households' ChatGPT use from surrounding internet activity and find that households primarily employ it for productive non-market tasks. Together, these results suggest that generative AI frees up leisure time by raising the efficiency of productive digital activities. Interpreting these findings through a standard time-allocation model implies economically large productivity gains from generative AI at home.
\vspace{2cm}

\end{abstract}
\end{titlepage}


\doublespacing

\newpage 
\noindent 
The prevailing view of the economic impact of generative artificial intelligence (GenAI) centers on the workplace — how it reshapes jobs, firms, and labor markets. However, GenAI may be at least as transformative at home as it is at work.  As GenAI diffuses rapidly into everyday life, it raises several first-order economic questions: Which households adopt it, for which tasks, and with what consequences for household productivity, leisure, and inequality? In this paper, we document that households that adopt ChatGPT increase their leisure time online, even though they mostly use ChatGPT for productive (non-leisure) online tasks. Our findings are consistent with ChatGPT use making household adopters' productive online tasks substantially more efficient, allowing them to reallocate the freed-up time to leisure.

Our focus on generative AI's impact ``at home'' complements existing empirical work on its impact ``at work'' (e.g. \cite{brynjolfsson2023}, \cite{yotzov_firm_2026}) and highlights one of the key margins of economic adjustment, since Americans are more likely to use GenAI tools for private purposes than for their jobs: \cite{bick2024} find in a representative survey fielded in August and November 2024 that 33.7\% of all respondents had used generative AI outside of work, while only 26.4\% of employed respondents had used it at work.\footnote{A Pew Research survey in February 2024 found that 23\% of all adults had used ChatGPT, 17\% had used it for entertainment, and 17\% to learn something new, while only 12\% of all adults (20\% of those employed) had used it for a work task. A recent report by OpenAI confirmed that 69\% of messages in the consumer version of the app were non-work messages \citep{openai2026}.} 

In spite of this greater prevalence of private uses, there is so far little empirical evidence of the broader set of private activities that GenAI is used for.  While emerging evidence from surveys and laboratory experiments provides early snapshots of private GenAI use, the related studies do not address the consequences for household behavior and non-market household productivity. In this paper, we address this gap using detailed Internet browsing microdata from over 200{,}000 U.S. households’ home devices from 2021 to 2024 to provide the first large-scale evidence on the adoption and impact of generative AI on households' non-market activity. 

We make five contributions. First, we measure household ChatGPT adoption over time and document a widening ``generative AI divide'' across U.S. households by income and age, with little signs of convergence. Second, using our browsing data, we also construct a household-level measure of ex-ante exposure to generative AI based on pre-ChatGPT Internet browsing patterns and show that it strongly predicts subsequent adoption of ChatGPT. Third, using pre-existing exposure as an instrument for household ChatGPT adoption in a long-difference empirical design, we show that ChatGPT adoption substantially increases leisure browsing on home devices while leaving time spent on productive digital tasks unchanged. Fourth, using the browsing context surrounding ChatGPT site visits, we infer the purpose of household generative AI use and find that private ChatGPT use mainly happens in the context of productive online activities (e.g.\ education, job search, informational research), rather than during leisure browsing. Fifth, we interpret these findings using a quantitative model of household time allocation and find that generative AI approximately doubles the efficiency of productive online tasks:  our preferred calibration suggests efficiency gains in productive digital tasks of 76\%-176\% for adopters.

To study the impact of generative AI tools on private household behavior, we  utilize a novel source of data for studying generative AI use by households: Comscore's web browsing panel that allows us to comprehensively document the impact of generative AI on household online activities. We observe all browsing on household-owned computers. This data allows us to observe all browsing behavior of a large, diverse panel of U.S. households. A key advantage of this data over information on generative AI generated from surveys, lab settings, or chatbot conversation logs is that it allows us to track household-level changes in behavior over time---before and after the release of ChatGPT---and to observe changes in households' broader digital consumption patterns.

Our first key finding establishes that ChatGPT adoption among private households has been rapid — yet adoption is far from uniform. High-income and younger households adopt generative AI substantially faster than their low-income and older counterparts. Moreover, complementing the static adoption heterogeneity documented in prior survey evidence \citep{humlum2025}, we show that this gap is widening rather than converging over time, giving rise to a substantial and growing "generative AI divide" among U.S. households.

  
Second, we show that ChatGPT adoption can be in part predicted by households' pre-ChatGPT browsing patterns.  To show this, we first develop a measure of the ex ante usefulness of generative AI to a household based on the composition of its online browsing \textit{before} the release of ChatGPT. We use webscraping and LLMs to systematically classify the overlap between chatbot capabilities and the activities on a website, for essentially the universe of websites visited by U.S. households. We aggregate this website-level overlap across a household's complete browsing activity in January--December 2021, i.e., \textit{before} the release of ChatGPT, to arrive at our ex ante measure of potential benefits---the household's generative AI ``exposure''. A household has a higher exposure if a higher share of its browsing time was spent on sites with a high overlap with ChatGPT capabilities. We find that a household's expected benefit of using ChatGPT strongly predicts whether they actually adopt ChatGPT: a 1 SD higher exposure predicts a 2.5pp higher rate of having used ChatGPT by December 2024.

Third, our main finding shows how ChatGPT adoption \textit{changes} household behavior and welfare. Using LLMs to classify browsing activities, we categorize websites into those that are predominantly ``leisure'' sites and those that are associated with (non-market) ``productive'' digital activities, such as tasks related to building human capital (e.g., education and job search).  With this data, we can shed light on which types of online tasks households spend more or less time on after adopting ChatGPT. However, ChatGPT adoption may be endogenous to life events and household task demand. We address this concern by instrumenting adoption with the pre-ChatGPT exposure measure derived from 2021 browsing. Conditional on detailed demographic-by-region fixed effects and controls for broad browsing composition, this exposure provides plausibly exogenous variation in adoption.

We find that ChatGPT adoption causes households to spend more time on digital leisure activities while leaving the total time spent on productive online activities (including any time spent using ChatGPT) unchanged. In long-difference IV estimates, ChatGPT adoption raises total leisure browsing time by roughly 150 log points and increases the leisure share of browsing duration by about 30 percentage points. This finding suggests that using ChatGPT for digital household production tasks may substitute for time spent on productive activities on other websites, while also making households more efficient in completing those tasks. This leads to a reallocation of freed-up browsing time towards digital leisure activities, which suggests that ChatGPT adoption has positive welfare effects on households.

Fourth, to provide support for this mechanism, we analyze the high-frequency browsing context in which ChatGPT is used, based on the intuition that we can plausibly deduce what activities a person is doing inside ChatGPT by observing their online activities right before and after accessing the chatbot. Comparing the 30-minute browsing intervals around ChatGPT use to the browsing patterns of demographically similar households that never use ChatGPT, we establish that households predominantly utilize ChatGPT in the context of online activities and websites that are related to productive tasks (rather than leisure-oriented).  Moreover, we also document that the declines in  browsing time as a result of ChatGPT adoption are concentrated in website categories, such as search or news, that are highly exposed to substitution by generative AI. This evidence supports the idea that ChatGPT makes productive online tasks more efficient, which then allows households to reallocate freed-up browsing time to leisure. 


Fifth, because we observe time reallocation but not the ``output'' from online tasks, we need a model to translate browsing time shifts into implied productivity gains. We develop a quantitative model of household time allocation across digital tasks, adapted from \cite{aguiar2021}, that maps our empirical estimates into implied changes in household productivity. This mapping requires the Engel curve elasticities for leisure and productive digital activities as an input, which we estimate using an IV approach: we use exogenous variation driven by local precipitation shocks to estimate the response of leisure and browsing time to changes in total browsing time.

The resulting estimates imply about a doubling of the efficiency of productive digital tasks following household ChatGPT adoption (our preferred parameter inputs imply an efficiency gain of 76\%-176\%). These effects exceed the general magnitude of productivity effects documented in the workplace GenAI literature,\footnote{E.g., about 25\%-56\% improvement depending on job tasks and workers as shown in \cite{noy2023}, \cite{brynjolfsson2023}, \cite{dell2023}, and \cite{peng2023}. See details in Section \ref{subsec:efficiency_gain}.} consistent with the fact that household tasks are often performed by novices and may therefore have more potential upside in efficiency than market work done by specialists. It may also stem from the fact that many home internet activities involve extensive multi-site browsing that ChatGPT can largely substitute with a single conversational query. 


We note two limitations of our findings. First, our data captures households’ internet activity on home devices, but we do not observe offline, non-internet activities, and currently do not include activity on mobile devices. Although internet use on the included devices likely accounts for a substantial share of at-home activities—particularly entertainment, information acquisition, and other digital tasks—one should take caution when extrapolating our findings to households' total activities at home. Second, our analysis focuses on generative AI use ``at home'' and does not connect to workplace exposure. While recent survey evidence suggests that generative AI use outside of work is at least as prevalent as use on the job, understanding how home and workplace adoption interact remains a highly promising avenue for future research.

Our findings have important implications for the societal effects of generative AI. The use of ChatGPT in productive online activities, together with the associated increase in leisure time, provides compelling evidence that ChatGPT enhances household non-market productivity in addition to the labor market productivity effects documented in prior studies. Combined with the scale and speed of private household adoption that we document, these household-level benefits are likely to constitute a substantial share of the overall economic impact of generative AI and should be taken into account when designing regulation and quantifying the welfare effects of this new technology.

The size of the implied productivity changes and the growing generative AI divide that we document, highlight the importance of better understanding the dynamic effects of using generative AI: if use of the technology outside of work complements workplace productivity or affects human capital development, inequality in initial private adoption can exacerbate disparities in who benefits from generative AI. Our findings highlight the importance of targeting technology literacy efforts, training, or subsidies at groups with low take-up rates (such as older households) to enable a more even distribution of both the market and non-market productivity benefits of the technology.


\paragraph{Related literature.} Analyzing the household setting complements the existing literature in several ways. First, it provides revealed-preference evidence on what households actually do, complementing early insights on reported generative AI use in surveys (e.g., \cite{bick2024}) and on interactions with the technology in laboratory settings (e.g., \cite{kosmyna2025your}). Second, our focus on generative AI's impact ``at home'' complements a growing body of empirical work on its adoption and impact ``at work'' and within firms (e.g., \cite{brynjolfsson2023, schubert2025organizational, yotzov_firm_2026}). While some of these studies document a substantial disruption to jobs, including unequal productivity gains and job displacement,\footnote{See, for example, \cite{eisfeldt2023}, \cite{eisfeldt2024}, \cite{humlum2025}, \cite{chatterji_how_2025}, among others.} 
households' responses to these shocks—such as rebuilding human capital or reallocating time toward leisure—likely occur to a large extent at home. Third, our paper complements other work that has hypothesized that many of the most transformative impacts of generative AI lie \textit{outside} the workplace: it may reshape education \citep{khan2024}, social media interactions \citep{ghani2022}, entertainment content \citep{zhang2025}, and household production tasks such as financial planning \citep{lo2024} and shopping \citep{lei2025}.

As we document, ChatGPT affects household demand for digital services provided on other websites. Several recent studies examine changes in browsing behavior following the release of ChatGPT. \cite{lyu2025wikipedia} find that Wikipedia articles more similar to chatbot output experienced larger declines in views after ChatGPT’s release. \cite{padilla2025impact} show that informational Google searches decline after LLM adoption, while navigational searches do not; they also document declines in online ad exposure and visits to educational websites and Stack Overflow. Declines in Stack Overflow engagement are similarly documented by \cite{del2024large} and \cite{burtch2024consequences}, who interpret these patterns as evidence that users resolve programming questions directly with ChatGPT. Our study complements this literature by examining households’ overall browsing reallocation and documenting an increase in the leisure share of online activity.

Our findings on non-market household productivity complement evidence on generative AI’s productivity effects in labor markets and firms. In an experiment granting workers access to ChatGPT for professional writing tasks, \cite{noy2023} show that lower-performing workers benefit disproportionately. Management consultants \citep{dell2023}, software developers \citep{peng2023}, and call center employees \citep{brynjolfsson2023} all exhibit productivity gains, again with larger effects for initially less productive workers. Some studies document productivity effects outside traditional workplaces: \cite{yu2024}, for example, find improvements in academic writing proficiency among university students after ChatGPT’s release. \cite{jiang2025ai} show that generative AI exposure in \textit{market} work may reduce leisure, as more-exposed workers work longer hours. In contrast, we study private household use and its implications for non-market productivity and leisure.

Concerns about unequal adoption also arise in earlier generations of AI. \cite{mcelheran2024} document that firm-level AI adoption in 2018 was highly concentrated across cities and regions, describing an ``AI divide'' across geographies. \cite{humlum2025} find in Denmark that younger and less-experienced workers are more likely to use ChatGPT, while women and lower-earning workers are less likely to adopt. We extend this literature by documenting a ``generative AI divide'' in private household adoption, identifying  and measuring household exposure as a predictor of gaps in adoption, and quantifying the implications for non-market productivity differences.

Our findings suggest that ChatGPT adoption improves household welfare by raising non-market productivity and enabling greater leisure consumption online. Our conceptual framework for time reallocation in response to task-level productivity shocks builds on \cite{aguiar2021} and relates to empirical work by \cite{brynjolfsson2025gdp}, who measure the consumer surplus generated by free digital goods.

The remainder of the paper proceeds as follows. Section \ref{sec:data} describes the Comscore browsing data and our LLM-based website classification and exposure measures. Section \ref{sec:measure_adoption} documents adoption patterns, the predictive power of pre-ChatGPT exposure, and the generative AI divide. Section \ref{sec:Impact} estimates the causal impact of adoption on browsing behavior using an IV long-difference design, and Section \ref{sec:mechanism_empirics} provides mechanism evidence from high-frequency browsing context around ChatGPT use. Section \ref{sec:conceptual-framework} maps the reallocation into implied productivity gains in a simple time-allocation framework, and Section \ref{sec:conclusion} concludes.

\section{Data and Measurement} \label{sec:data}

\subsection{Comscore Internet browsing data}  \label{subsec:Comscore_data}

Studying the impact of generative AI tools on private household behavior has to overcome severe data constraints: Ideally, we would want data for a broad set of households that is informative about a broad range of behaviors \textit{before} adopting generative AI, the timing and nature of generative AI usage, and changes in behavior \textit{after} generative AI adoption. Typical data sources used to evaluate household economic impacts tend to be limited in one or more dimensions as a result of data collection methods and privacy restrictions: Surveys and guided experiments give detailed information on usage, but only at specific moments in time or hard-to-extrapolate contexts---and tend to collect limited information on what people do when not using generative AI tools. Moreover, respondents may underreport their AI use due to social desirability bias \citep{ling2025}. Similarly, microdata collected by generative AI companies, such as recent studies by Anthropic \citep{handa2025} and OpenAI \citep{chatterji_how_2025} provide a wealth of information about what people are doing inside the chatbots, but with little information about their broader behaviors and outcomes when not using a chatbot.

Our analysis uses the microdata for a large panel of U.S. households' Internet browsing activity provided by Comscore. Comscore is a U.S. based, publicly-traded media measurement and analytics company that specializes in characterizing the online behavior and demographics of different websites' user bases. To do so, Comscore reaches out to a large number of active Internet-using households in the U.S. who, in exchange for certain incentives, allow the company to collect comprehensive data on their browsing activities.\footnote{These incentives can entail free software, applications, or utilities, see \href{https://assets.pewresearch.org/wp-content/uploads/sites/13/2014/03/comScore-Media-Metrix-Description-of-Methodology.pdf}{Comscore Media Metrix Description of Methodology}. Each participating household has an application installed on its computer that tracks Internet usage. Comscore utilizes these data to produce estimates of online usage patterns as well as for other studies of online user behavior conducted internally or by their clients.}

This data allows us to observe all browsing behavior of a large, diverse panel of U.S. households. Three particular features make it uniquely suited to studying generative AI's impacts on households: (1) It is \textit{high-frequency and comprehensive}, containing all of a household's browsing activity on a given machine, which, importantly, includes activities before and after interacting with generative AI tools. (2) It is a \textit{multi-year panel} spanning both pre-adoption and post-adoption household behavior, allowing us to more plausibly estimate underlying drivers and the impact of adoption. (3) It includes large sub-samples of households across the income and age distribution. This allows us to characterize \textit{heterogeneity} in the take-up and effects of generative AI tools.

The microdata provides detailed Internet browsing activities of tens of thousands of households per year, where each household is identified by a unique identifier \textit{machine\_id}. To capture active meaningful human visits to a website, Comscore filters out short visits that are less than 3 seconds, times out sessions after 30 minutes of inactivity, and removes non-human visits using a multi-layered approach to detect sophisticated bots and fraud. For each qualified website visit, the data provides the timestamp, the \textit{duration} of the visit in seconds, the URL of the website, a classification code of the URL, and a unique browsing session id that groups the website visit with other adjacent visits in a continuum of browsing period. The data also provides regularly-updated demographic information of the households, including the combined income of the household in 8 bins, the age of the household head in 11 bins, the geographic location in cities, and the number of members in the household.\footnote{See data manual at \href{https://wrds-www.wharton.upenn.edu/documents/1008/Comscore_Web_Behavior_Database_August_2022.pdf}{Comscore Web Behavior Database}. Compared to the synthesized Comscore web behavior data available on WRDS and used in prior studies, our raw microdata has three advantages: First, the synthesized data includes a curtailed set of about 50,000 households each year, while our raw data includes all households. Second, the synthesized data groups multiple adjacent website visits into a browsing session and provides only the main domain of the browsing session, while our raw data shows all website URLs clicked within the browsing session. Third,  our raw data timely reflects any update on households' demographics, such as income, while the synthesized data does not.}

\paragraph{Sample.} Our baseline sample includes household-level browsing data from 2021 to 2024, covering periods before and after the release of ChatGPT on November 30, 2022. We set the twelve months from January to December 2021 as the \textit{benchmarking period}.\footnote{We define the benchmarking period as ending in December 2021 to ensure households' browsing behavior is not contaminated by any anticipation of the ChatGPT release. This also prevents short-run household shocks from affecting both the benchmarking period and the post-ChatGPT browsing activity.}  To be included in our baseline sample, we require the household to have non-missing information on income and age, have browsing activities for at least six months during the benchmarking period, and have browsing activities for at least one month after the release of ChatGPT. We focus on private household-owned computers by further removing a small set of machines that are labeled as work-owned. Internet Appendix Table \ref{tab:sample_restrictions} provides more details on the impact of each filter on the sample size.


\paragraph{Sample representativeness.} While Comscore designed the recruiting process to capture a representative sample of Internet users in the U.S., we note that maintaining a perfectly representative sample at a particular time, such as during the benchmarking period in our study, can be challenging. Table \ref{tab:sample-distribution} compares the composition of our baseline sample with the U.S. Internet-using population from the 2022 American Community Survey (ACS) database. We observe that our sample tends to over-represent households at the low (<\$60K) and high ends (>\$200K) of the income distribution, and under-represent middle-income (\$60K-200K) households. In terms of age, our sample somewhat over-represents the upper middle-age households (45-54 years old) and under-represents younger households (25-44 years old). 

To account for these sample discrepancies, we construct a weight factor for each household in our baseline sample so that the joint income and age distribution of our sample households in November 2022 matches that of the 2022 ACS, and we present all statistics using this weight. In addition, to avoid our measures of household demographics being endogenously influenced by household adoption of ChatGPT, we keep households' demographics fixed as of November 2022 in most of our analysis unless otherwise noted.

\begin{table}[htbp]
    \caption{\\ \centering  \textbf{Distribution by income and age, Comscore sample vs. ACS}}

\vspace{-0.1cm} \label{tab:sample-distribution} \footnotesize  This table shows the distribution with respect to household income and age in both our main Comscore sample (first column), where income and age are from the most recent month prior to November 2022 in which the panelist updated their demographic information, and in the U.S. Census American Community Survey (second column) among all Internet-using households in the 2022 survey.
\vspace{.1cm}

    \centering
 \begin{adjustbox}{max width=\textwidth}   
\begin{tabular}{@{}lcc@{}}
    \toprule
    \textbf{} & \textbf{Comscore sample} & \textbf{ACS} \\
    \midrule
    HH Income (\% of Panelists) \\ 
    \hspace{3mm} $<$\$25K & 21.25 & 13.94 \\
    \hspace{3mm} \$25K - \$40K & 21.38 & 10.45 \\
    \hspace{3mm} \$40K - \$60K & 13.99 & 14.33 \\
    \hspace{3mm} \$60K - \$75K & 4.67 & 9.89  \\
    \hspace{3mm} \$75K - \$100K & 7.34 & 13.38 \\   
    \hspace{3mm} \$100K - \$150K & 9.96 &  17.50 \\     
    \hspace{3mm} \$150K - \$200K & 2.49 & 9.19 \\       
    \hspace{3mm} $>$\$200K & 18.93 & 11.33 \\
    \\
    Head of HH's Age (\% of Panelists) \\
    \hspace{3mm} 18 - 24 & 7.19 & 4.09 \\
    \hspace{3mm} 25 - 34 & 9.78 & 16.14 \\
    \hspace{3mm} 35 - 44 & 14.06 & 18.50 \\
    \hspace{3mm} 45 - 54 & 38.63 & 17.69 \\
    \hspace{3mm} 55 - 64 & 16.68 & 18.64 \\
    \hspace{3mm} 65 and over & 13.53 & 24.93 \\    
    \bottomrule
    \end{tabular}
\end{adjustbox}
\end{table}

\subsection{Measuring household browsing behavior}
Here, we introduce the main outcome measures of interest that we examine in this study. We begin by introducing two underlying measures of browsing intensity at the website level. Our data includes millions of websites. To draw economic inferences, we aggregate the intensity measures to a broader category level in certain analyses, and we further develop our own categorization of websites into productive versus leisure groups for other main analyses.

\paragraph{Browsing intensity measures.} Most of our analyses focus on households' Internet browsing behavior at the monthly frequency. We focus on two measures of a household's browsing intensity on a website in a month. The first is the total number of seconds household $i$ spends on website $j$ during month $t$, labeled as \textit{duration}$_{j,i,t}$. This measure captures both intensive and extensive margins of a household's visit to the website. Our second measure of browsing intensity counts the number of times a household visits the website during the month, labeled as \textit{visit}$_{j,i,t}$. In some illustrations, we aggregate the browsing intensity measures of websites to broad website categories provided by Comscore, which capture the broad topic of the website's content (e.g., ``social media'').


\paragraph{Categorizing productive vs. leisure websites.} 
A key focus of our study is on whether technology changes households' productivity in different \textit{tasks} that can be accomplished through Internet browsing. In particular, in line with the common breakdown of time use into `market work', leisure and `non-market work' activities in time use studies (e.g., \cite{aguiar2021}), we want to distinguish whether the home browsing activity that we observe is for `productive' tasks (i.e., non-market work activity on the Internet) or `leisure'. To the best of our knowledge, there is no existing approach for labeling websites based on the household tasks that they are associated with, so we develop a novel methodology for classifying domains using a large language model (LLM). We focus on the top 160K domains by visits between 2022 and 2024, which account for nearly all browsing activity during this period.  

The methodology, as illustrated in Figure \ref{fig:llm} and detailed in the Internet Appendix \ref{ia.sec:appx_methodology}, involves the following three steps: First, for each domain, we obtain a description of each website by using a webscraper to try and access each domain address to retrieve the description and keywords that the website’s owner has embedded in ``meta tags’’ in the website’s source code. Second, we submit a prompt to an LLM (OpenAI’s GPT-4.1 mini model accessed through the Azure API) that asks the LLM to consider the domain, title of the site, and the description from the website's source code and come up with the 5 ``main activities'' that the website could be used for households.




Third, we submit the website descriptions and list of 5 key activities to an LLM with a prompt that asks for two labels: (1) A classification of the domain’s main usage into the categories ``Productive’’, ``Leisure’’, and ``Ad platform/ CDN’’ (content delivery network). (2) Whether the domain is ``mixed-use’’ or ``single-use’’ with regard to these usage types.\footnote{In this context, \textit{Mixed-use} domains will include, for instance, email providers and other platforms that are commonly used for both productive and leisure purposes.}  Our definition of productive and leisure activities closely follows the categorization of general (non-Internet) activities in \cite{aguiar2013}: productive uses are those potentially relating to market work, education, childcare, non-market work, civic activities, shopping and personal health care, while leisure includes gaming, social activities, TV, movies, and reading for personal interests.

\begin{figure}[!hbt]
\caption{\textbf{Website duration shares by assigned browsing purpose category}} 
\label{fig:leisure_prod_shares} \footnotesize This figure shows the share of browsing visits and duration in each assigned website purpose category in 2021 Comscore data.

    \centering 
\includegraphics[trim={0.5cm 0 0.5cm 0},clip, width=0.4\textwidth]{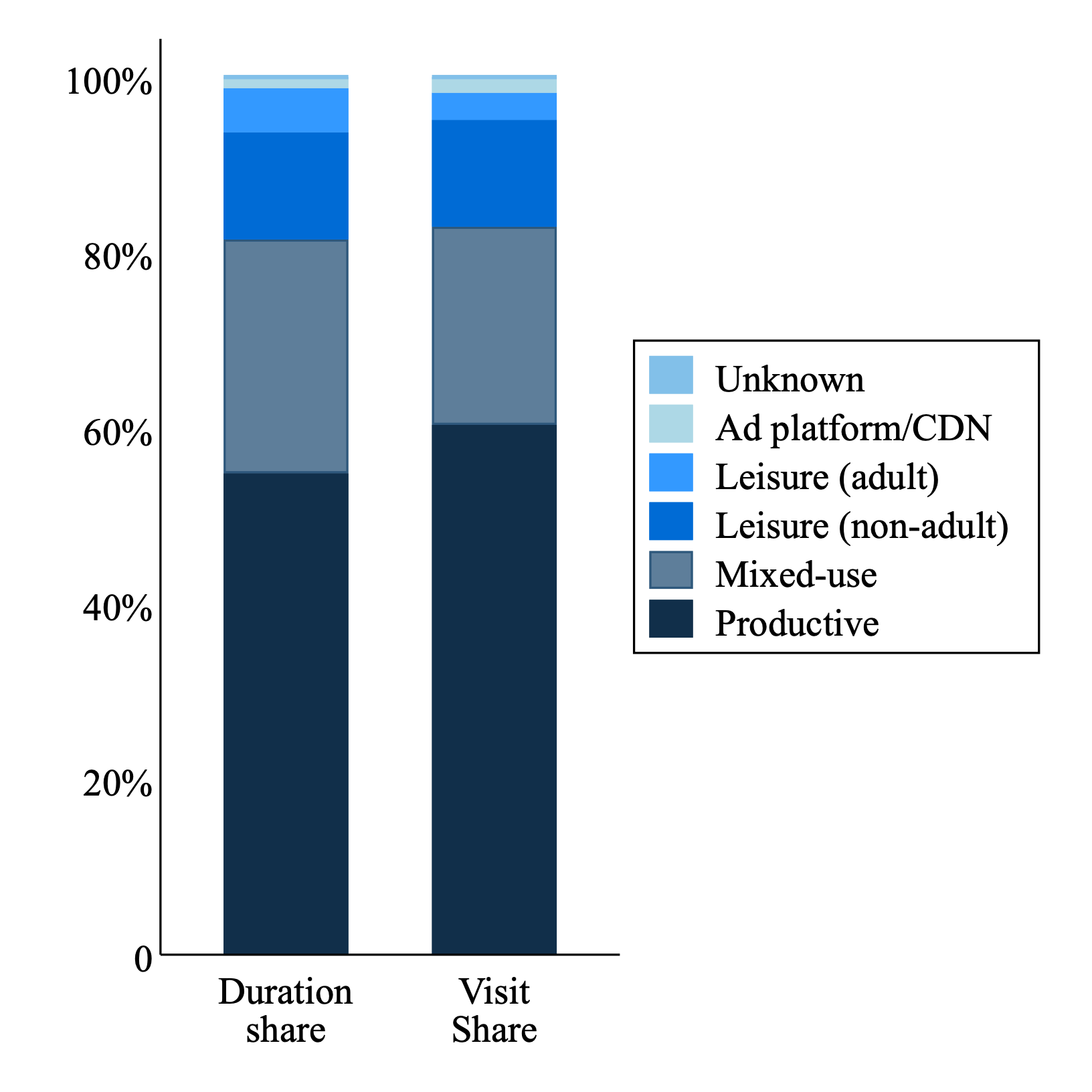} 
\end{figure}

Importantly, we refined and validated our prompt to ensure that these labels are applied as expected: Two of the authors manually labeled the top 70 domains by page views, and we adjusted the prompt language until the LLM’s answers substantially agreed with our own labels on this validation set. Based on this manual labeling, we also adjusted the classification categories to capture ad platforms and content delivery networks as a separate category. These are not customer-facing sites in the sense that we are interested in, and they mechanically generate a large number of page views, so by putting them into a separate category, we ensure that they do not get mislabeled as productive or leisure uses of a household’s time.

After scaling this process to the full set of 160K domains, we derived a final label for each domain as having a main use that is \textit{Productive}, \textit{Leisure}, \textit{Mixed} or \textit{Ad platforms/CDN}. Based on this categorization, we can measure each household's browsing intensity in productive versus leisure websites within the month, which is the key focus of our study on economic outcomes.   The distribution of total browsing activity over these purpose categories is shown in Figure \ref{fig:leisure_prod_shares}: both by visits and by browsing duration, the majority of browsing activity is classified as `productive', followed by `mixed-use' activity. For illustration, we also split out adult vs. non-adult content leisure sites (but otherwise combine these in the rest of the paper). To illustrate what these labels mean, we break out the top 5 websites by duration for the 3 largest of these categories in Table \ref{tab:top5-bycat}.


\begin{table}[!hbt]
\caption{\\ \textbf{Top 5 domains by duration by assigned browsing purpose category in 2021}} 
\footnotesize  This table shows the 5 biggest domains by duration in each of the three largest assigned browsing purpose categories in 2021 Comscore data. \label{tab:top5-bycat}

\centering \footnotesize
\begin{tabular}{@{} l c c c c >{\bfseries}c @{}}

 \addlinespace
 \toprule
 Domain  &  $\substack{\text{\% of}\\ \text{Category} }$  &  $\substack{\text{\% of}\\ \text{All Sites} }$ & Automated description  \\
                   \midrule
                   \addlinespace
 \multicolumn{4}{c}{\textbf{Productive Duration Top 5}} \\
\addlinespace
 google.com&24.97&13.69&widely used search engine \\
amazon.com&3.74&2.05&comprehensive online retail platform \\
bing.com&3.50&1.92&smart search engine \\
instructure.com&1.95&1.07& educational technology company \\
live.com&1.93&1.06&Microsoft-operated web portal \\
  
 \addlinespace
 \multicolumn{4}{c}{\textbf{Mixed-use Duration Top 5}} \\
 \addlinespace
 youtube.com&51.02&13.59&video-sharing platform \\
msn.com&33.78&9.00&comprehensive web portal \\
yahoo.com&14.24&3.79&comprehensive web portal \\
playtika.com&0.60&0.16&gaming company \\
lycos.com&0.15&0.04&comprehensive web portal \\
  
\addlinespace
 \multicolumn{4}{c}{\textbf{Leisure (non-adult) Duration Top 5}} \\
 \addlinespace
 facebook.com&15.48&1.89&social networking platform \\
instagram.com&4.09&0.50&social media platform \\
twitch.tv&1.91&0.23&live streaming platform \\
espn.com&1.87&0.23&comprehensive sports media website \\
twitter.com&1.48&0.18&social media platform \\
  
 \bottomrule
\end{tabular} 

\end{table}


\subsection{Measuring household exposure to GenAI}
\paragraph{Website activity exposure to generative AI.} To evaluate the degree to which a household's Internet use patterns are likely impacted by the release of ChatGPT, we adapt a methodology from prior work in labor market exposure measures, such as \cite{eisfeldt2023}, to the context of Internet browsing behavior. Specifically, we focus on the overlap between the activities enabled by websites and the functionalities of ChatGPT. First, we take the website descriptions and lists of activities that we previously generated and submit them individually to an LLM, together with a rubric that defines which types of website activities ChatGPT can be used for.\footnote{We develop this rubric by summarizing the clusters of user activities observed by \cite{handa2025} in their analysis of actual chatbot activities observed for users of Anthropic's Claude chatbot. This data provides an empirical starting point for what activities chatbots similar to ChatGPT are actually used for. To account for the additional capabilities of OpenAI's ChatGPT, we additionally incorporate multimodal capabilities that were  announced for ChatGPT in September 2023. See the announcement at: \url{https://openai.com/index/chatgpt-can-now-see-hear-and-speak/}}. The LLM assigns each activity a binary exposure label based on whether ChatGPT can perform or complement the website’s tasks. The distribution of the types of activities that tend to be labeled as ``exposed'' by this method is shown in Figure \ref{fig:exposuretypes}.

We then calculate a continuous exposure score by measuring the share of activities on each website that can be substituted by ChatGPT, with these scores ranging from 0 (no exposure) to 5 (high exposure). Table \ref{tab:browsesharesbyexposure} lists the websites that make up the highest browsing duration shares of each exposure category. The lowest exposure categories (0 or 1 out of a maximum of 5) are dominated by social media platforms like Facebook, Instagram, and Twitch; video streaming services like Netflix and YouTube, and other service sites like AOL, Craigslist, and Weather.com. The high end of the exposure score spectrum (4 or 5 out of 5) is dominated by education and knowledge sites, like Wikipedia, Quizlet,  or Edgenuity. The middle range of exposure (scores of 2 or 3) consists of many sites that often are multi-purpose (e.g., MSN, Yahoo), search engines  (e.g., Bing, Google), or platforms for shopping and trading, such as Amazon, Robinhood, and eBay.

\begin{table}[t!]
    \caption{\\ \centering  \textbf{Top 5 websites by GenAI exposure}}

\vspace{-0.1cm} \footnotesize  This table shows the top 5 websites by browsing duration in each GenAI exposure category defined by how many of the top 5 website activities are exposed to generative AI. 

\label{tab:browsesharesbyexposure}
\vspace{0.1cm}
\small
\centering
    \begin{tabular}{lc|lc|lc}%
    \multicolumn{2}{c}{\textbf{0/5 Exposure}} & \multicolumn{2}{c}{\textbf{1/5 Exposure}}  & \multicolumn{2}{c}{\textbf{2/5 Exposure}}   \\   
     \cmidrule(lr){1-2}  \cmidrule(lr){3-4}  \cmidrule(lr){5-6}  
    {Website} & {Share} &   {Website} & {Share} &   {Website} & {Share}  \\ \hline   
                       facebook.com&25.18&youtube.com&57.18&msn.com&33.84\\
weather.com&4.24&aol.com&4.34&yahoo.com&14.27\\
twitch.tv&3.11&instagram.com&2.10&amazon.com&7.71\\
pluto.tv&1.68&craigslist.org&1.14&bing.com&7.23\\
netflix.com&1.45&veryfast.io&0.89&robinhood.com&2.91\\
  \\
\midrule 
    \multicolumn{2}{c}{\textbf{3/5 Exposure}} & \multicolumn{2}{c}{\textbf{4/5 Exposure}}  & \multicolumn{2}{c}{\textbf{5/5 Exposure}}   \\   
    \cmidrule(lr){1-2}  \cmidrule(lr){3-4}  \cmidrule(lr){5-6}  
      {Website} & {Share} &   {Website} & {Share} &   {Website} & {Share}  \\ \hline
                       google.com&57.03&intuit.com&6.83&wikipedia.org&7.93\\
instructure.com&4.46&office.com&2.93&gingersoftware.com&5.45\\
live.com&4.40&edgenuity.com&2.93&wwnorton.com&3.06\\
ebay.com&2.82&indeed.com&2.91&quizlet.com&2.81\\
microsoft.com&1.33&reddit.com&2.28&mathxl.com&2.34\\
  
    \end{tabular}
\end{table}

 We show the distribution of these labels in Figure \ref{fig:exposureshares}: Panel A shows that 39\% of all household browsing activities are exposed to ChatGPT (weighting activities by the duration of time spent on their websites).\footnote{Implicitly, this assumes that all activities on a website are happening simultaneously, as we cannot allocate browsing duration within a website to different activities.} If we aggregate exposure to the website level, we obtain the distribution shown in Panel B: 16\% of website browsing duration involves no activities that ChatGPT could be useful for, and only 3\% of websites feature \textit{only} exposed activities. The majority of websites, by browsing duration (about 55\%), are exposed in 40-60\% of the activities on the website. Panel A of Appendix Figure \ref{fig:exposure_by_activity} shows the distribution of website types in each exposure category: the majority of unexposed websites are entertainment sites, while the more exposed sites are comprised of education,  lifestyle, and general information sites to a disproportionate extent, according to our methodology. Sites related to productive home activity (which include travel booking, shopping, health information, government services, home renovations, and technology information) make up a substantial share of all but the least exposed category. Panel B of Appendix Figure \ref{fig:exposure_by_activity} shows that most social media and entertainment websites have low generative AI exposure (score of 0 or 1), while most business, education, job search, or lifestyle sites are categorized as having high exposure (score of 4 or 5). Some categories show large variation in the exposure of websites in that group, with a lot of variation in exposure across sites within the general information, search, personal finance, and productive home activity categories.

\paragraph{Household exposure to generative AI.}
The exposure label for each household is then derived from the exposure of the domains they visit, weighted by the time spent on each site.
The key intuition behind this approach is that households whose browsing activities are more aligned with tasks that generative AI can assist with are more likely to benefit from or adopt such technologies. We compute household-level exposure by summing the activity-based exposure labels for each website visited by the household, adjusting for the proportion of total browsing time spent on each site. This method assumes that a higher share of activities on a website that generative AI can perform increases the likelihood that the household will find the technology beneficial. To ensure the accuracy of this exposure measure, we use browsing data from the January  to December 2021 period, before the public release of ChatGPT, to avoid any endogeneity issues from early adoption effects.

Specifically, we define a household's exposure to generative AI as the expected share of highly generative AI-exposed websites in its Internet browsing. That is, we compute
\begin{align} \label{eq:hhexposure}
    \text{HHGenAIExp}_i = \sum_j \phi_{ij} \mathbbm{1}[E_j \in \{4,5\}], 
\end{align}
where $E_j$ is the count of activities (out of 5) on website $j$ that are generative AI exposed, and $\phi_{ij}$ is the share of a household's browsing duration during a pre-period that is spent on website $j$. Here, we focus on websites with high exposure (4 or 5 out of 5 activities), as switching costs are likely to lead households to require a minimum level of substitutability (and implied productivity benefits) to want to use ChatGPT instead of their previous websites of choice.  Empirically, this measure allows us to examine the relationship between pre-existing household exposure to generative AI and subsequent adoption of ChatGPT.

\subsection{Measuring GenAI adoption}

To capture when a household first starts using generative AI technology, we focus on ChatGPT, the most popular LLM chatbot. ChatGPT use is a good proxy for generative AI adoption by households for several reasons. First, ChatGPT’s public release on November 30, 2022, marked a breakthrough that significantly enhanced the accessibility and capabilities of generative AI, as seen in widespread media attention and immediate market reactions \citep{eisfeldt2023}. This release provided a sharp break in households' ability to access sophisticated GenAI tools, making it a key event in our analysis of changes in browsing behavior before and after November 2022. Second, surveys such as \cite{bick2024} and our browsing data confirm that ChatGPT adoption spread rapidly, particularly outside the workplace, making it easy to identify its use consistently in the Comscore data. Lastly, while other GenAI tools like Anthropic’s Claude and Google’s Bard emerged after ChatGPT, we focus on ChatGPT for simplicity, as it remains the most widely used tool among our sample, with most panelists who use alternatives first trying ChatGPT.

We construct two measures of households' adoption of ChatGPT.  The first one captures whether a household has \textit{ever} used ChatGPT. That is, we want to know when a household tries out the technology for the first time, and we define a measure of having ever used ChatGPT that is defined by whether we have observed that household accessing the ChatGPT site at any point in the past. We deliberately do not require a household to continuously use ChatGPT, as it is possible that households eventually either switch to other generative AI tools, or access ChatGPT through means other than browsing on a home computer (e.g., through a cellphone app)---but we would not want to label those households as not using generative AI.

Our second measure of interest is the intensity of the household's usage of ChatGPT, which is the number of seconds the household browsed the ChatGPT website within the time period.

\section{Household Exposure and Generative AI Adoption} \label{sec:measure_adoption}

In this section, we first establish that ChatGPT adoption patterns over time in our Internet browsing data are consistent with evidence from other data sources, and vary substantially with household demographics. Then, we show that a substantial share of this heterogeneity in household adoption can be predicted by differences in the share of substitutable website activities in pre-ChatGPT browsing behavior.

\subsection{Household adoption of ChatGPT over time}

In this section, we document the extent and timing of generative AI adoption by households based on their browsing activity. Previous studies show a significant share of households use ChatGPT outside of work: \cite{bick2024} finds that 35.5\% of respondents used generative AI outside work in the second half of 2024, with 18\% using ChatGPT in the past week, while Pew Research reported 18\% of U.S. adults had used it in March 2023 \citep{pewresearch2025}. Our approach offers new evidence on adoption speed by tracking actual ChatGPT usage in household browsing, distinguishing between occasional and regular use, and linking usage patterns to other browsing behaviors and demographics. However, a limitation of our data is that it only captures use via websites on home computers, likely underestimating total generative AI usage.

\begin{figure}[!htb]
\caption{\textbf{Household ChatGPT adoption over time}} 

This figure shows the share of households  that have ever used ChatGPT (panel A) and the share of total browsing duration that is spent using ChatGPT (panel B) based on the Internet browsing activity in Comscore data. 

\label{fig:ts_useadopt} 
      \centering
\begin{subfigure}{.49\textwidth}
  \centering
    \includegraphics[width=\linewidth]{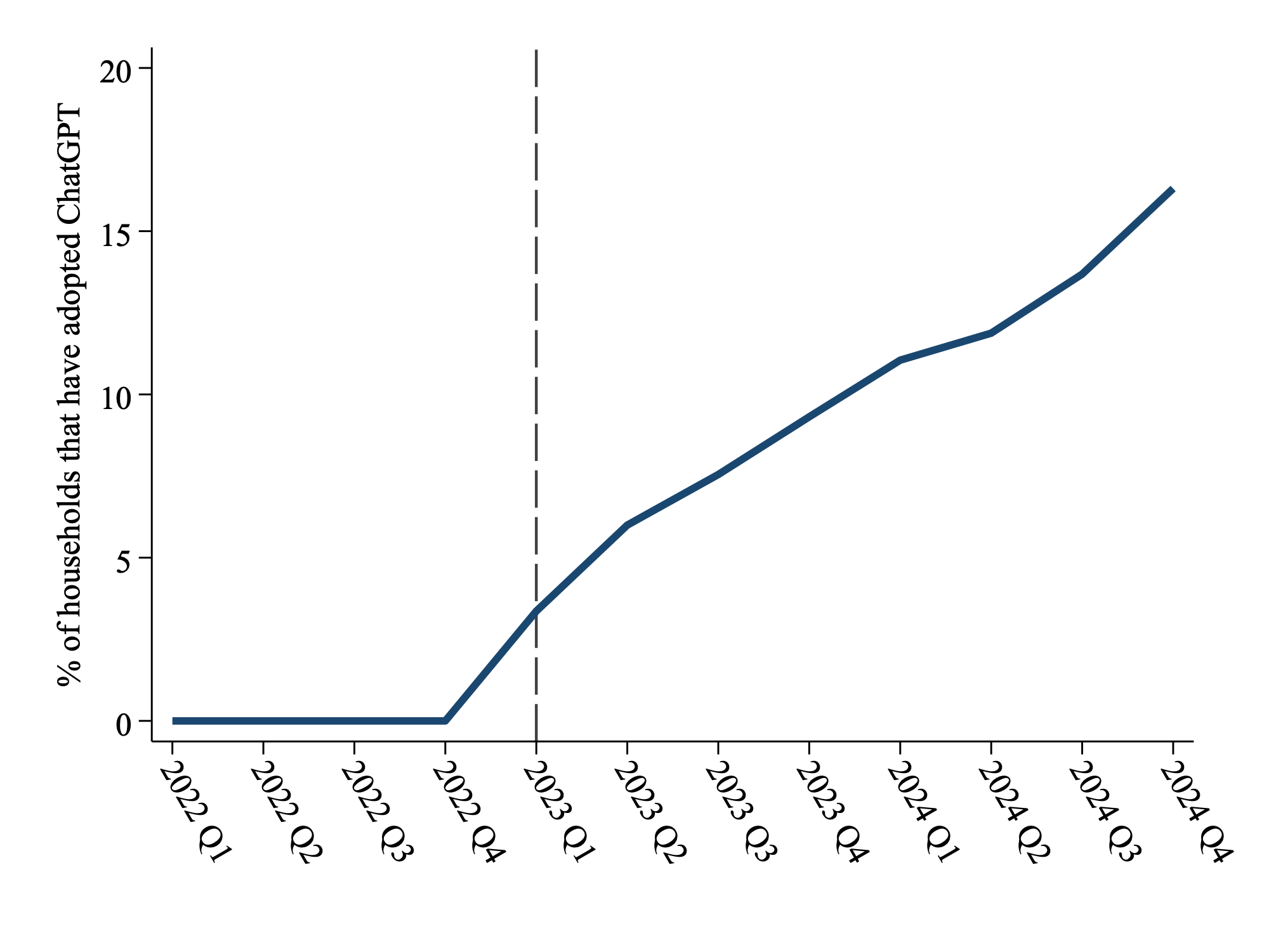}
  \caption{Share that has used ChatGPT}
\end{subfigure} 
\begin{subfigure}{.49\textwidth}
  \centering
  \includegraphics[width=\linewidth]{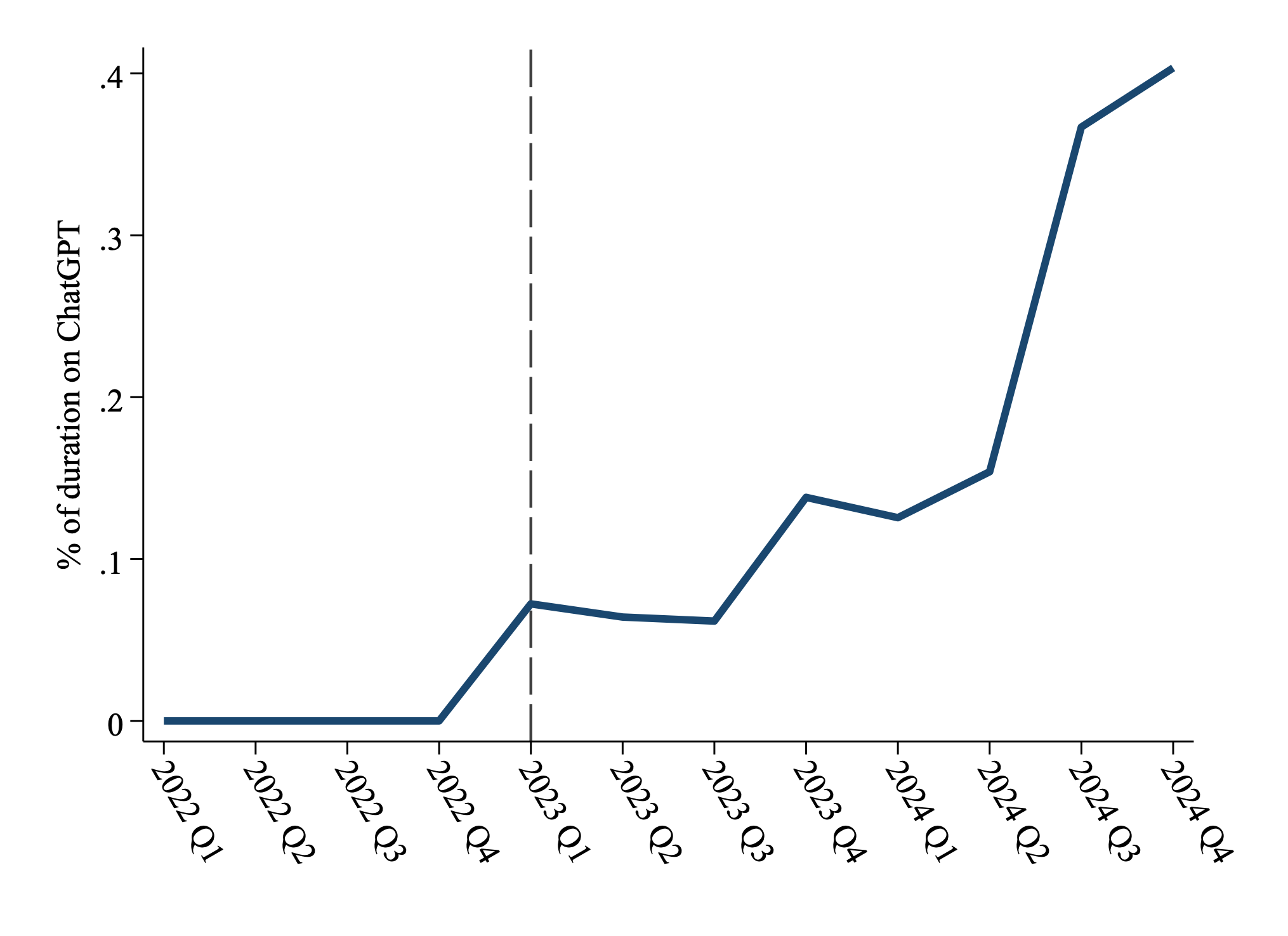}
  \caption{Share of browsing duration on ChatGPT}
  \label{fig:sub2}
\end{subfigure}
\end{figure}

\paragraph*{Household ChatGPT adoption over time.} We plot the time series of ever using ChatGPT and the share of browsing time spent using the chatbot in Figure  \ref{fig:ts_useadopt}: the share of all households that have ever used ChatGPT rises from zero to 9.3 pp in Q4 2023  to 16.3 pp by Q4 2024, almost doubling within a year. The share of total browsing time spent on ChatGPT also captures the intensive use margin and shows an even steeper increase during 2024, rising from 0.14 pp in Q4 2023 to 0.40 pp by Q4 2024 (see Table \ref{tab:use_by_dem} for summary statistics).

\paragraph*{Comparison to other estimates of adoption.} An important validation of our data is whether our time series provides estimates that are comparable to snapshots of household use gleaned from surveys.
Our estimated rates of ChatGPT adoption are comparable to other sources based on surveys:  In a February 2024 survey, Pew Research found that 23\% of all adults had used ChatGPT, 17\% of all adults had used it for entertainment, and 17\% to learn something new, while only 12\% of all adults (20\% of those who were employed ) had used it for a work task.  This marked an increase from July 2023, when the same survey found 18\% of all adults had used the tool, 15\% for entertainment and 7\% at work  (12\% of the employed ). This is consistent with our measure of use representing both a lower bound in terms of all possible channels through which a household might access ChatGPT, and also focusing on household use outside of work, which is necessarily a subset of all use, as not all households that use ChatGPT at work will also use it privately. The fact that the measure of having used ChatGPT shows both similar rapid growth patterns and orders of magnitude of usage is reassuring.

\begin{table}[htbp]
\caption{\\ \centering \textbf{ChatGPT exposure and use over time in browsing activity: by age \& income}}

\vspace{-0.1cm} \label{tab:use_by_dem} \footnotesize  The table shows average rates of ChatGPT adoption for all households, and by age and income groups. See text for definitions of income and age categories. The sample for this analysis is the same as the regression sample in Table \ref{table:2sls-bycat}. \vspace{.2cm}
 
 \centering
\begin{adjustbox}{max width=\textwidth}
\begin{tabular}{@{}l*{8}{c}@{}}
\toprule
 & &\multicolumn{3}{c}{By Income} &  \multicolumn{3}{c}{By Age} \\
   \cmidrule(lr){3-5} \cmidrule(lr){6-8}
   \addlinespace

   & All & Low & Middle & High & Young & Middle & Old \\
      \cmidrule(lr){2-2} \cmidrule(lr){5-5}  \cmidrule(lr){3-3} \cmidrule(lr){4-4}  \cmidrule(lr){6-6}  \cmidrule(lr){7-7} \cmidrule(lr){8-8}
                                        &\multicolumn{1}{c}{(1)}   &\multicolumn{1}{c}{(2)}   &\multicolumn{1}{c}{(3)}   &\multicolumn{1}{c}{(4)}   &\multicolumn{1}{c}{(5)}   &\multicolumn{1}{c}{(6)}   &\multicolumn{1}{c}{(7)}       \\ 
                   \midrule
\addlinespace
 HHGenAIExp \%&        10.7&        10.0&        10.5&        11.2&        11.7&        11.2&         9.9\\
  
 \addlinespace
\emph{Used ChatGPT \%}  &&&& \\
 Q4 2023     &         9.3&         7.1&         8.1&        12.0&        14.1&        11.3&         6.3\\
  
  Q2 2024     &        11.9&         8.8&        10.4&        15.3&        18.0&        14.5&         7.9\\
  
   Q4 2024     &        16.3&        13.3&        14.3&        20.3&        24.1&        21.0&        10.6\\
  
 \bottomrule
\end{tabular}
\end{adjustbox}
\end{table}

\begin{figure}[!hbt]
\caption{\textbf{ChatGPT adoption by household income and age}} 

\small This figure shows the share of households in each demographic group  that have ever used ChatGPT. \\
\label{fig:ts_use_by_demo} 

    \centering
\begin{subfigure}{.49\textwidth}
  \centering
  \includegraphics[width=\linewidth]{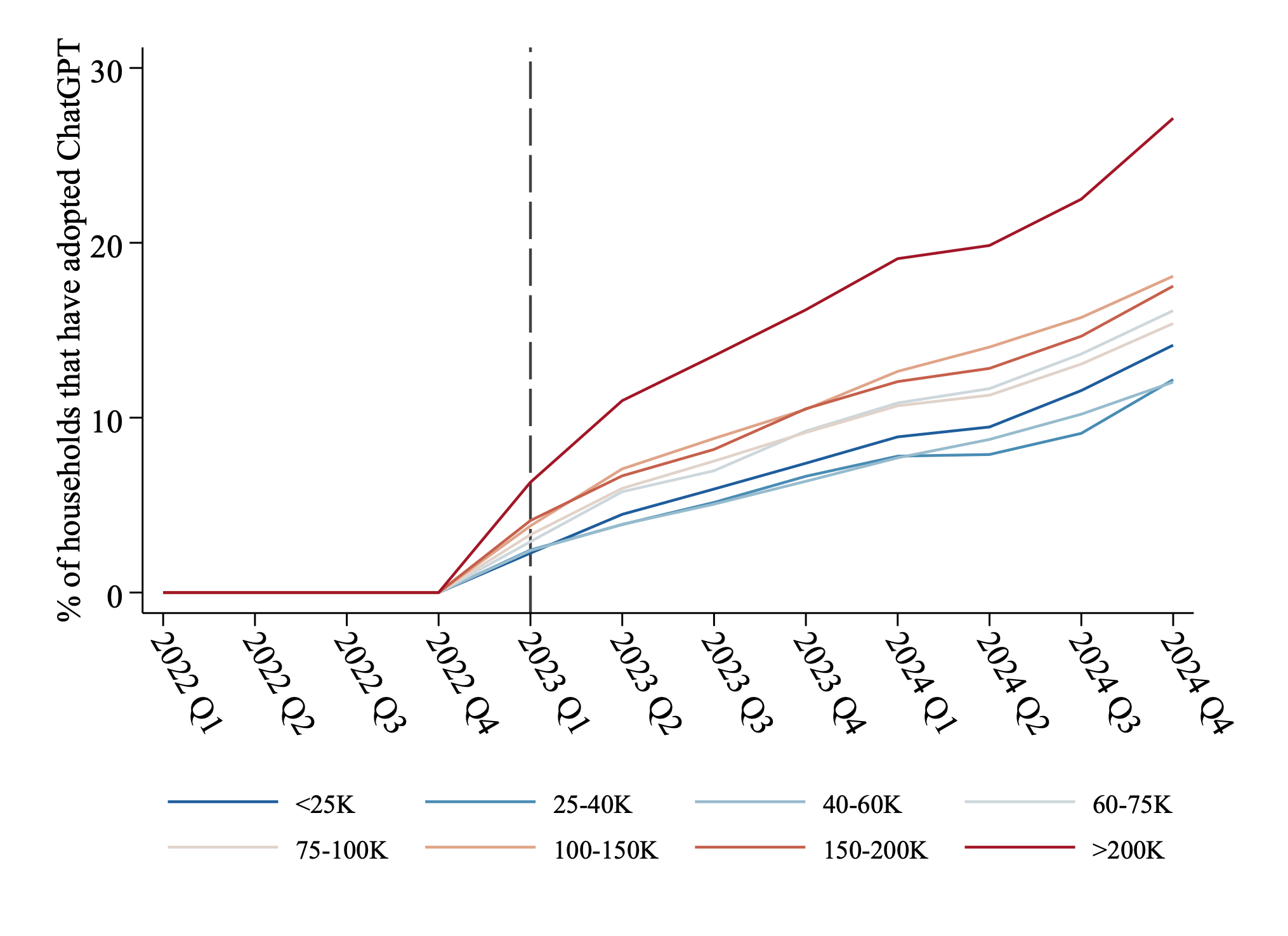}
  \caption{ChatGPT adoption by income bucket}
\end{subfigure} 
\begin{subfigure}{.49\textwidth}
  \centering
  \includegraphics[width=\linewidth]{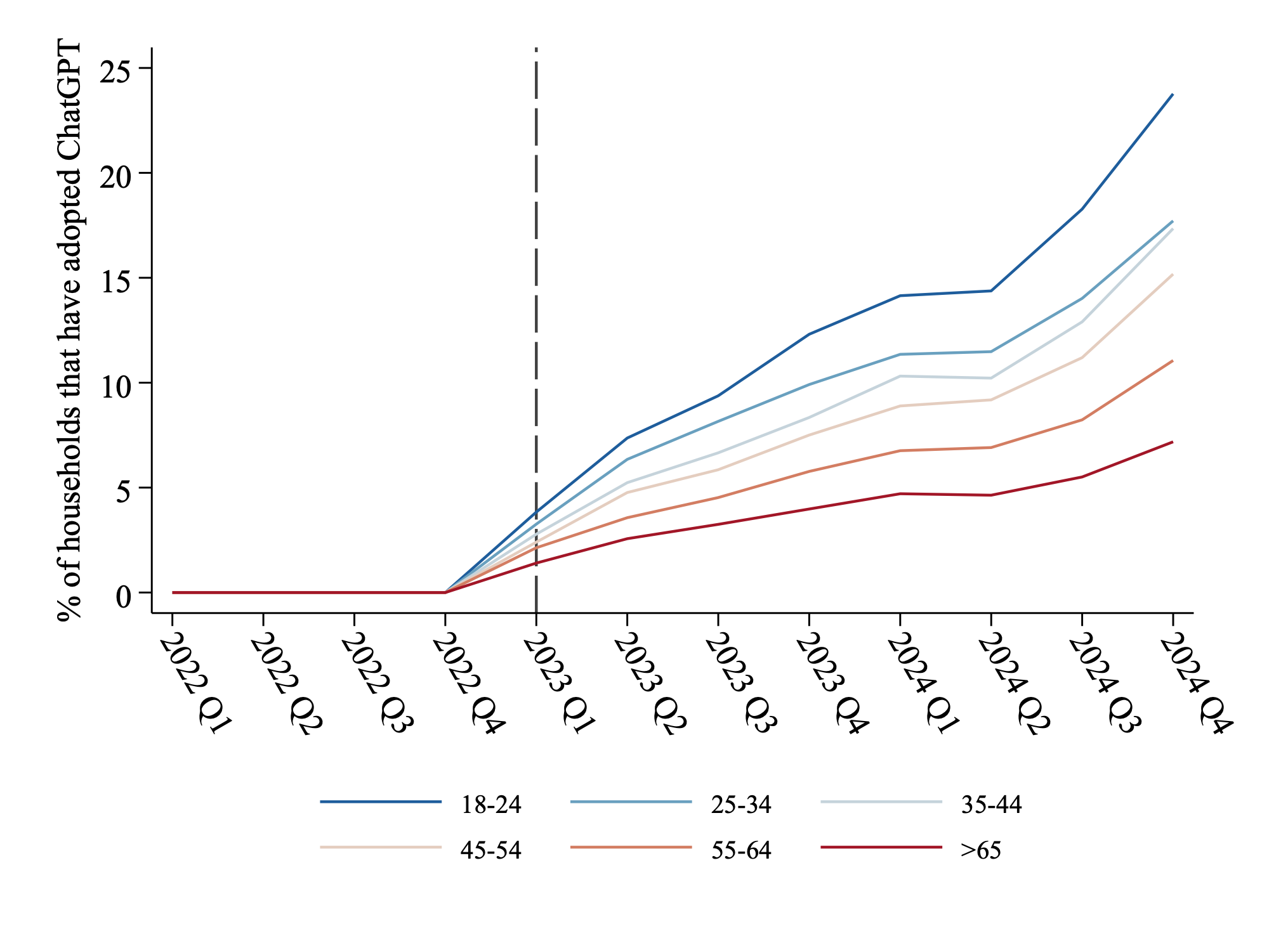}
  \caption{ChatGPT adoption by age bucket}
  \label{fig:sub2}
\end{subfigure}
\end{figure}

\paragraph*{Household ChatGPT adoption by income and age.} Our data also allows us to compare generative AI use across different demographic groups. Figure \ref{fig:ts_use_by_demo} shows how the share of households ever having used ChatGPT varies across income categories (panel A) and by age of the household head (panel B). 
In panel A, we find that higher income households ($>\$200$K) are substantially more likely to have used ChatGPT since its release.  While the relationship is not precisely monotonic, it is generally true that higher income categories tend to have higher rates of adoption. This pattern is violated mainly by the lowest-income categories, which is likely due to the fact that many students and recent graduates have high adoption rates and are likely to fall into the lower-income categories. Panel B of Figure \ref{fig:ts_use_by_demo} shows that age strongly predicts ChatGPT adoption patterns: the younger the household head, the more likely is it that they have used ChatGPT, with an adoption gap that increases over time. 

\paragraph*{The generative AI divide.} The gap in the use of generative AI documented in Figure \ref{fig:ts_use_by_demo}  recalls concerns over similar patterns of a ``digital divide'' associated with previous waves of information and communication technologies between ``those people who have access and use of digital media and those who do not'' (\cite{van2020digital}, p. 1). Just as access to past technological improvements was mediated by skills, motivation, and socio-economic contexts \citep{van2020digital}, generative AI access is likely impacted by these characteristics. An important new insight from our time series data is that a similar gap exists in the use of generative AI at the household level and low adoption is associated with other indicators of socioeconomic disadvantage, such as low income or old age. Large differences in generative AI adoption among households raise the specter of a persistent ``Generative AI Divide''. Moreover, the Comscore data allows us to track time trends in this adoption gap for a comparable population of households: the gap in adoption between high and low income households, and between young and old households, is, if anything increasing over time. This increasing divide means that any productivity benefits of generative AI are also unequally distributed---which motivates our attempt to quantify the productivity effect in people's private lives.

\subsection{Household exposure and ChatGPT adoption}

To understand whether households that are more likely to benefit from using ChatGPT are more likely to adopt the technology, we first consider whether greater exposure based on a household's pre-ChatGPT browsing behavior predicts greater adoption. This evidence motivates the empirical approach in the next section for estimating the causal effect of adoption on household behavior.

\begin{figure}[!hbt]
\caption{\textbf{ChatGPT adoption by household GenAI exposure}} 
\label{fig:ts_adoption_by_exposure} This figure shows the share of households in each generative AI exposure tercile---with the 3rd tercile having the highest exposure---that have ever used ChatGPT (Panel A), and also the share of total browsing duration that is spent using ChatGPT in each tercile of exposure (Panel B).

    \centering 
\begin{subfigure}{.49\textwidth}
  \centering
  \includegraphics[width=\linewidth]{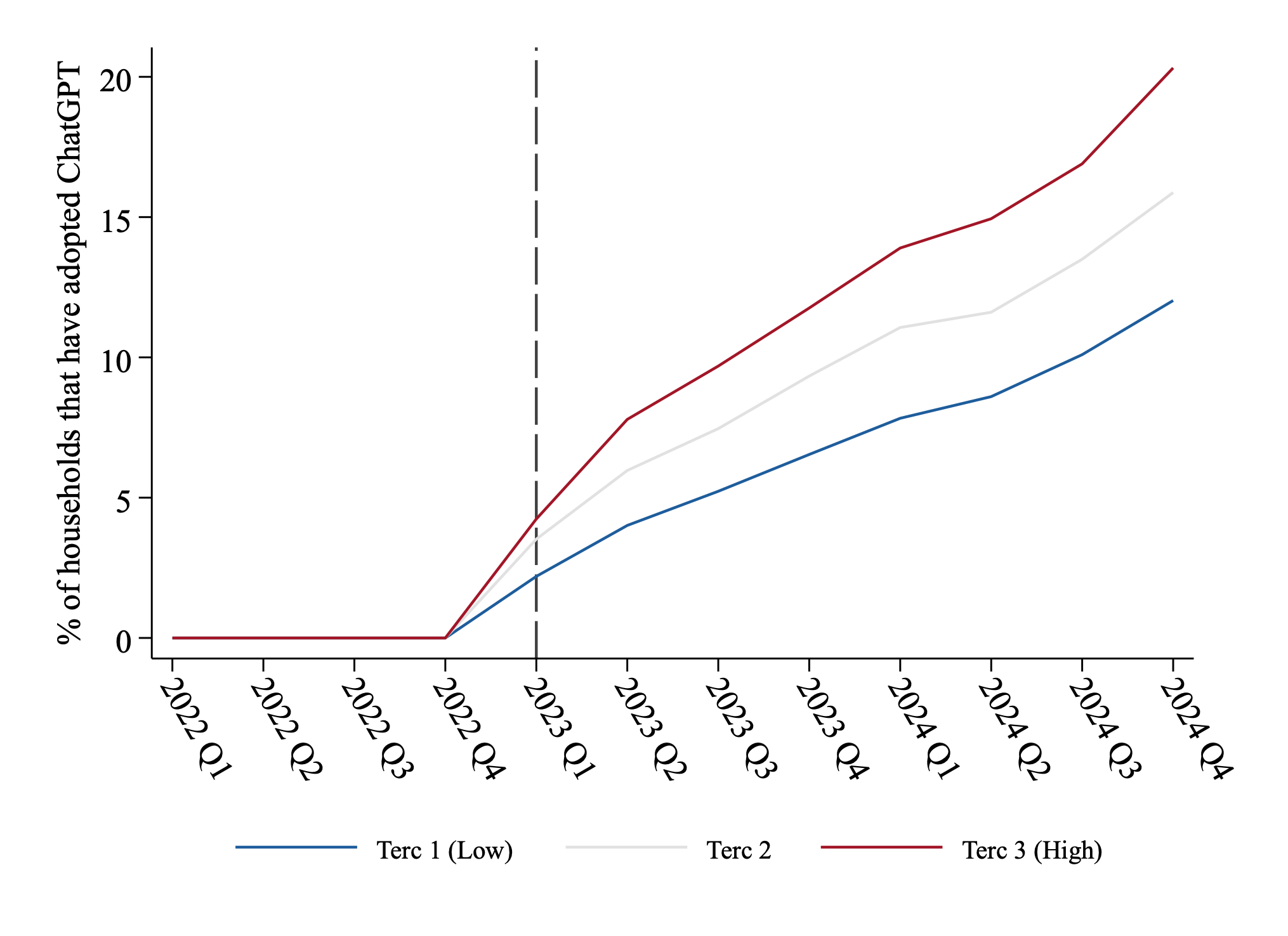}
  \caption{Share that has used ChatGPT}
\end{subfigure} 
\begin{subfigure}{.49\textwidth}
  \centering
  \includegraphics[width=\linewidth]{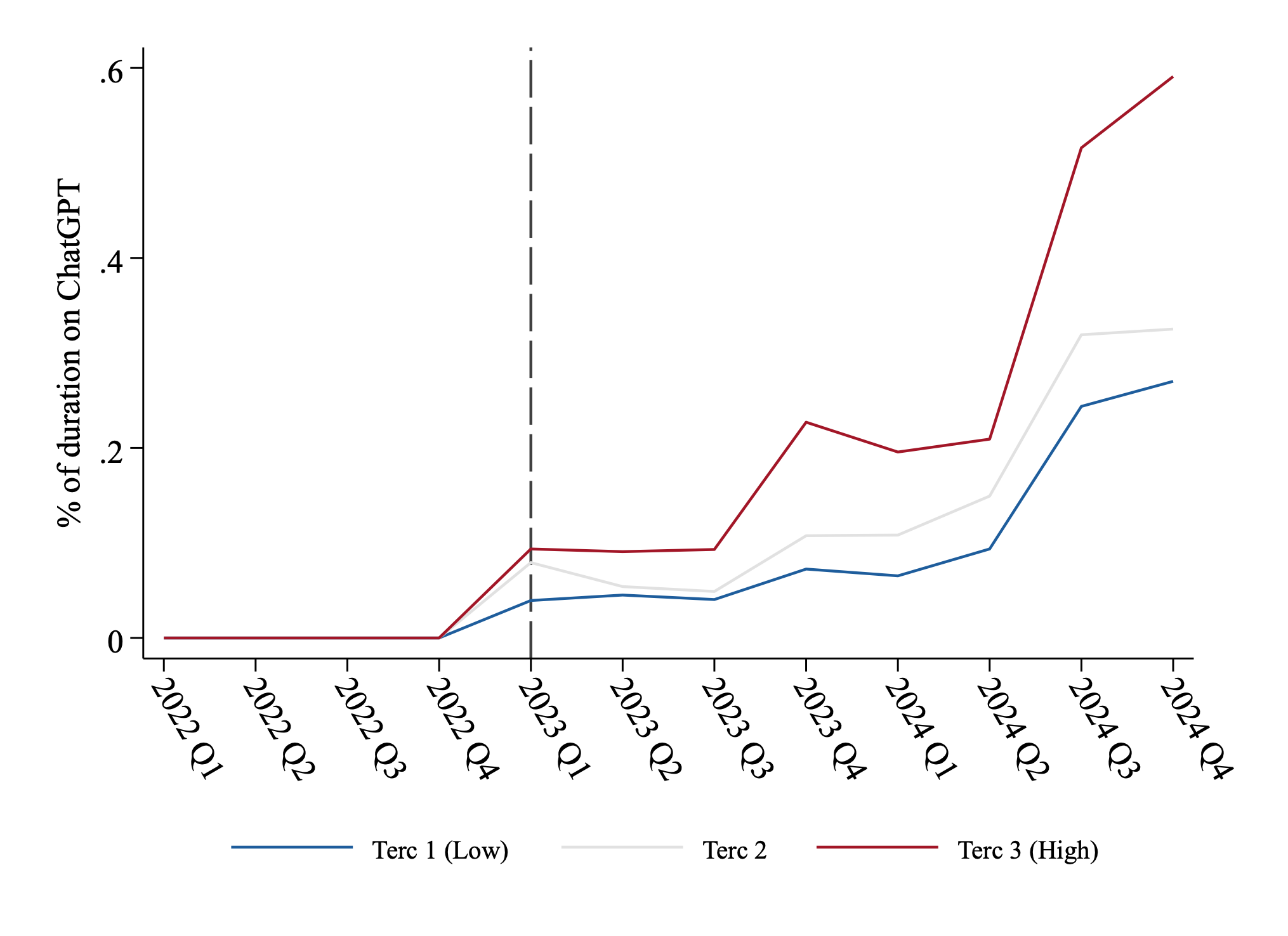}
  \caption{Share of browsing duration on ChatGPT}
  \label{fig:sub2}
\end{subfigure}
     
\end{figure}

\textbf{Time patterns by exposure.} The relationship between predicted household benefits based on browsing behavior---``household GenAI exposure''---and adoption of ChatGPT can be seen in the raw time series: Figure \ref{fig:ts_adoption_by_exposure}  plots adoption rates by tercile of exposure of the household: as the Figure shows, households with greater exposure adopt ChatGPT more rapidly, starting right after its release in November 2022, and continue to show higher adoption rates until the end of our data in December 2024. This relationship between exposure and adoption is monotonic, with adoption speed and levels across terciles aligning with the ordering of the terciles by exposure. The cumulative gap in having used ChatGPT is large: the likelihood of having used ChatGPT is 20pp by Q4 2024 in the highest exposure tercile households compared to 12pp in the lowest exposure tercile. Similarly, the share of browsing time spent using ChatGPT is 0.59pp for highly exposed households compared to 0.27pp among low exposure households in Q4 2024. That is, the adoption gap by exposure increases over time, with little sign of convergence: at least through 2024, it is true that households with greater exposure do not just adopt earlier than other households but also at a higher rate.

This empirical pattern validates that the potential benefit to households from adopting generative AI does, in fact, drive differential adoption, which both validates our measure of household exposure and provides, to our knowledge, the first systematic empirical evidence that household characteristics with regard to browsing behavior can explain inequality in adoption of generative AI among private households.

\textbf{Effect of household exposure on GenAI adoption.} To estimate the size of the effect of exposure on both the intensive and extensive margin of ChatGPT adoption, we estimate regressions of the form
\begin{equation}
\text{ChatGPT Use}_{j,t} = \theta\; \ln(\text{HHGenAIExp}_{j}) + X_{j,t}'\xi + \text{FE}_{j,t} +\varepsilon_{j,t} \label{eq:use}
\end{equation}
where $\text{GenAIExp}_{j}$ is the household's pre-ChatGPT benchmark period (Jan.-Dec. 2021) exposure to generative AI, and $X_{j,t}$ is a vector of household characteristics. We include saturated interactions of income (8 categories), age (6 categories), and metro area  as fixed effects in all cross-sectional household level regressions to capture the effect of different demographics and regional differences. We also control for the topic-based exposure to generative AI of the household - the component that is explained by the average exposure of websites in particular content categories, and the share of the household's total browsing activity in the benchmark period for which we can compute exposure scores.

We measure \textit{extensive margin} adoption effects by using a dependent variable that reflects whether the household has ever used ChatGPT by 2024, such that the estimated coefficient on exposure indicates the degree to which exposure leads to adoption. To quantify the \textit{intensive margin} of ChatGPT adoption, we estimate the effect on the average share of total browsing duration spent using ChatGPT during 2024.

The results are shown in Table \ref{table:2sls-first-rf}, where columns (1) and (2) show the extensive and intensive margin effects of household GenAI exposure on adoption. The coefficients in columns (1) and (2) show that a 1 SD ($\sim$1.3 unit) change in log household exposure is associated with about a 2.5pp increase in the likelihood of having used ChatGPT by 2024, and a 0.07pp increase in  the share of total browsing time spent using ChatGPT. Note that the flexible demographic controls mean that these estimates are holding the observable demographics constant, i.e., they cannot be explained by the differences in adoption patterns across demographic groups that we previously documented.

The next section considers whether these changes in GenAI use as a result of a household's exposure lead to changes in a household's behavior that suggest productivity gains.

\begin{table}[htbp]
\caption{\\ \centering \textbf{Effect of household GenAI exposure on ChatGPT adoption and browsing by purpose}}  

\setlength{\abovedisplayskip}{0pt}
\setlength{\belowdisplayskip}{0.5pt}
\setlength{\abovedisplayshortskip}{0pt}
\setlength{\belowdisplayshortskip}{0.5pt}
\vspace{-0.1cm} \label{table:2sls-first-rf} \footnotesize  This table shows estimates of long difference reduced-form regressions comparing 2024 browsing characteristics to 2022 browsing characteristics, in an estimation of the form:
$$\Delta\text{BrowsingOutcome}_{i} = \gamma\; \text{ln(HHGenAIExp}_{i})  + \lambda_{FEs} + X_i'\xi + \varepsilon_{i}.$$
The dependent variable in Column (1) is a dummy for whether household $j$ has ever used ChatGPT by the end of 2024. The independent variable is the log of the panelist household's predicted exposure to ChatGPT in 2021, based on the household's share of browsing on highly substitutable websites. The dependent variable in Column (2) is the share of the household's total browsing duration that is spent on ChatGPT; in Columns (3),(4), (5) and (6) it is the change in the log of browsing duration on all, leisure, productive, and mixed-use sites. All regressions include fixed effects for income bin X age bin X MSA. All regressions also include control variables for the generative AI exposure predicted by the households 2021 browsing composition by Comscore content category, and for the sum of the browsing share of websites with a GenAI exposure label in 2021. T-statistics based on heteroskedasticity-robust standard errors clustered at the income bin X age bin X MSA level are shown in parentheses: * p$<$0.10, ** p$<$0.05, *** p$<$0.01. \vspace{0.1cm} 

\footnotesize
 \centering \setlength{\tabcolsep}{3pt}
\begin{adjustbox}{max width=\textwidth}
\begin{tabular}{@{}l*{7}{c}@{}} 
\toprule
\textit{Specification:} &  \multicolumn{6}{c}{\textit{Long differences: 2024 vs. 2022}}\\
\cmidrule(lr){2-7} 

&  \multicolumn{2}{c}{\textit{GenAI adoption measures}} &  \multicolumn{4}{c}{\textit{Browsing activity by purpose}} \\
\cmidrule(lr){2-3} \cmidrule(lr){4-7} 
\textit{Dep. var.} &  $\mathbb{1}[\text{Used ChatGPT}]$  &   $\substack{\text{ChatGPT Duration} \\ \text{Share \%} }$  &  
$\substack{\Delta \text{ Log Duration} \\ \text{All sites}}$ & $\substack{\Delta \text{ Log Duration} \\ \text{Leisure}}$ &  $\substack{\Delta \text{ Log Duration} \\ \text{Productive}}$   &  $\substack{\Delta \text{ Log Duration} \\ \text{Mixed}}$   \\
   \cmidrule(lr){2-2}   \cmidrule(lr){3-3}   \cmidrule(lr){4-4}   \cmidrule(lr){5-5}   \cmidrule(lr){6-6}   \cmidrule(lr){7-7}  
 &  (1) & (2) & (3) & (4) & (5)  & (6) \\
\midrule \addlinespace ln(HHGenAIExp)&       0.019***&       0.053***&       0.024** &       0.029** &       0.000   &      -0.006   \\
            &    (10.180)   &     (5.404)   &     (2.421)   &     (2.087)   &     (0.021)   &    (-0.452)   \\
\midrule \addlinespace Observations&      42,886   &      42,886   &      42,886   &      42,886   &      42,886   &      42,886   \\

\addlinespace
 	Estimation & OLS & OLS & OLS & OLS & OLS  & OLS   \\ 
Inc. X Age X MSA FEs     & X & X	& X & X	& X    & X     \\	
Bartik share control     & X & X	& X & X	 	& X  & X    \\	
Website category exposure    & X & X	& X & X	& X  & X    \\	
    \bottomrule
\end{tabular}
\end{adjustbox}
\end{table}


\section{The Household Impact of Generative AI Adoption} \label{sec:Impact}

The previous section established that differences in the potential benefit from generative AI based on pre-ChatGPT browsing patterns can drive differences in ChatGPT adoption post-2022. In this section, we estimate the \textit{effect} that ChatGPT adoption has on households' overall browsing behavior: which types of online tasks do households spend more or less time on after they start using generative AI?

\subsection{Empirical design}

We are interested in estimating the impact that adopting ChatGPT has on the quantity of digital tasks whose productivity ChatGPT plausibly increases, such as tasks related to acquiring information or learning new skills, relative to the impact on the quantity of digital consumption, such as streaming videos for entertainment. In theory, adopting ChatGPT could lead households to tilt the composition of their digital activities either away from or towards the productive websites that ChatGPT displaces. Intuitively--and as we formally model in the following section--if the output that households attain via these websites are ``necessary goods,'' then adopting a tool that allows them to be completed more efficiently may lead a household to spend \textit{less} time on them. Conversely, to the extent that websites that provide households with leisure are ``luxury goods,'' adopting ChatGPT may lead a household to spend more of their home computing time consuming these goods. 

\paragraph{Empirical specification.} In general, we want to estimate empirical specifications similar to the following expression:
\begin{align}
  \Delta\text{BrowsingOutcome}_{i,t} &= \gamma\; \text{ChatGPTUse}_{i,t}  + \lambda_{FEs} + X_{i,t}'\xi + \varepsilon_{i}  \label{eq:2ndstage}
\end{align}
Here, the dependent variable in each regression might, for example, be a measure of the household's browsing activity in different categories. We want to estimate the effect of a household $i$'s use of ChatGPT on the characteristics of the household's browsing activity.

\paragraph{Instrumental variable identification.} We want to obtain an estimate of the effects of generative AI use in this empirical setting that is plausibly causal: among households that use ChatGPT, what is the amount and composition of their \textit{actual} home browsing activity 
relative to their \textit{counterfactual} activity if they had not adopted ChatGPT?  The key identification concern is that adoption of ChatGPT may be endogenously driven by recent life events that may then also change online browsing behavior: for example, a recent job loss may lead an unemployed worker to use ChatGPT for drafting cover letters, while also devoting more time to online job search and to online education to brush up on in-demand skills, but the latter changes in browsing behavior should not be causally attributed to generative AI adoption.\footnote{Similarly, among households with the same income and age, those who use ChatGPT for home browsing may be more likely to have been introduced to the technology at work. To the extent that GenAI-exposed jobs are in occupations that employ workers with different skills and/or productivity levels \citep{eisfeldt2013}, this would lead to bias in our estimate of the effects of ChatGPT use.} 

To arrive at a causal estimate  of the effects of generative AI technology, we therefore need to identify variation in adoption that is plausibly exogenous with regard to other short-term changes in a household's life circumstances that might be driving browsing behavior.

Our analysis in the previous section suggests a potential instrument: the exposure of a household to ChatGPT's release given their ex ante browsing behavior. This measure is, as documented in the previous section, strongly related to households' actual take-up of ChatGPT. Moreover, it plausibly satisfies the exclusion restriction: Once we condition on observable sociodemographic characteristics and broad differences in pre-ChatGPT release browsing behavior, variation in GenAI exposure is the result of differences among otherwise similar households in the precise tasks done through online browsing, which are unlikely to drive short-run life events years later when the household starts using ChatGPT.

We therefore implement this IV approach by using the household-level GenAI exposure measure defined in equation \eqref{eq:hhexposure} computed over the period from January 2021 to December 2021. This historical browsing behavior predicts which households utilize ChatGPT following its November 2022 release. Under the plausible assumption that the 2021 browsing behavior on which our measure of household GenAI exposure is based, is exogenous with regard to other post-2022 shocks to browsing behavior, this IV approach identifies the causal effects of generative AI on browsing activity (conditional on household-level fixed effects and detailed demographic controls).

We estimate IV regressions corresponding to the two-stage system in which the second-stage equation is given by equation \ref{eq:2ndstage} and the first stage corresponds to equation \ref{eq:use} introduced in the previous section.

\paragraph{Long difference sample.} In order to smooth out some of the noise from short-run fluctuations in browsing activity and account for the fact that Section \ref{sec:measure_adoption} showed that ChatGPT use increased substantially only by 2024, we estimate long-difference regressions, comparing household browsing activity (i.e., machines) in 2024 to activity in 2022 before ChatGPT was released. This focuses our analysis on broad changes in behavior rather than short-run fluctuations and further reduces the possibility that the measurement period for our instrument (Jan.-Dec. 2021) is affected by household life event shocks that also drive digital activity in 2024.

\paragraph{Control variables.} The long difference estimation includes age bucket$\times$income bucket$\times$region fixed effects. All regressions also include control variables for the generative AI exposure predicted by the households' 2021 browsing composition by Comscore content category, and for the sum of the browsing share of websites with a GenAI exposure label in 2021 - the 'Bartik share control'.

\paragraph{Dynamic effect estimates.} To show how the effect of ChatGPT adoption changes over time and address the `pre-trends' concerns often associated with this event study-style estimation, we also  estimate  dynamic quarterly panel IV regressions of the form 
\begin{center}$
\text{Browsing outcome}_{j,t} = \Sigma_t \gamma^{IV}_t \; \ln\left(\text{HHGenAIExp}_{j,t}\right) \times \mathbb{1}[\text{Quarter}=t]_t  + \lambda_j + \lambda_t + X_{j,t}'\xi + \varepsilon_{j,t},$
\end{center}
where the left-hand side is a time-varying browsing outcome of household $j$ in quarter $t$, \textit{HHGenAIExp}$_{j,t}$ is the household's 2021 browsing exposure to GenAI. The dynamic effects estimation estimates a reduced-form effect because it would otherwise not be possible to estimate effects before ChatGPT is released (as the endogenous ChatGPT adoption variable is zero for everyone). We use Q4 2022 as the pre-release reference period.\footnote{For quarterly regressions, we cannot precisely capture the pre- vs. post-period around the November 30, 2022 release. We pick 2023Q1 as the first post-release quarter because, in practice, users of ChatGPT in our sample do not appear to start using ChatGPT, and demonstrate different behavior across other websites, until (and generally after) the first months of 2023. However, all of our results are quantitatively and qualitatively similar if we instead define 2022Q4 as the first post-release quarter.} These panel regressions include household fixed effects,  as well as quarter$\times$age bucket$\times$income bucket$\times$region fixed effects.

\subsection{Main findings: household browsing effects of generative AI adoption}

\begin{figure}[t]
\caption{\textbf{ChatGPT adoption effects of Household GenAI exposure over time}} 
\setlength{\abovedisplayskip}{0pt}
\setlength{\belowdisplayskip}{0.1pt}
\setlength{\abovedisplayshortskip}{0pt}
\setlength{\belowdisplayshortskip}{0.1pt}
\label{fig:rf-dynamic-gpt-use} \footnotesize This figure shows the results of estimating specifications of the form
\begin{center}$
\text{Browsing outcome}_{j,t} = \sum_{h=-4}^{8} \gamma^{RF}_h \; \ln\left(\text{HHGenAIExp}_{j}\right) \times \mathbb{1}\{ t = 2022Q4 + h \}   + \lambda_j + \lambda_t + X_{j,t}'\xi + \varepsilon_{j,t},$
\end{center}
where the dependent variable in panel A is an indicator for whether the household has ever used ChatGPT by that time period, and in panel B it is the share of total browsing time spent using ChatGPT.

     \centering
\begin{subfigure}{.49\textwidth}
  \centering
  \includegraphics[width=\linewidth]{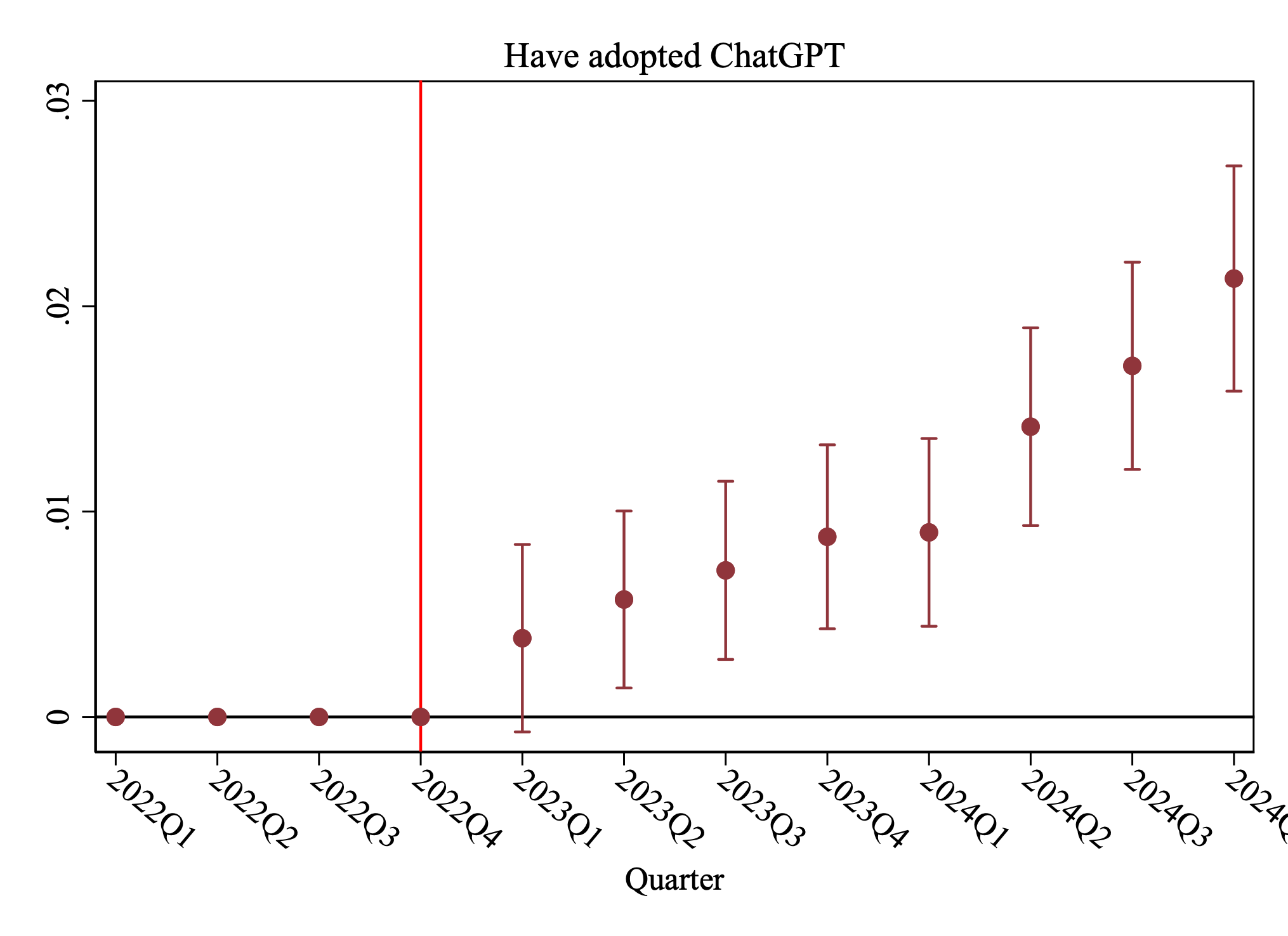}
  \caption{Effect on share that has used ChatGPT}
\end{subfigure}
\begin{subfigure}{.49\textwidth}
  \centering
  \includegraphics[width=\linewidth]{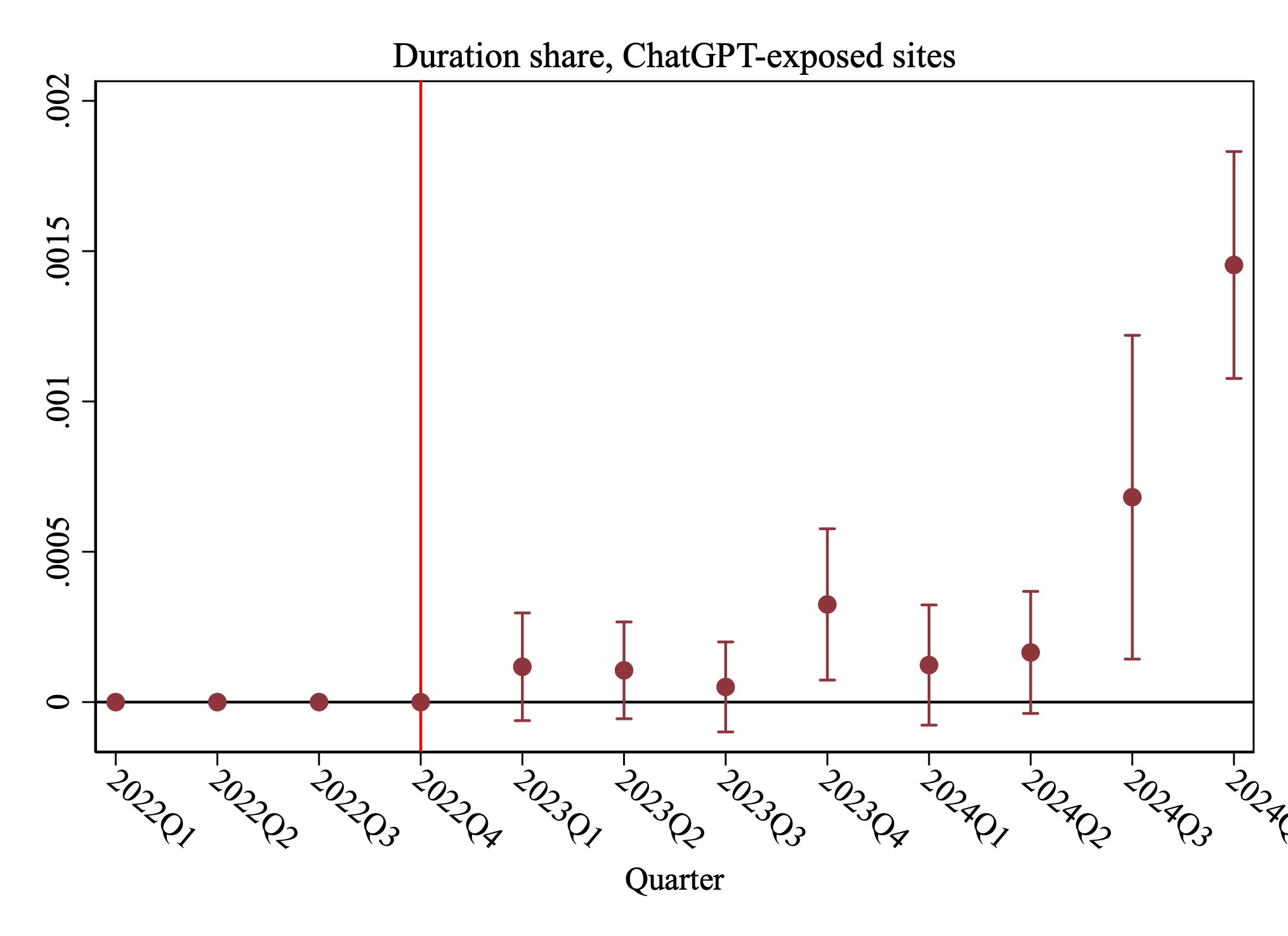}
  \caption{Effect on duration share of ChatGPT use}
\end{subfigure} 
\end{figure}

\paragraph{Reduced-form dynamic effects.} To understand the time pattern of ChatGPT effects on households, we first estimate a version of the reduced-form dynamic effect estimation that corresponds to the `first-stage' of the IV regression: Figure \ref{fig:rf-dynamic-gpt-use} shows the effects of household GenAI exposure on the use of ChatGPT over time. As the figure shows, households with greater exposure are increasingly more likely to have used ChatGPT, and also spend an increasing amount of time  on the ChatGPT site. However, as the time patterns show---and in line with the raw adoption patterns in Section \ref{sec:measure_adoption}---the effects only become more pronounced by the end of 2024. This is the reason why we will estimate the main behavioral effects parameters using the long difference design.

\begin{figure}[!ht]
\caption{\textbf{Browsing behavior effects of household GenAI exposure over time}} 
\setlength{\abovedisplayskip}{0pt}
\setlength{\belowdisplayskip}{0.1pt}
\setlength{\abovedisplayshortskip}{0pt}
\setlength{\belowdisplayshortskip}{0.1pt}
\label{fig:rf-dynamic-log-time} \footnotesize This figure shows the results of estimating specifications of the form
\begin{center}$
\text{Browsing outcome}_{j,t} = \sum_{h=-4}^{8} \gamma^{RF}_h \; \ln\left(\text{HHGenAIExp}_{j}\right) \times \mathbb{1}\{ t = 2022Q4 + h \}   + \lambda_j + \lambda_t + X_{j,t}'\xi + \varepsilon_{j,t},$
\end{center}
where the dependent variable in panel A is the log of the household's total browsing time, and in panels B, C and D, it is the log of total browsing time spent on leisure, productive, or mixed purpose sites. Confidence intervals shown correspond to 95\% CIs based on heteroskedasticity-robust standard errors clustered at the age bucket$\times$income bucket$\times$region level. These panel regressions include household fixed effects,  as well as quarter$\times$age bucket$\times$income bucket$\times$region fixed effects.

\centering
\begin{subfigure}{.49\linewidth}
  \centering
  \includegraphics[width=\linewidth]{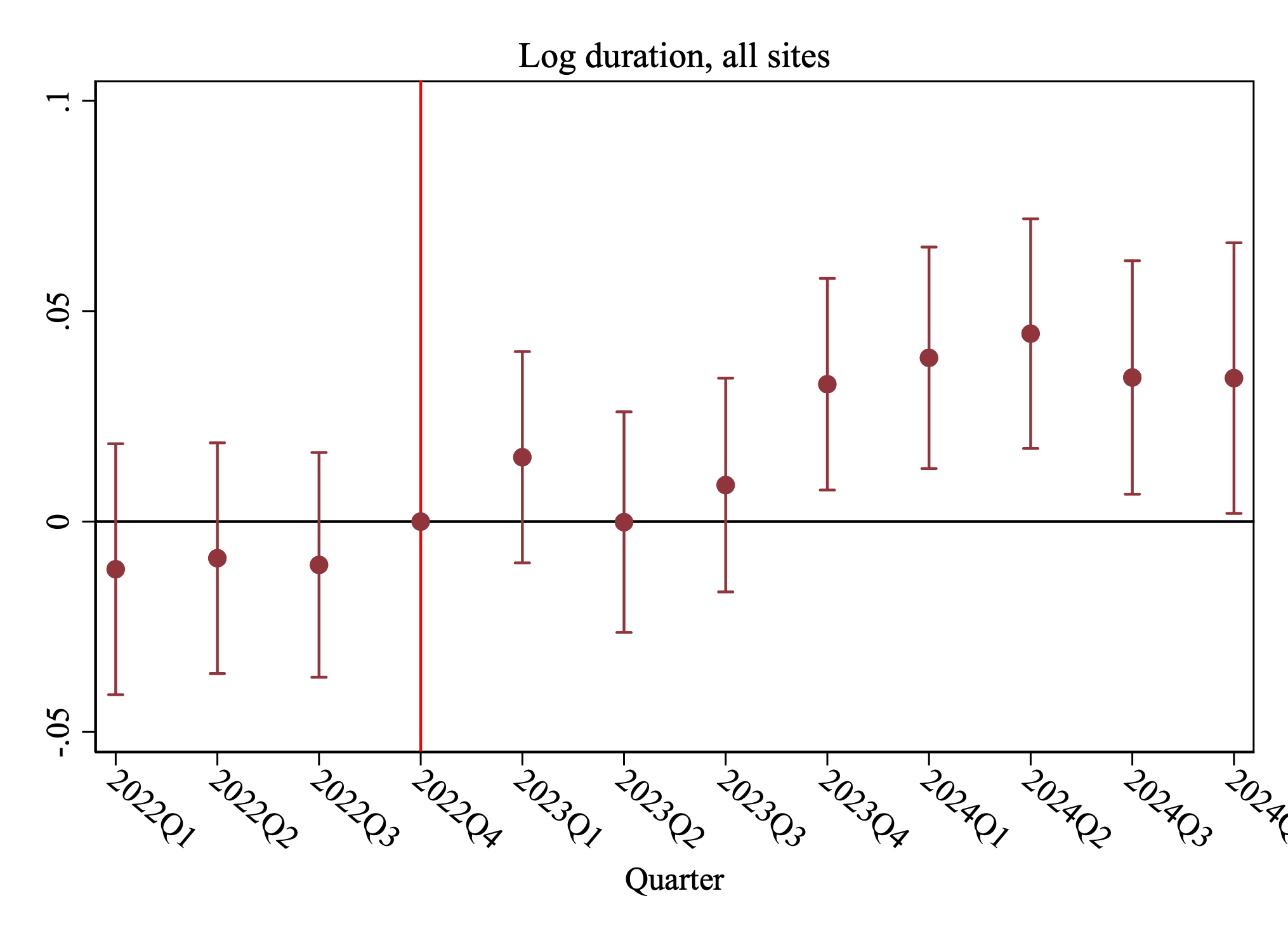}
  \caption{Log browsing duration, all sites}
\end{subfigure}
\begin{subfigure}{.49\linewidth}
  \centering
  \includegraphics[width=\linewidth]{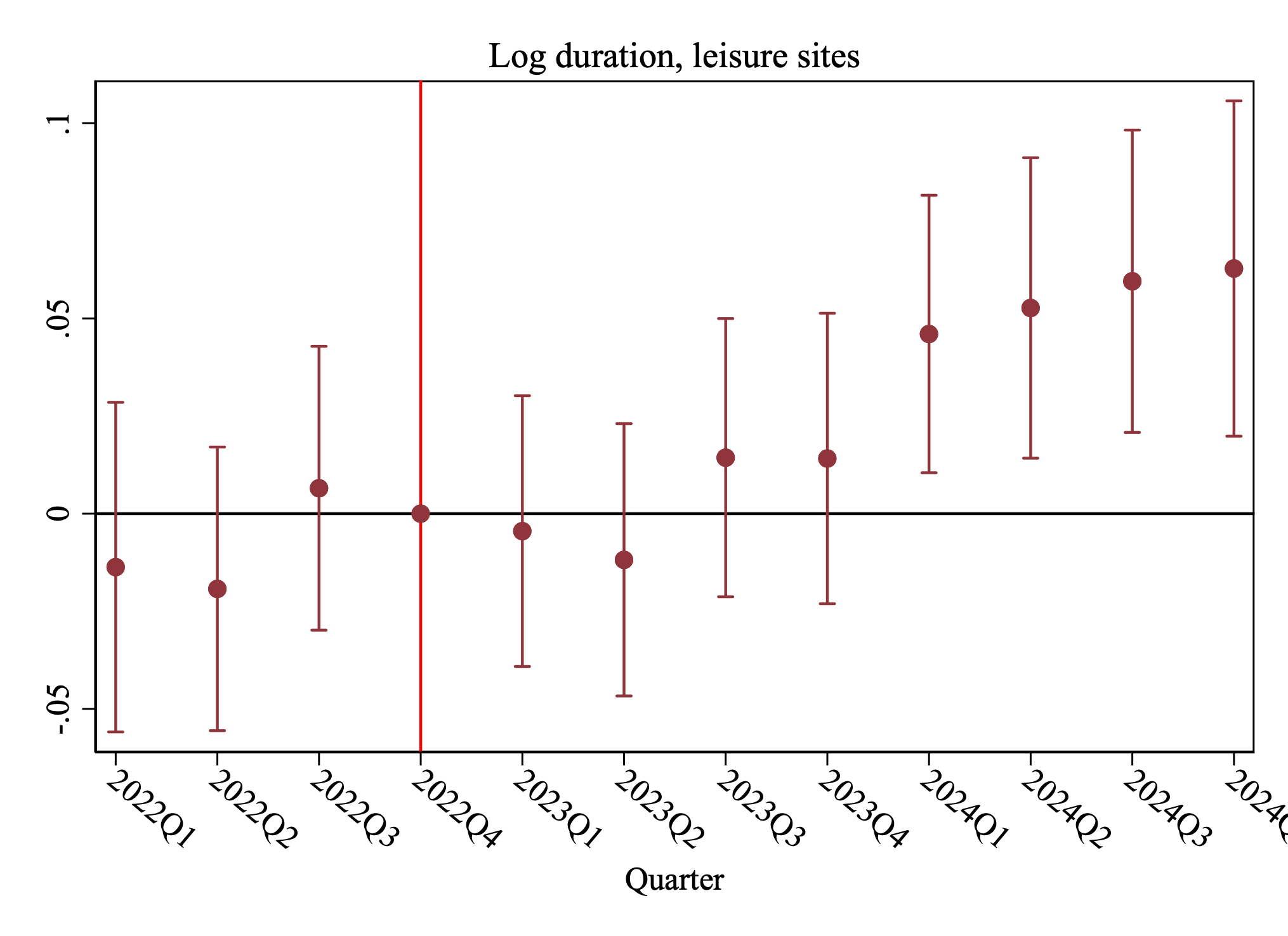}
  \caption{Log browsing duration, leisure sites}
\end{subfigure}\\
\begin{subfigure}{.49\linewidth}
  \centering
  \includegraphics[width=\linewidth]{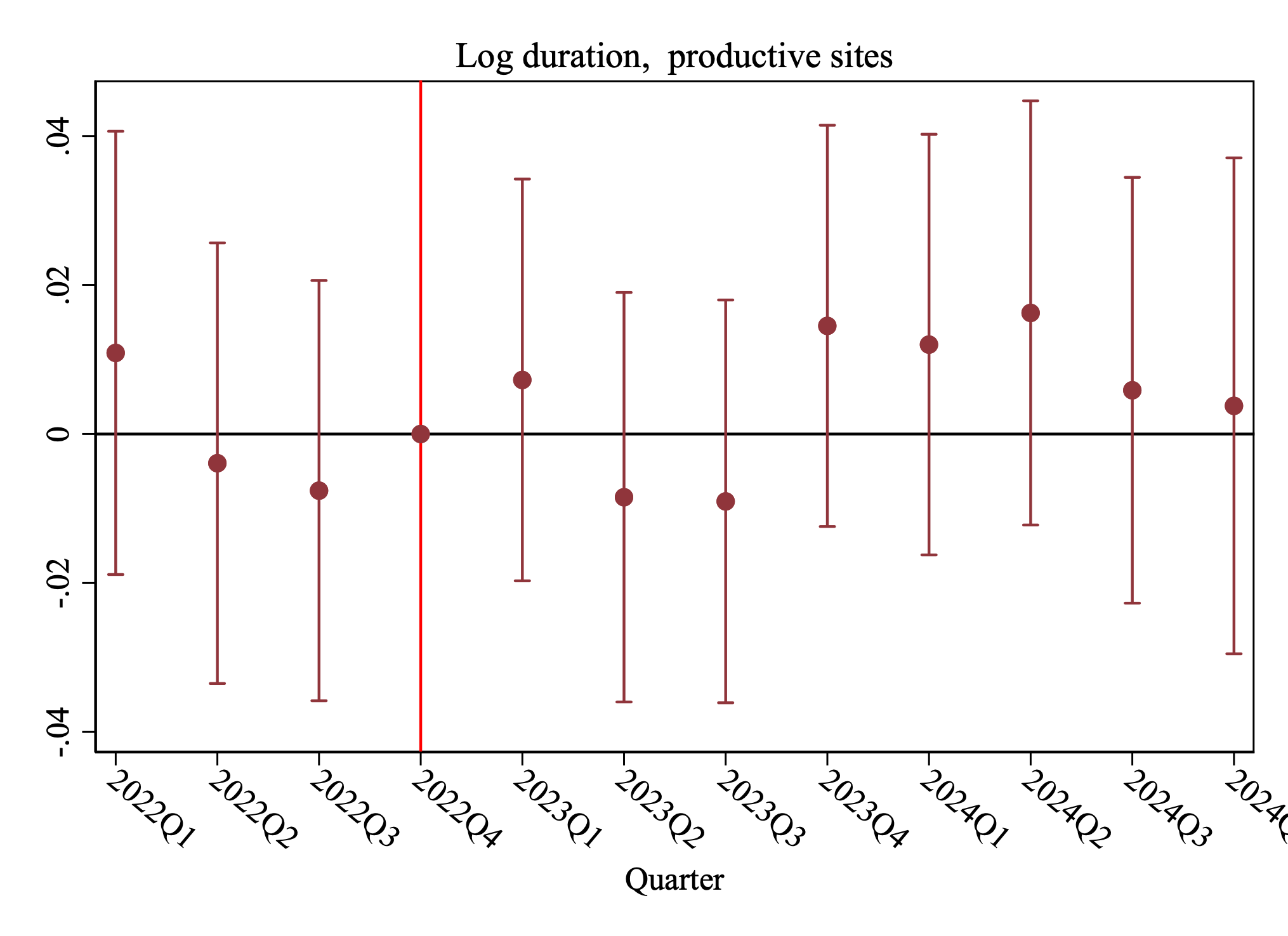}
  \caption{Log browsing duration, productive sites}
\end{subfigure}
\begin{subfigure}{.49\linewidth}
  \centering
  \includegraphics[width=\linewidth]{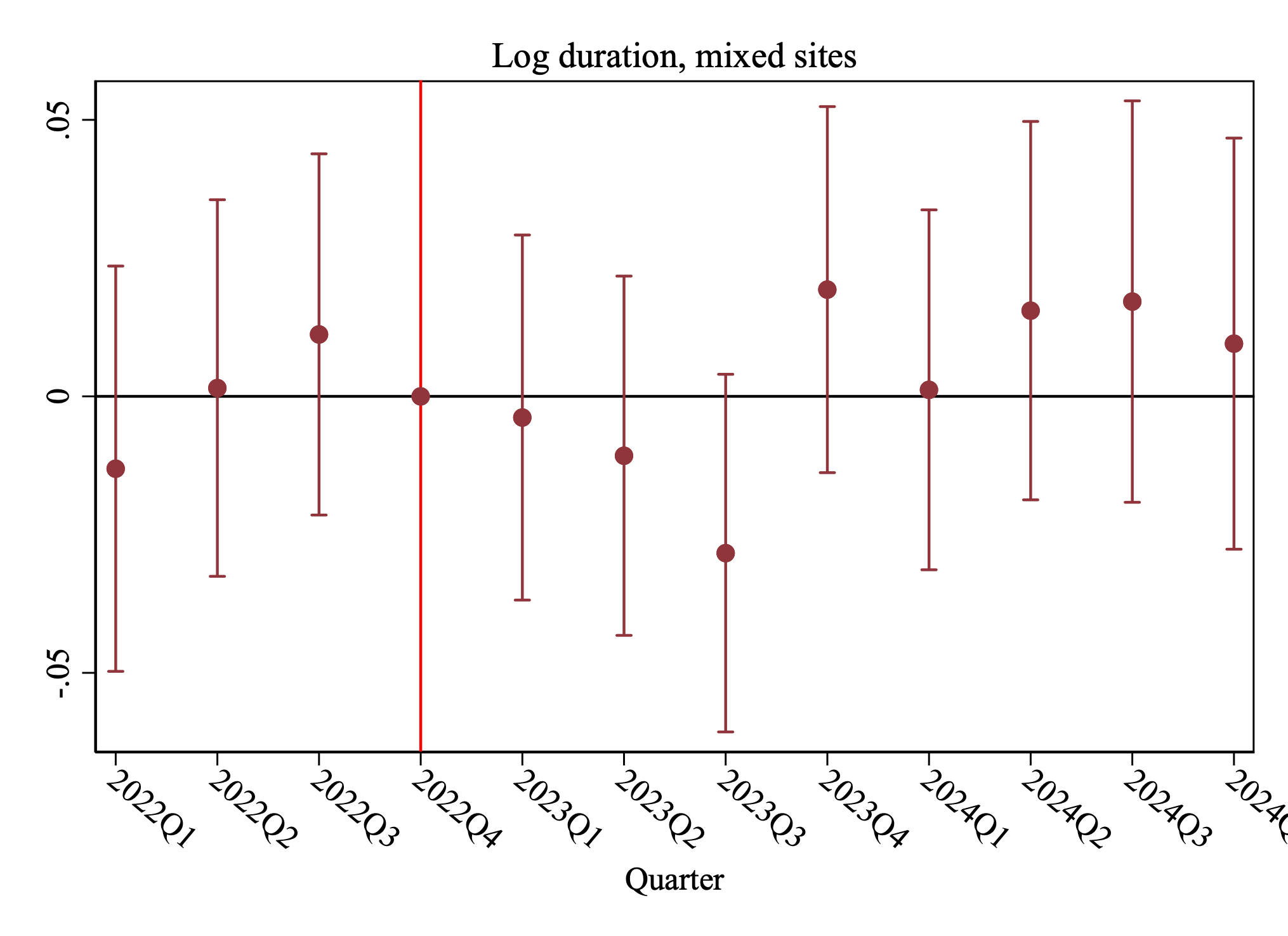}
  \caption{Log browsing duration, mixed sites}
\end{subfigure}
\end{figure}

How does household browsing activity change in response to household GenAI exposure? Figure \ref{fig:rf-dynamic-log-time} shows the effects of household GenAI exposure on the log of browsing time in different purpose categories, focusing on leisure vs. productive online activities.
Panel (A) shows that there is an increase in total duration.
Panel (B) shows that households substantially increase their browsing time on leisure-oriented websites. This is especially the case starting in 2024, consistent with the pattern of when ChatGPT adoption in response to exposure surges as well. In comparison, Panels (C) and (D) show that exposure to generative AI does not lead to a similar increase in browsing on productivity-oriented websites and on mixed purpose sites.  These dynamic effect estimates align with the reduced-form long-difference effects in columns (3)-(6) of Table \ref{table:2sls-first-rf}, which also show increases in total browsing and leisure browsing and no significant changes in time spent on productivity-related or mixed-purpose sites. To interpret the magnitudes, this suggests that a 1 SD change in log exposure increases total browsing by $0.024*1.3\approx 3\%$
and leisure browsing by $0.029*1.3\approx 4\%$.
\paragraph{Pre-trends.} The dynamic effects are also reassuring with regard to the possibility that more GenAI-exposed households may have unobservable different trends in browsing behavior that precede the ChatGPT release: all 4 panels of Figure \ref{fig:rf-dynamic-log-time} show no significant differences in browsing behavior for exposed households in the year \textit{before} ChatGPT is released. Moreover, the changes in behavior found after the ChatGPT release concentrate in the later quarters and do not appear in the initial quarters after the release when actual adoption rates were still low.

\paragraph{Baseline effect estimates.}
Table \ref{table:2sls-bycat} presents our long-difference IV regressions to estimate the effect of ChatGPT adoption on changes in browsing behavior between 2022 and 2024. In line with the dynamic reduced-form estimates in  Figure \ref{fig:rf-dynamic-log-time}, the amount of browsing activity devoted to leisure-oriented sites and to total browsing increases in an economically and statistically significant way in 2024. At face value, the estimates suggest that total leisure-related digital time increases by a factor of $\sim 4.5$ after households adopt ChatGPT.\footnote{Computed as $exp(1.512)\approx4.5$.} We will discuss how to interpret this magnitude more in the next section. Here, the relevant takeaway is that ChatGPT adoption significantly increases leisure browsing and total browsing, but productive and mixed-purpose browsing levels (columns 3 and 4) do not change significantly. Columns (5)-(7) show that this implies a significant increase in the \textit{share} of browsing time spent on leisure sites, a decrease in the share of time spent on productive sites, and a negative but non-significant decline in the share spent on mixed-purpose sites.

In the next section, we adapt a household time allocation model to show that these estimates can be used to infer the size of the productivity effect of ChatGPT on households' different digital activities.

\begin{table}[htbp]
\caption{\\ \centering \textbf{Effect of ChatGPT adoption on browsing activity by category}}  

\setlength{\abovedisplayskip}{0pt}
\setlength{\belowdisplayskip}{0.5pt}
\setlength{\abovedisplayshortskip}{0pt}
\setlength{\belowdisplayshortskip}{0.5pt}
\vspace{-0.1cm} \label{table:2sls-bycat} \footnotesize  This table shows estimates of long difference IV regressions comparing 2024 browsing characteristics to 2022 browsing characteristics, in an estimation of the form
$$\Delta\text{BrowsingOutcome}_{i} = \gamma\; \text{ChatGPTUse}_{i}  + \lambda_{FEs} + X_i'\xi + \varepsilon_{i}.$$
The endogenous variable is a dummy for whether household $j$ has ever used ChatGPT by the end of 2024. The instrument is the panelist household's predicted exposure to ChatGPT in 2021, based on the household's share of browsing on highly substitutable sites. The dependent variable in each regression is a measure of the household's browsing activity in different categories. In Column (1) it is the quarterly log of browsing duration across all sites; in Columns (2),(3) and (4) it is the quarterly log of browsing duration on  leisure, productive, and mixed sites; and in Columns (5), (6), and (7) the change in the share of the household's total browsing that is  on  leisure, productive, and mixed sites. All regressions include fixed effects for income bin X age bin X MSA. All regressions also include control variables for the generative AI exposure predicted by the households' 2021 browsing composition by Comscore content category, and for the sum of the browsing share of websites with a GenAI exposure label in 2021 - the 'Bartik share control'. T-statistics based on heteroskedasticity-robust standard errors clustered at the income bin X age bin X MSA level are in parentheses: * p$<$0.10, ** p$<$0.05, *** p$<$0.01.  \vspace{0.1cm}

\footnotesize
 \centering \setlength{\tabcolsep}{2pt}
\begin{adjustbox}{max width=\textwidth}
\begin{tabular}{@{}l*{7}{c}@{}} 
\toprule
\textit{Specification:} &  \multicolumn{7}{c}{\textit{Long differences: 2024 vs. 2022}}\\
\cmidrule(lr){2-8} 
\addlinespace
\textit{Dep. var.} &  $\substack{\Delta \text{ Log Duration} \\ \text{All}}$  &   $\substack{\Delta \text{ Log Duration} \\ \text{Leisure}}$ &  $\substack{\Delta \text{ Log Duration} \\ \text{Productive}}$   &  $\substack{\Delta \text{ Log Duration} \\ \text{Mixed}}$   &  $\substack{\Delta \text{ Share} \\ \text{Leisure}}$  &  $\substack{\Delta \text{ Share} \\ \text{Productive}}$  &  $\substack{\Delta \text{ Share} \\ \text{Mixed}}$ \\
   \cmidrule(lr){2-2}   \cmidrule(lr){3-3}   \cmidrule(lr){4-4}   \cmidrule(lr){5-5}   \cmidrule(lr){6-6}   \cmidrule(lr){7-7}   \cmidrule(lr){8-8} 
 &  (1) & (2) & (3) & (4) & (5) & (6) & (7) \\
   \addlinespace
   \midrule
\addlinespace ChatGPT Use&       1.243** &       1.512** &       0.011   &      -0.285   &       0.307***&      -0.215***&      -0.092   \\
            &     (2.409)   &     (2.106)   &     (0.021)   &    (-0.449)   &     (4.103)   &    (-2.738)   &    (-1.609)   \\
\addlinespace Observations&      42,886   &      42,886   &      42,886   &      42,886   &      42,886   &      42,886   &      42,886   \\
1st-stage KP F-stat.&         104   &         104   &         104   &         104   &         104   &         104   &         104   \\

\midrule
 	Estimation & IV & IV & IV & IV & IV & IV & IV   \\ 
Inc. X Age X MSA FEs     & X & X	& X & X	& X & X	& X      \\	
Bartik share control     & X & X	& X & X	& X & X	& X      \\	
Website category exposure    & X & X	& X & X	& X & X	& X      \\	
    \bottomrule
\end{tabular}
\end{adjustbox}
\end{table}

\section{Mechanism: What Do Households Use ChatGPT For?} \label{sec:mechanism_empirics}

In this section, we provide further evidence of the mechanism for \textit{why} ChatGPT adoption would increase the leisure share of household browsing while leaving the productive task share unchanged. We provide support for our interpretation of this empirical pattern as showing that ChatGPT primarily raises the efficiency of productive (non‑market) digital tasks, freeing up time that is reallocated toward leisure digital tasks. Importantly, this interpretation assumes that the leisure increase does not result from ChatGPT itself mainly being used to support leisure browsing.

We provide analyses to better understand this mechanism that answer the following questions: (1) Which types of websites are households using less or more after they adopt generative AI? (2) What types of websites and digital tasks (productive vs. leisure) are households engaging with in the context of generative AI use (right before or after using ChatGPT)? 

\subsection{Effects of ChatGPT adoption on browsing by website content category}
 This section shows which categories of websites, based on their Comscore content category, experienced an increase or decrease in their share of browsing activity as a result of ChatGPT adoption. The results are shown in Table \ref{table:2sls-bycontent}. As the table shows, households that adopt ChatGPT decrease their browsing time spent on search-related websites (column 1). They also decrease the time spent on news sites (column 11), while increasing the time spent on gaming sites (column 13).  Note that these findings directly align with our proposed mechanism: categories of websites that are likely used for productive Internet activities, such as search or news consumption, see a decline in their relative share. Importantly, these are the types of websites that generative AI chatbots are most likely to substitute for.  At the same time, the increase in website activity that we observe for gaming sites is unlikely to result from chatbots being directly useful for increasing the productivity of leisure activities related to gaming, as there is not an obvious overlap between basic ChatGPT capabilities and gaming tasks.

\begin{table}[htbp]
\caption{\\ \centering \textbf{Effect of ChatGPT adoption on browsing activity by Comscore category}}  

\vspace{-0.1cm} \label{table:2sls-bycontent} \footnotesize  This table shows estimates of long difference IV regressions in the same specification as in Table \ref{table:2sls-bycat}, but with the change in the share of browsing in different Comscore content categories from 2022 to 2024 as the dependent variable. T-statistics based on heteroskedasticity-robust standard errors clustered at the income bin X age bin X MSA level in parentheses: * p$<$0.10, ** p$<$0.05, *** p$<$0.01. 

\vspace{0.1cm}
 \centering \setlength{\tabcolsep}{2pt}
\begin{adjustbox}{max width=\textwidth}
\begin{tabular}{@{}l*{14}{c}@{}} 
\toprule
\textit{Dep. var.:} &  \multicolumn{7}{c}{\textit{Long difference change in browsing share: 2024 vs. 2022}}\\
\midrule
 \multicolumn{8}{l}{\textit{Panel A: Productivity-related content categories}} \\
 &  (1) & (2) & (3) & (4) & (5) & (6) & (7)  \\
            &      Search   &     Educat.   &    Shopping   &  Technology   & Information   &    Finances   &$\substack{\text{Job} \\  \text{Search}}$   \\
\midrule \addlinespace ChatGPT Use&      -0.177** &       0.053   &      -0.024   &       0.029   &       0.035*  &       0.052   &      -0.022   \\
            &    (-2.200)   &     (1.330)   &    (-0.634)   &     (0.893)   &     (1.843)   &     (1.370)   &    (-1.250)   \\

\midrule
 \multicolumn{8}{l}{\textit{Panel B: Leisure-related content categories}} \\
 &  (8) & (9) & (10) & (11) & (12) & (13) & (14)  \\
            &   Communic.   &$\substack{\text{Social} \\  \text{Media}}$   & Promotional   &        News   &$\substack{\text{Enter-} \\  \text{tainment}}$   &       Games   &       Other   \\
\midrule \addlinespace ChatGPT Use&      -0.031   &       0.014   &       0.242   &      -0.039***&       0.080   &       0.099***&      -0.041   \\
            &    (-1.280)   &     (0.397)   &     (1.451)   &    (-3.069)   &     (1.447)   &     (2.636)   &    (-0.587)   \\
\midrule \addlinespace Observations&      42,886   &      42,886   &      42,886   &      42,886   &      42,886   &      42,886   &      42,886   \\
1st-stage KP F-stat.&         104   &         104   &         104   &         104   &         104   &         104   &         104   \\

 	Estimation & IV & IV & IV & IV & IV & IV & IV   \\ 
Inc. X Age X MSA FEs      & X & X	& X & X	& X & X	& X      \\	
Bartik share control     & X & X	& X & X	& X & X	& X      \\	
Website cat. exposure    & X & X	& X & X	& X & X	& X      \\	
    \bottomrule
\end{tabular}
\end{adjustbox}
\end{table}

\begin{figure}
\caption[.]{\textbf{Share of GenAI-exposed website activity by type}}\label{fig:exposuretypes}

\vspace{-0.1cm} \small  This figure shows the share by duration of different website activities among those that are classified as `exposed' to generative AI by the methodology detailed in Appendix \ref{ia.sec:appx_methodology}. 

\centering
\includegraphics[width=0.65\textwidth]{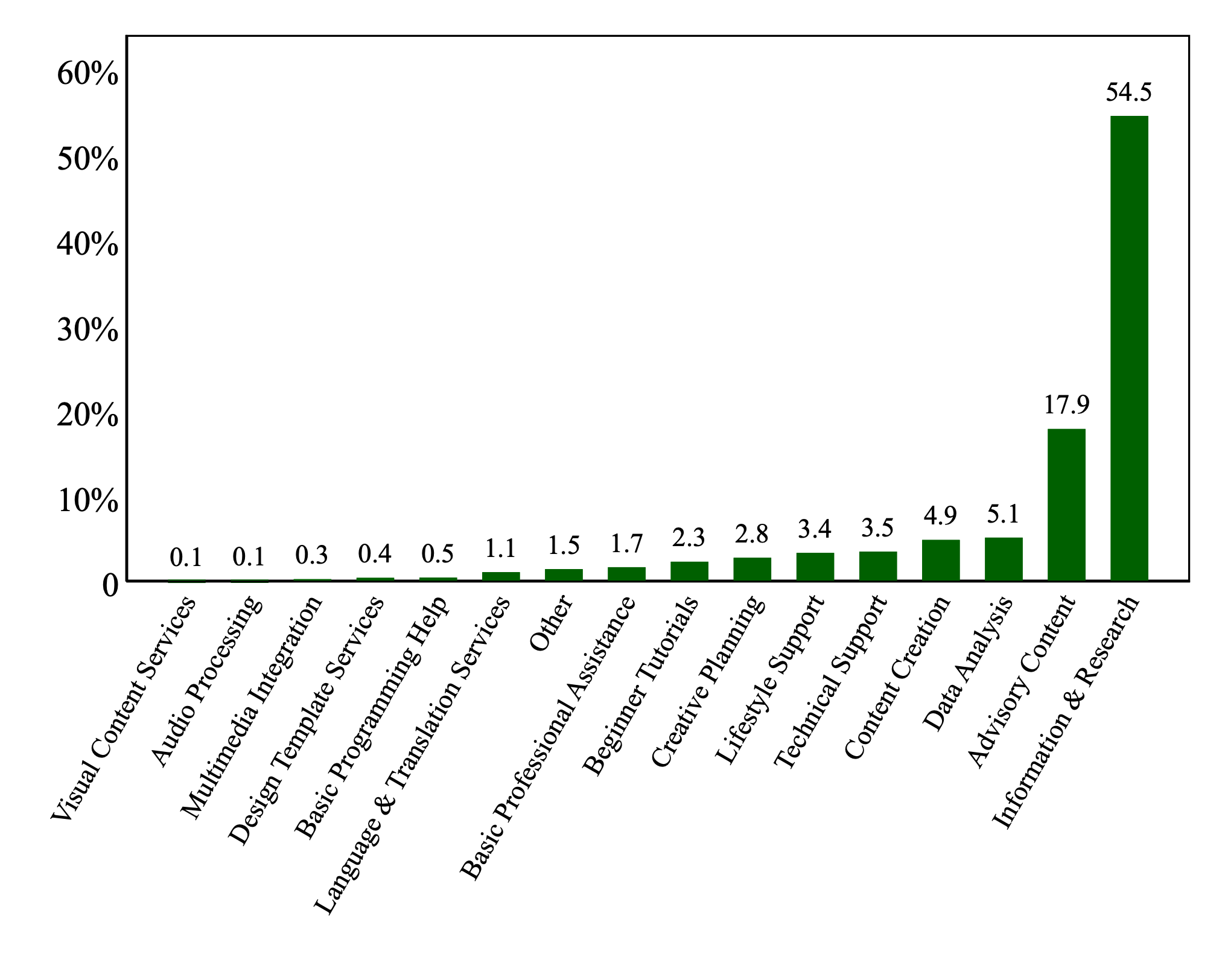} 
\end{figure}

 \paragraph{GenAI-exposed activities.} For comparison, we also assign each activity on a website that is labeled as `exposed' by our LLM classification pipeline into a broader category based on the type of activity that was labeled. The distribution of activities labeled as exposed over these categories is shown in Figure \ref{fig:exposuretypes}. As the graph shows, more than half of the activities that we determine ex ante to be most exposed to the substitution by generative AI fall into the category of `Information \& Research', which closely aligns with the search and news website categories for which we estimate that ChatGPT adoption actually causes a drop in browsing share.

\subsection{ChatGPT use context in high-frequency browsing data}

\noindent ChatGPT, as a powerful general-purpose AI tool, can potentially be used to fulfill many tasks and needs of households. An important empirical question in the context of our study is what \textit{tasks} households actually choose to use ChatGPT for. While observing households' questions to ChatGPT would provide some information on this (see \cite{chatterji_how_2025} for evidence on this), we do not have access to chatbot interaction data. However, such interaction data could only provide a partial answer as it would lack the context of the broader task the household is performing on the Internet: any task that involves websites other than a chatbot would only be partially captured this way.
For example, an inquiry like ``\textit{Can you give me a simple explanation of the Black-Scholes formula?}'' inside ChatGPT may reflect a student preparing for an exam in finance, a financial analyst preparing for a client presentation, or a retail trader evaluating an investment. What is lacking for evaluating the task that is being done is the \textit{context} of the chatbot interaction.

Conversely, in this section, we argue that one can use this intuition and infer households' purpose for using ChatGPT based on the browsing context of the chatbot interaction. We leverage our high-frequency browsing data to observe the websites that users visit in the minutes before or after they use ChatGPT (the ``GPT-window''). This allows us to identify users' purpose of using ChatGPT. We also observe the websites that users visit outside the ChatGPT window (``NonGPT-window''), which allows us to assess how the usage of ChatGPT impacts households' task allocation in general. 

Our approach starts with identifying about 3.6 million 30-minute intervals of our sample households' Internet visitation from 2022 to 2024. For each interval, we have the duration of the household's visitation of productive, leisure, mixed, and ad/CDN websites and a label of whether the household visited openai.com or chatgpt.com in the interval. On average, about 0.2\%  and 1.1\% of the intervals are labeled as being a GPT-window in 2023 and 2024, respectively.  For several of the analyses below, we want to compare browsing activity in the intervals around ChatGPT use to what is `normal'. As a benchmark, we match ChatGPT use intervals to browsing activity by households that never use ChatGPT and are in the same age and income categories. Moreover, for general comparisons between users and non-users we also match browsing activity that occurs on the same day of the week and at the same time of day, which avoids, for example, bias from comparing weekend browsing to work day evening browsing.

\paragraph{Websites associated with ChatGPT use.} To provide some intuition for what kinds of Internet tasks are associated with ChatGPT use, we first consider which websites show the greatest  difference in browsing shares in the intervals around ChatGPT use relative to matched browsing intervals by non-users. Table \ref{tab:topsites-gptwindow} shows the 30 sites with the largest gap. That is, these are the sites for which browsing time is unusually high around ChatGPT use. We find that search on google.com is  unusually high. Moreover, unusually prevalent are also education sites such as Pearson.com or Instructure.com; the graphic design platform Canva; writing assistant platform Quillbot.com; and job search platforms/professional networks Linkedin and Indeed. This shows that ChatGPT use tends to be associated with sites commonly used for productive digital tasks. In Table \ref{table:bottomsites-gptwindow} in the Appendix, we show that the sites that are unusually lacking from ChatGPT use intervals include general entertainment/streaming sites like YouTube, MSN.com, and Yahoo; the social network Facebook; and also some shopping sites like Amazon and eBay, as well as several adult content sites. Moreover, Wikipedia also has unusually low prevalence in ChatGPT use intervals, showing that some sites with high exposure are substituted by chatbot use. Overall, these simple descriptive patterns in our data align closely with our interpretation that ChatGPT use tends to occur in the context of websites used for productive digital tasks, and not in the context of leisure-oriented entertainment and social media sites. This supports our assumption that ChatGPT itself is more likely to be used for these productive tasks.

\begin{table}[htbp]
\caption{\\ \centering \textbf{Websites with highest relative prevalence around ChatGPT use.}}  

\vspace{-0.1cm} \label{tab:topsites-gptwindow} \footnotesize  This table shows websites with the {highest} prevalence in 30 minute intervals with ChatGPT use compared to non-user browsing matched on time of day and age bin $\times$ income bin demographics.

\vspace{0.2cm}

\centering
\begin{adjustbox}{max width=\textwidth}   
\begin{tabular}{@{}l*{4}{c}@{}}
\toprule
Rank & Domain & $\Delta$ Browsing \% \\
\midrule
      \addlinespace
                   1&google.com&16\\
2&instructure.com&4.7\\
3&canva.com&2.2\\
4&quillbot.com&1.3\\
5&discord.com&1.3\\
6&gist.build&.75\\
7&quizlet.com&.48\\
8&indeed.com&.45\\
9&pearson.com&.44\\
10&brightspace.com&.44\\
11&blackboard.com&.42\\
12&linkedin.com&.38\\
13&cengage.com&.36\\
14&webassign.net&.35\\
15&uworld.com&.34\\
  
                   \bottomrule
                   \end{tabular}
\end{adjustbox}
\begin{adjustbox}{max width=\textwidth}
\begin{tabular}{@{}l*{4}{c}@{}}
\toprule
Rank & Domain & $\Delta$ Browsing \% \\
\midrule
      \addlinespace
                   16&mheducation.com&.34\\
17&grammarly.com&.29\\
18&udemy.com&.24\\
19&googleapis.com&.24\\
20&schoology.com&.24\\
21&edgenuity.com&.23\\
22&eesysoft.com&.22\\
23&coinbase.com&.2\\
24&edmentum.com&.2\\
25&github.com&.19\\
26&wwnorton.com&.17\\
27&chegg.com&.17\\
28&character.ai&.14\\
29&sharepoint.com&.13\\
30&office365.us&.11\\
  
                   \bottomrule
                   \end{tabular}
\end{adjustbox}
\end{table}

\begin{figure}[t]
\caption{\textbf{Website content types during ChatGPT sessions}}

\label{fig:hf-comscorecats} \footnotesize
This figure plots the level (panel A) and difference (panel B) of browsing duration shares by websites' Comscore content categories during a ChatGPT use session and matched  never-user browsing sessions, matched on time and demographics.
\smallskip

\centering
\begin{subfigure}{.49\textwidth}
  \centering
  \includegraphics[width=\linewidth, trim={0.2cm 0.4cm 0.2cm 0.2cm},clip]{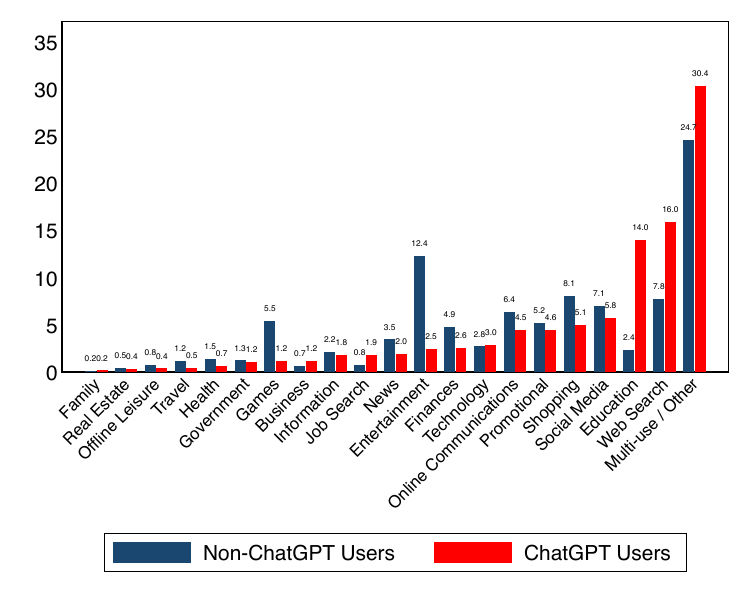}
  \caption{Browsing shares by website content category}
\end{subfigure}
\begin{subfigure}{.49\textwidth}
  \centering
  \includegraphics[width=\linewidth, trim={0.2cm 0.5cm 0.2cm 0.2cm},clip]{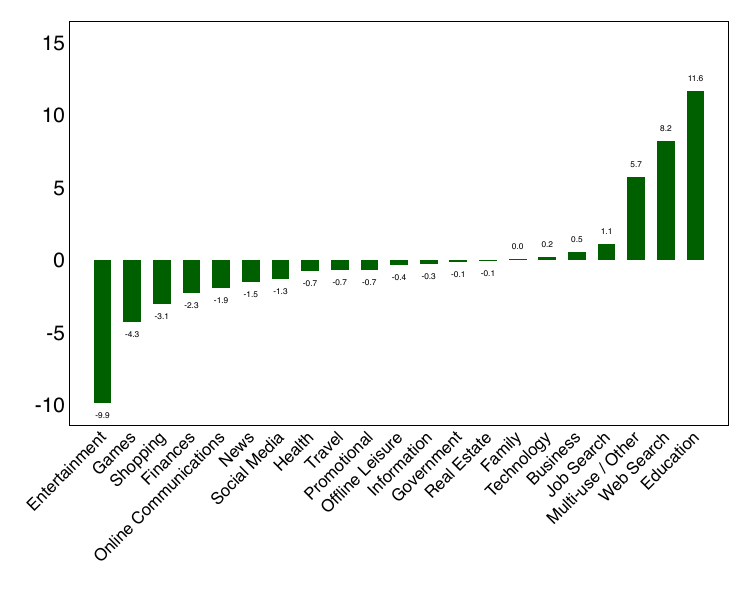}
  \caption{Difference between GPT windows and never-users}
\end{subfigure} 

\end{figure}

\paragraph{Website content types associated with ChatGPT use.} We can compare ChatGPT use intervals and matched non-user browsing behavior more systematically by considering which types of Comscore website categories are over- or under-represented in browsing behavior around ChatGPT uses. Figure \ref{fig:hf-comscorecats} shows that both ChatGPT use intervals and the matched browsing behavior by never-users show a large prevalence of browsing on multi-use sites, search engines, and on education sites (panel A). However, the ChatGPT use intervals show, in relative terms, a much larger prevalence of these three website categories, and a much lower prevalence of browsing on entertainment, gaming, and shopping sites (panel B).

\begin{figure}[t]
\caption{\textbf{Household browsing activities during ChatGPT session}}
\label{fig:hf-leisureprod}

This figure plots the difference between households' browsing activities in leisure, productive, and other categories during the ChatGPT session relative to outside of the ChatGPT session. See more details in Section \ref{sec:mechanism_empirics}. 
\bigskip

\centering
\begin{subfigure}{.3\textwidth}
  \centering
  \includegraphics[width=\linewidth]{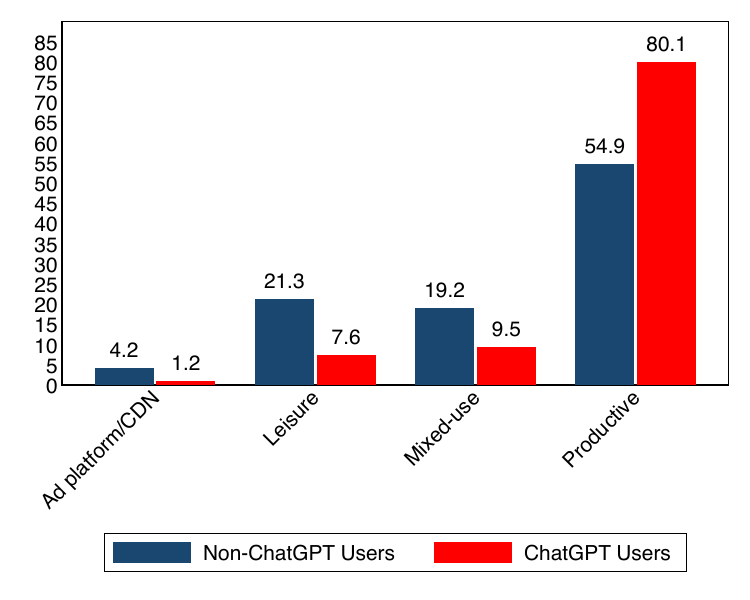}
  \caption{GPT window: \\ browsing shares}
\end{subfigure}
\begin{subfigure}{.3\textwidth}
  \centering
  \includegraphics[width=\linewidth]{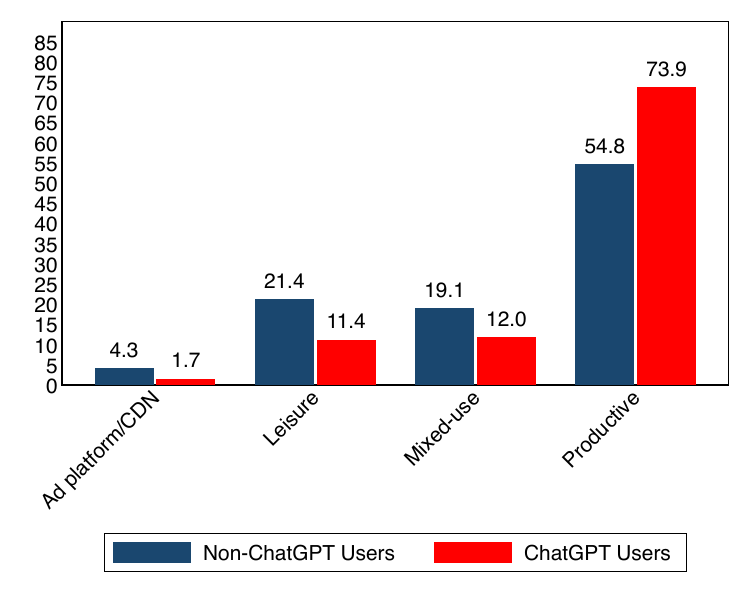}
  \caption{GPT day, outside window: \\ browsing shares}
\end{subfigure} 
\begin{subfigure}{.3\textwidth}
  \centering
  \includegraphics[width=\linewidth]{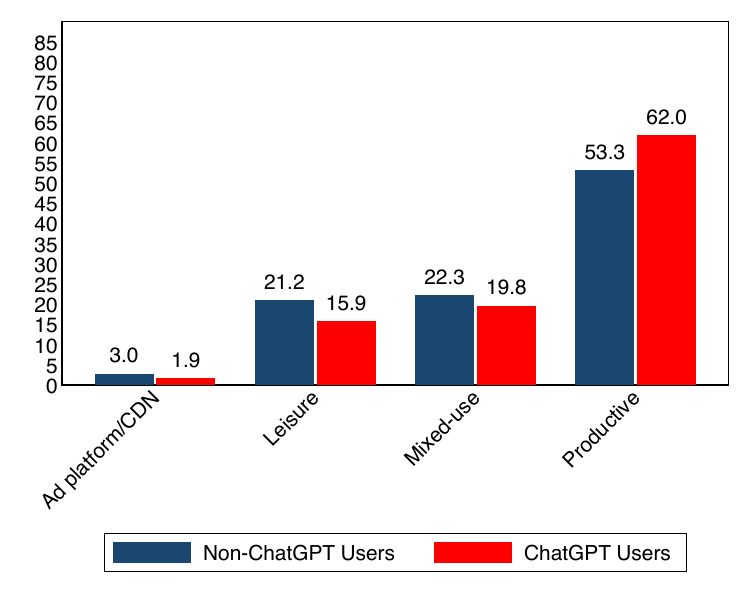}
  \caption{Pre-GPT use: \\ browsing shares}
\end{subfigure} 
\begin{subfigure}{.3\textwidth}
  \centering
  \includegraphics[width=\linewidth]{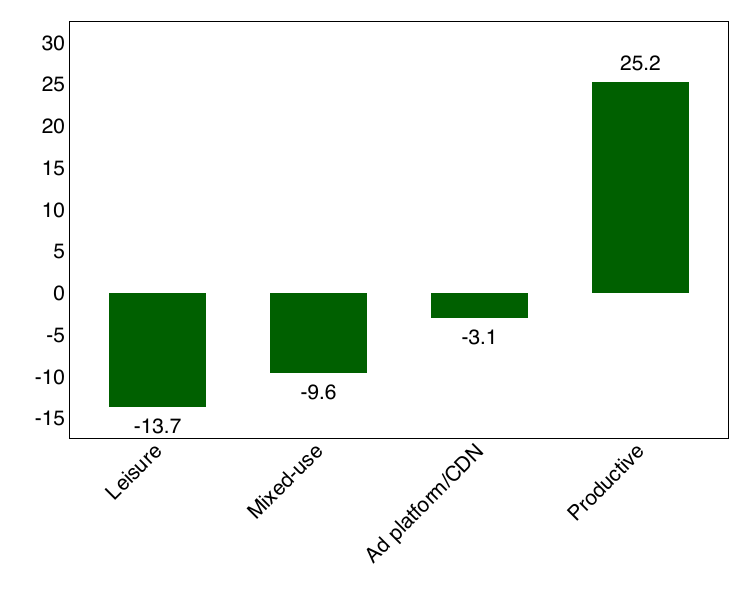}
  \caption{GPT window: \\ differences in shares}
\end{subfigure}
\begin{subfigure}{.3\textwidth}
  \centering
  \includegraphics[width=\linewidth]{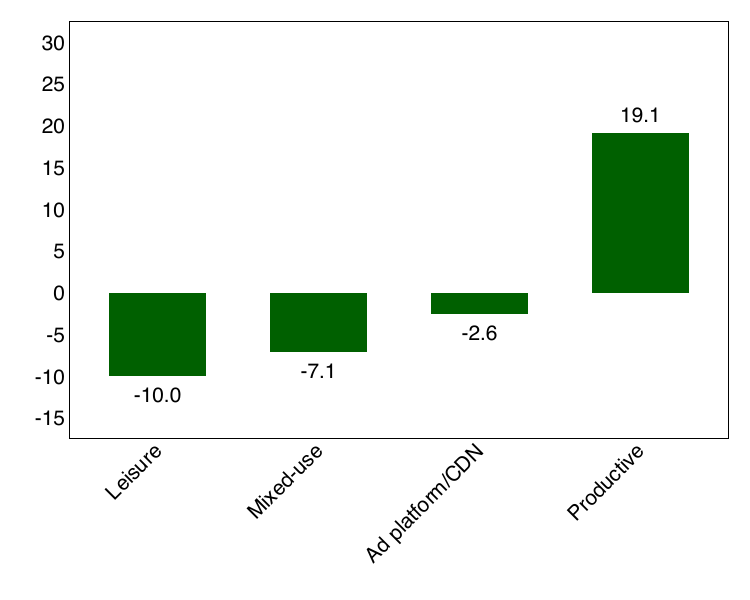}
  \caption{GPT day, outside window: \\ differences in shares}
\end{subfigure} 
\begin{subfigure}{.3\textwidth}
  \centering
  \includegraphics[width=\linewidth]{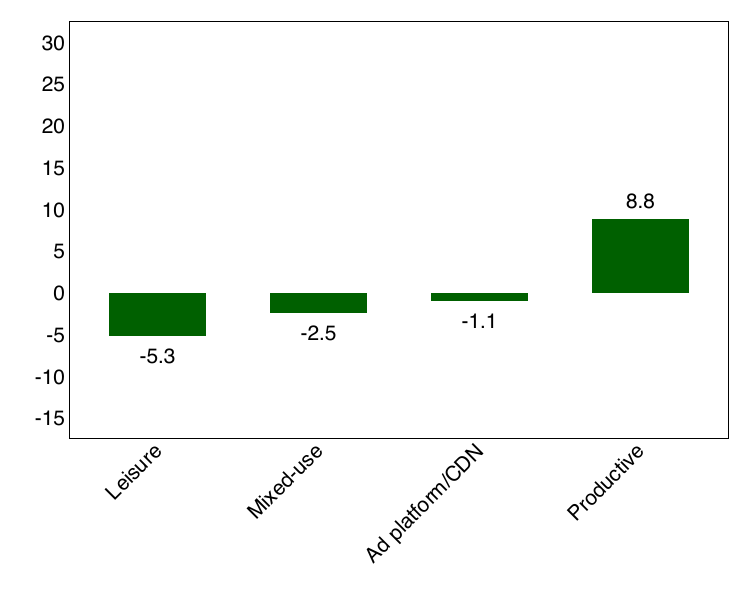}
  \caption{Pre-GPT use: \\ differences in shares}
\end{subfigure} 
\end{figure}

\paragraph{Task purposes associated with ChatGPT use.}
Finally, we can map the behavior around ChatGPT uses to our distinction between sites associated with productive and leisure digital tasks (as well as mixed-use and ad/CDN sites for completeness). Figure \ref{fig:hf-leisureprod} shows the distribution of the four types of website purposes during the GPT-window relative to time- and demographics-matched non-user browsing behavior. On average, during the 30-minute window when households use ChatGPT, the tasks they focus on are 80.1\% productive, 9.5\% mixed-use, 7.6\% leisure, and 1.2\% ad/CDN sites (panel A). Relative to never-user browsing (see panel D), GPT-window browsing focuses substantially more on productive tasks---a 25.2pp gap---and substantially less on leisure tasks, with a -13.7pp gap.

We explore further differences in the distribution of browsing task purposes between ChatGPT users and never-users in the other of Figure \ref{fig:hf-leisureprod}. Panels (B) and (E) show that ChatGPT users are generally more likely to do productive digital tasks than never-users on days when they use ChatGPT. However, this behavioral gap is not as large as the gap during the ChatGPT use interval itself. Panels (C) and (F) show that there is even a slight bias towards doing more productive digital tasks before ChatGPT is released among ChatGPT users---in fact, differences of this type underlie the intuition why our pre-ChatGPT household exposure measure is able to predict ChatGPT adoption. However, this bias is much smaller than the latter gap in productive task browsing associated with ChatGPT use intervals. This evidence  confirms that productive digital tasks are much more likely to be done in browsing sessions that involve ChatGPT use, and that this association cannot be explained by a selection effect with regard to on what days households use ChatGPT or which households choose to use ChatGPT at all.

Together, the mechanism evidence in this section provides an empirical justification for interpreting our baseline IV estimates as reflecting a productive digital task-biased productivity shock that frees up time which is re-allocated to leisure digital tasks.


\paragraph{Methodological limitations.} We end this section by discussing the premises and limitations of this high-frequency browsing interval approach. Our approach is based on two premises: First, we view ChatGPT as a \textit{tool} for fulfilling an ongoing task the household is working on. Hence, if we observe proxies for the tasks the household is performing on the Internet around her use of ChatGPT, e.g., writing emails or visiting personal relationship therapy websites, we can approximately infer the tasks the household uses ChatGPT for. Our second premise assumes that the household performs associated tasks within a short time window on the Internet. Together, these premises allow us to approximately infer the tasks that households perform using the generative AI tool by examining high-frequency data on browsing activity in short browsing intervals in which ChatGPT is used. 

Our approach faces two potential sources of noise: First, households may resort to ChatGPT for tasks without visiting other websites. For example, as noted above, searching for factual information may no longer involve Wikipedia as ChatGPT becomes the first and only website to consult---in which case no trace of the completed task remains in the browsing context of ChatGPT. Second, households may be multi-tasking when visiting ChatGPT, making the surrounding websites less informative about the household's purpose of using ChatGPT. While this noise may lead to bias in the magnitudes found in our analysis, we would still expect our results to qualitatively reflect the average purpose of ChatGPT usage by households.

\section{Quantifying Household Productivity Gain} \label{sec:conceptual-framework}
In this section, we develop a simple conceptual framework that adapts the model of \cite{aguiar2021} to understand household choices of online activity. This framework allows us to map our empirical measures of changes in households' time spent on productive and leisure activities to the impact of generative AI as a productivity-enhancing tool.  The conceptual framework shows that to translate these reallocations into an implied productivity improvement for digital tasks, we need to combine
 two ingredients:
\begin{enumerate}
    \item {Causal adoption effects on time use across categories.} These are our empirical results that identify how adopting ChatGPT reallocates households' \emph{digital time} across productive and leisure browsing (see the IV estimates in Table \ref{table:2sls-bycat}).
    \item {Time-demand (``Engel curve'') elasticities} that govern how category-level time
    responds to shocks to total digital time (estimated below in Table \ref{tab:engel_regs_main}).
\end{enumerate}

Below, we (1) derive expressions for changes in digital time allocation in response to ChatGPT adoption; and then (2) show how we can estimate ``Engel curve'' elasticities for digital time allocation from our Internet browsing data; (3) Finally, we  combine these estimates into an implied magnitude of the household productivity change as a result of ChatGPT adoption.

\begin{figure}[!hbt]
\caption[.]{\textbf{Illustration: Generative AI and Household Production}} 

\vspace{-0.1cm}  \small This figure illustrates how a change in the productivity of productive household activities would be expected to change household time allocation.  \\
\label{fig:model} 

    \centering
  \includegraphics[width=0.5\linewidth]{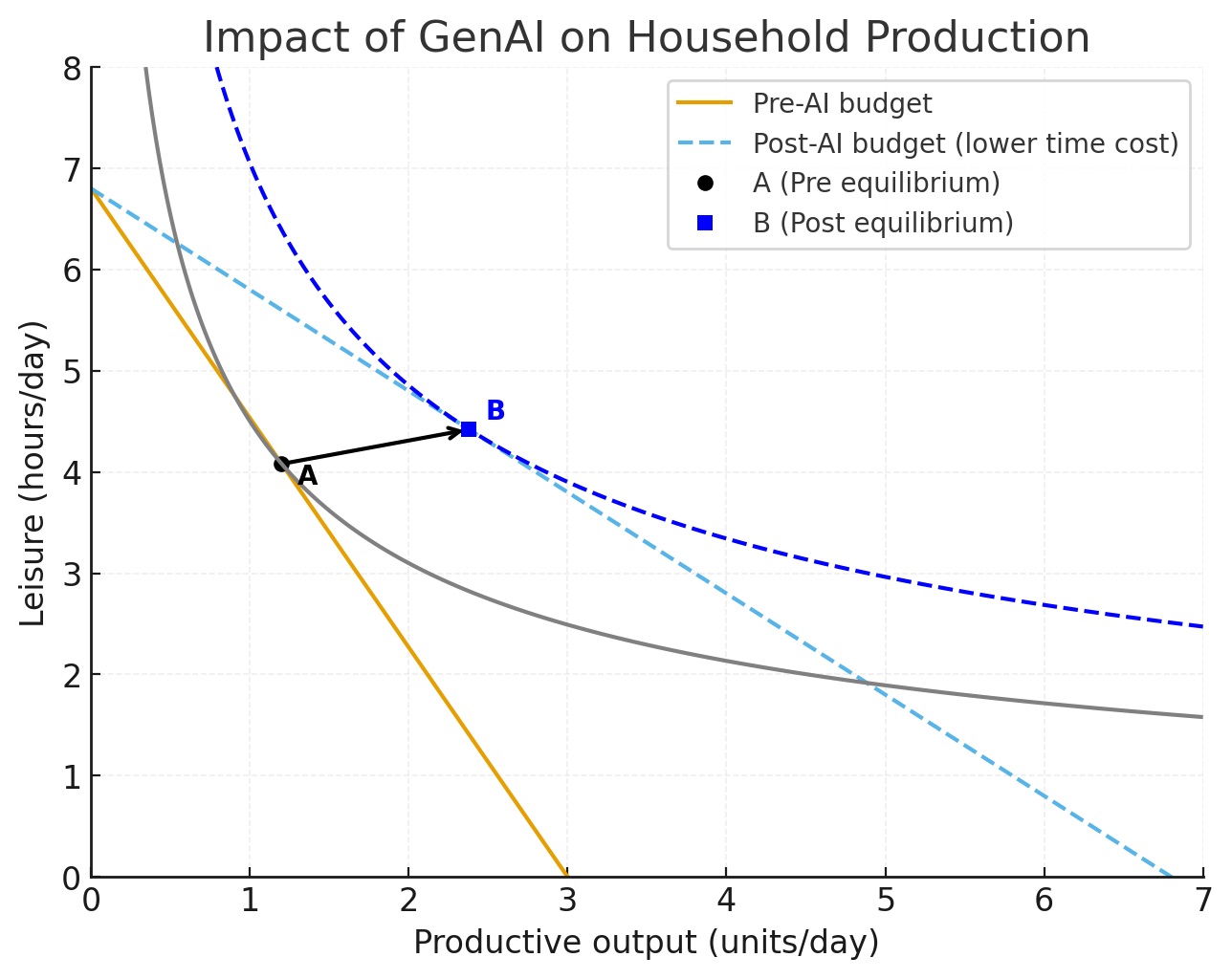}
\end{figure}

\textbf{Model intuition.} Before presenting the full model below, the core intuition is illustrated in Figure \ref{fig:model}: we consider a household that faces a budget constraint with regard to how to allocate effective browsing time between leisure purposes (effectiveness normalized to one and output shown on the vertical axis) and productive purposes (horizontal axis). The solid budget line represents the household's time budget before generative AI. The household chooses the time allocation where the effective time budget is tangent to the iso-utility curve. As generative AI makes it faster to complete productive tasks, this expands the household's effective time budget to the dotted line. The impact on relative time allocation depends on the elasticity of substitution between leisure time and productive time spent browsing. The illustrated case shows how the expansion of the effective time budget due to the effect of generative AI on productive tasks increases leisure browsing time. Note that the horizontal axis shows effective time units after accounting for productivity. As less time is needed post-GenAI to produce a given effective productive digital task output, even the substantial increase in effective time output from productive digital tasks that is shown can be produced with a smaller `clock time' input of browsing duration, so productive digital task time allocation falls. The full model below captures this overall mechanism in a more general setting.

\subsection{Model setup and household digital time allocation}

\paragraph{Household utility and time allocation problem}

We consider a household $i$ that, at each time $t$, spends a fixed amount of time (normalized to one) engaged in digital activities at home. The household allocates an amount of time $\ell_{it}$ to \textit{digital leisure consumption} (e.g., streaming movies, scrolling social media) and $z_{it}$ to \textit{digital home‑production} tasks (paying bills, searching for information, planning, and other chores that could be sped up by ChatGPT), so the period-by-period time budget is 
\begin{equation}
\label{eq:stylized-model-time-constraint}
    z_{it} + \ell_{it} \le 1.
\end{equation}
In the quantification below, we relax the normalization and allow total digital time to vary, so the model can map directly to observed variation in total browsing time.

The household optimally allocates time between these two digital activities to maximize flow utility, given by the power utility function 
\begin{equation} \label{eq:aggregator}
    v_{it} = \frac{(\theta_{it}^{\ell} \xi_{it}^{\ell} \ell_{it})^{1-(1/\eta^{\ell})} }{1-(1/\eta^{\ell})}  + \frac{(\theta_{it}^{z} \xi_{it}^{z} z_{it})^{1-(1/\eta^{z})} }{1-(1/\eta^{z})},  
\end{equation}
where $\theta^\ell_{it}$ and $\theta^z_{it}$ are taste shifters (preference weights) for leisure and productive digital activities,
and $\xi^\ell_{it}$ and $\xi^z_{it}$ are activity-specific efficiency shifters which affect how much utility-relevant output per unit time the activity generates.
The curvature parameters $\eta_\ell, \eta_z > 0$ govern diminishing marginal utility of time within each activity.
When $\eta_a > 1$, activity $a$ is a \emph{time luxury} (analogous to \cite{aguiar2021} ``leisure luxuries''): higher effective quality $\theta^a \xi^a$ increases the optimal time
allocated to $a$. When $0 < \eta_a < 1$, activity $a$ is a \emph{time necessity}: higher $\theta^a \xi^a$ reduces time allocated
to $a$, because the household reaches diminishing returns quickly and reallocates time elsewhere.
The knife-edge case $\eta_a = 1$ corresponds to log utility in activity $a$. The model generalizes to more than two digital activities (e.g., to incorporate the `mixed' sites from our empirical design), but we simplify the exposition of the conceptual framework here by focusing on leisure and productive purposes. We consider the effect of adding mixed sites when calibrating the productivity shock estimates below.

\paragraph{First order conditions}
This optimization problem for each activity $a\in\{z,\ell\}$ yields a first-order condition which can be written as an activity time allocation equation (after taking logs of both sides) in the form
\begin{equation}
\label{eq:stylized-model-productive-activity-foc}
     \text{ln}(a_{it}) = (\eta^a - 1) \times \big(  \text{ln}(\theta_{it}^a) +  \text{ln}(\xi_{it}^a) \big) - \eta^{a} \text{ln}(\omega_{it})   \qquad a\in\{z,\ell\}
\end{equation}
where $\omega_{it}$ is the shadow price of the time constraint. This equation shows that households allocate more time to an activity when its taste/productivity parameter is high  and less when the shadow cost of time is high---as long as the activity behaves like a ``leisure luxury'', i.e., $\eta > 1$. If $\eta < 1$, these effects reverse:   households allocate less time to an activity when its taste/productivity parameter increases. If leisure and productive digital activities differ in their behavior---e.g., because $\eta^z < \eta^{\ell}$, as we show empirically below---then even an identical productivity change in both activities can lead to a relative change in the share of browsing time allocated to them. See the Internet Appendix for details on the derivation.

\subsection{The Release of ChatGPT and the choice to adopt}

We now consider how, within this basic framework, the release of ChatGPT affects the household's time allocation constraints and decisions. We model ChatGPT's release as being a task-biased technological shock to the household's digital home activities. In particular, we analyze the impact of a potential shock to the efficiency of the household's completion of digital production tasks, shifting $\xi^z$ while leaving $\xi^{\ell}$, $\theta^z$ and $\theta^{\ell}$ unchanged.

For simplicity, assume that there are just two periods, $t = pre$ and $t = post$, and that ChatGPT is released in between (i.e., after $t=pre$ and before $t=post$). At the start of $t = post$, the household chooses whether or not to pay some cost to adopt ChatGPT. Adopting ChatGPT increases the efficiency $\xi_{it}^{z}$ of the household's digital home production. In particular, assume that this object's time $pre$ value equals $ \xi_{i,pre}^{z} = \underline{\xi}$
while its time $post$ value is 
\begin{equation}
    \xi_{i,post}^{z} = \underline{\xi} \times \left( 1 +  \delta \cdot \text{Adopt}_{i,post} \right)
\end{equation}
for $\text{Adopt}_{i,post} \in \{0,1\}$ an indicator for whether the household chooses to adopt ChatGPT and for $\delta > 0$ the magnitude by which ChatGPT improves efficiency on the productive digital tasks. Note that if the household does not adopt ChatGPT, then their efficiency on these tasks does not change across the two periods. In what follows, it is also convenient to work with the log technology shock (conditional on adoption)
$\Delta \ln \xi^z \equiv \ln(1 + \delta)$, which is approximately $\delta$
for small productivity gains.

The household chooses to adopt ChatGPT if 
\begin{equation}
    v_{i,post}^{\text{Adopt}} - v_{i,post}^{\text{No adopt}} \ge c 
\end{equation}
for $ v_{i,post}^{\text{Adopt}}$ and $ v_{i,post}^{\text{No adopt}} $ the indirect utility functions if, respectively, the household adopts or does not adopt. We use superscripts `Adopt' and `No adopt' to denote the optimal choices under the two technology regimes, holding fixed all other shocks in period \(t=\text{post}\). Here $c$ is measured in units of period utility, so the left‑hand side is the utility gain from faster productive tasks and the right‑hand side is the utility cost of learning and adopting ChatGPT.

A first-order approximation assuming small productivity effects delivers a simple adoption rule: if adopting ChatGPT raises the efficiency
of productive digital tasks by a factor $1 + \delta$, and adoption entails a fixed time cost $c_{\text{time}}$,
then adoption is optimal when the implied time savings relative to the counterfactual exceed the learning/setup time. Formally,
adoption occurs when
\[
\delta \cdot z^{\text{No adopt}}_{i,\text{post}} \geq c_{\text{time}},
\]
with derivations in the Appendix. This maps directly to our empirical ``exposure'' measure: households
with higher baseline intensity of GenAI-substitutable productive browsing have larger potential gains
and should adopt at higher rates.

\subsection{Browsing effects of ChatGPT adoption}

Denote the treatment effect of adopting ChatGPT on the household's time spent on productive digital tasks as $\beta^{GPT}_{z} \equiv \text{ln}(z_{i,post}^{\text{Adopt}}) - \text{ln}(z_{i,post}^{\text{No adopt}})$, where $z_{i,post}^{\text{Adopt}}$ and $z_{i,post}^{\text{No adopt}}$ are the household's optimal time allocated to productive digital tasks under, respectively, the decision to adopt or not adopt ChatGPT. Denote $\beta^{GPT}_{\ell} \equiv \text{ln}(\ell_{i,post}^{\text{Adopt}}) - \text{ln}(\ell_{i,post}^{\text{No adopt}})$ as the analogous object for time spent on leisure tasks. 

By plugging the time constraint (\ref{eq:stylized-model-time-constraint}) into the FOCs (\ref{eq:stylized-model-productive-activity-foc}) and differencing the household's allocation of time across the two types of digital activities, we can characterize how the household's relative time allocation changes due to the adoption of ChatGPT. The relative impact on time allocated to productive tasks versus leisure tasks is 
\begin{equation}
\label{eq:stylized-model-effect-difference}
\frac{\beta^{GPT}_{z}}{\eta^z} - \frac{\beta^{GPT}_{\ell} }{\eta^{\ell}} = \frac{\eta^z - 1}{\eta^z} \times \ln(1+ \delta_z )
\end{equation}
while the impact on time allocated to productive tasks is
\begin{equation}
\label{eq:stylized-model-effect-productive}
\beta^{GPT}_{z} = \frac{(\eta^z - 1 )}{1 + \left(\eta^{z} / \eta^{\ell} \right)\cdot \left(z_{i,post}^{\text{No adopt}} / \ell_{i,post}^{\text{No adopt}}\right)}  \times  \ln(1+ \delta_z )
\end{equation}
Equation (\ref{eq:stylized-model-effect-productive}) shows that the effect of adoption on the amount of time spent on productive digital activities--i.e., the tasks that are directly made more efficient by ChatGPT--can be either positive or negative. In particular, ChatGPT adoption leads to more (less) time spent on productive tasks, $\beta_z^{GPT} > 0$ ($\beta_z^{GPT} < 0$) if the elasticity of utility with respect to productive browsing, $\eta^z$, is greater than (less than) one. Intuitively, if $\eta^z < 1$, then the positive substitution effect from adopting ChatGPT--productive tasks becoming cheaper (in terms of foregone time) to do--is dominated by a negative income effect--the marginal productive task being less valuable the more time that the household has to devote to home computing. The income effect is negative when $\eta^z < 1$, and gets more negative the lower $\eta^z$ is, due to the decreasing returns that the household gets from completing additional productive tasks. 

If there is no direct impact on digital leisure efficiency as a result of ChatGPT use, the effect on the time allocated to leisure tasks takes a form that is similar but in which $\eta^z$ enters with the opposite sign: 
\begin{equation}
\label{eq:stylized-model-effect-leisure}
\beta^{GPT}_{\ell}  =  \frac{(1-\eta^z) \cdot \frac{\eta^{\ell}}{\eta^z}}{1 + \left(\eta^{\ell} / \eta^z \right)\cdot \left(\ell_{i,post}^{\text{No adopt}} / z_{i,post}^{\text{No adopt}}\right)}  \times  \ln(1+ \delta_z )
\end{equation}
Adopting ChatGPT leads a household to spend more (less) time on leisure tasks if $\eta^z < 1$ ($\eta^z > 1$). The intuition for why is the same as with productive tasks, but in the opposite direction: when $\eta^z < 1$, a boost in the efficiency of productive tasks means that the household runs into decreasing returns in the value of such tasks, leading them to tilt their time allocation towards leisure tasks. Note that although this positive effect on leisure time is attenuated as $\eta^{\ell}$ itself goes down, the effect remains positive regardless of whether leisure is a luxury ($\eta^{\ell} > 1$) or an inferior  / necessity good ($\eta^{\ell} < 1$).

The key implication for our empirical results is straightforward. When productive digital tasks behave as a `necessity' ($\eta^z < 1$), the model predicts that a productivity shock to productive tasks reduces time spent on productive browsing and increases time spent on leisure. This is exactly the pattern we estimate: we find that ChatGPT adoption leads to a statistically significant rise in the leisure share of browsing and a decrease in the share of time spent on productive browsing activities  (see Table \ref{table:2sls-bycat}).

In the next section, we discuss how we obtain the ``Engel curve'' elasticities $\eta^a$ that we need to translate the IV estimates of changes in browsing behavior into estimates of the size of the productivity shock.

\subsection{Estimating Engel elasticities} \label{sec:engelest}

The previous sections showed how to derive an expression that relates  the causal effects of ChatGPT adoption on digital task consumption to the size of the productivity shock and the parameters $\eta^a$. In this section, we show how the parameters $\eta^a$ can be estimated.

Let total digital time in period $t$ be $H_{it}$, allocated across productive, leisure, and mixed
(``other'') digital activity:
\begin{equation}
    H_{it} = z_{it} + \ell_{it} + o_{it}.
\end{equation}

Under the additively separable isoelastic aggregator in equation  \ref{eq:aggregator}, optimal time allocations imply
constant-elasticity ``demands'' for time in each category. A key implication is that each activity
has an \emph{Engel elasticity in time},
\begin{equation}
    \beta_a \equiv \frac{\partial \ln h^a_{it}}{\partial     \ln H_{it}} = \frac{\eta^a}{\bar{\eta}}, 
    \qquad a \in \{z,\ell,o\},
\end{equation}
that summarizes whether activity $a$ behaves as a \emph{time luxury} ($\beta_a > 1$) or a
\emph{time necessity} ($\beta_a < 1$) as total digital time expands. The Engel elasticities $\{\beta^a\}$ identify $\{\eta^a\}$ only up to a common scale. To obtain {levels} of $\eta^a$, we need to impose (or calibrate) a
value for the average curvature $\bar{\eta}$ to compute
\begin{equation}
    \eta^a = \beta^a \,\bar{\eta}.
\end{equation}
Conditional on a chosen $\bar{\eta}$, the Engel-curve estimates then deliver $\eta^\ell$, $\eta^z$, and $\eta^o$.

\paragraph{Engel elasticity specification.} We estimate the elasticities $\{\beta^a\}$ using aggregated group panel regressions of the form
\begin{equation} \label{eq:engeleq}
    \ln h^a_{g,t} = \delta^a_t + \beta_a \ln H_{g,t} + u^a_{g,t},
\end{equation}
where $g$ indexes age--income--region cells and $\delta^a_t$ are quarter fixed effects. Here, $\ln H_{it}$ is the log of total browsing time, and $\ln h_{it}^{a}$ is the log of the browsing time of activity $a$ (e.g., leisure or productive activity).  All regressions include fixed effects that capture the interaction between income and age categories, and also fixed effects in the time dimension (at the year-quarter level).

\paragraph{Instrument for total browsing time.} One concern in the Engel elasticity estimation is that total browsing time is likely endogenous with regard to other shocks to households' activity that may also change their need to complete productive vs. leisure tasks online. To address this endogeneity, we use an instrumental variable approach based on plausibly exogenous local precipitation shocks that discourage offline outdoor activity and increase total online browsing time. 

\paragraph{Engel elasticity estimates.} The Engel elasticity estimates are shown in Table \ref{tab:engel_regs_main}. The estimates in panel B indicate that online leisure browsing is a time luxury ($\beta^{\ell}>1$) while productive digital activity is a time necessity ($\beta^{z}<1$), consistent with the reallocation patterns we
observe after generative AI adoption.  Intuitively, this suggests that households do not substantially expand the amount of productive online work they do in their private lives when they spend more total time online. Rather, they tend to allocate more total time spent online towards leisure activities such as social media or streaming. Moreover, to capture the full range of online activities we also estimate the response of other mixed-purpose browsing, which, based on our classification algorithm, combines leisure and productive tasks, and which, according to the estimates in panel C, has an Engel elasticity that lies about halfway between the estimates for leisure and productive browsing. The IV estimates result in somewhat more differentiated coefficients than the OLS estimates, but both tell a qualitatively similar story. \cite{aguiar2021} conduct a similar estimation exercise using time shares  as the dependent variables (rather than the log of activity time) to avoid dropping cells with zero activity, and then approximate the elasticity using the estimated semi-log coefficients. We show in \ref{tab:engel_regs} that a similar approach in our setting results in qualitatively and quantitatively comparable estimates as those shown in Table \ref{tab:engel_regs_main}---likely because our estimation panel does not contain many cells with zero browsing activity.

\begin{table}[htbp] \renewcommand{\arraystretch}{0.95}
\caption{\\ \centering  \textbf{Engel elasticity estimation results}}  

\vspace{-0.1cm} \footnotesize    This table shows estimates of age-income-region-level panel regressions of the form shown in equation \ref{eq:engeleq}. See Section \ref{sec:engelest} for details. The instrument in column 2 is log precipitation in the household's metro area in the year-quarter. T-statistics based on standard errors double clustered at the age-income-region level and at the year-quarter level in parentheses: * p$<$0.10, ** p$<$0.05, *** p$<$0.01. 
\footnotesize

\begin{center}

\begin{tabular}{@{}l*{6}{c}@{}}
\toprule
\textit{Dep. var} &   \multicolumn{2}{c}{Log Activity Time} \\
   \cmidrule(lr){2-3}
   	Estimation & OLS & IV  \\ 

      \cmidrule(lr){2-2}  \cmidrule(lr){3-3} 
                                        &\multicolumn{1}{c}{(1)}   &\multicolumn{1}{c}{(2)}       \\ 
                   \midrule
\emph{Panel A: Productive digital activities}  && \\
                   
                   Log(Total Browsing Time) &       0.991***&       0.931***\\
            &   (530.879)   &    (29.244)   \\
Observations&     129,146   &      99,430   \\
1st-stage KP F-stat.&               &          54   \\
  
\midrule
\emph{Panel B: Leisure digital activities}  && \\
                   
                   Log(Total Browsing Time) &       1.130***&       1.374***\\
            &   (241.949)   &    (15.823)   \\
Observations&     120,785   &      94,691   \\
1st-stage KP F-stat.&               &          48   \\
  
\midrule
\emph{Panel C: Other digital activities}  && \\
                   
                   Log(Total Browsing Time) &       1.026***&       1.110***\\
            &   (276.169)   &    (17.912)   \\
Observations&     125,862   &      97,739   \\
1st-stage KP F-stat.&               &          51   \\
  
\midrule
	Age x Income Group FEs & X & X	     \\		
	Time FEs & X & X	   \\		
    \bottomrule
\end{tabular}
\label{tab:engel_regs_main} 
\end{center}
\end{table}


\subsection{Implied generative AI efficiency gain} \label{subsec:efficiency_gain}

With the estimates of $\beta^a$ in hand, we can use our estimates of the causal effect of ChatGPT use on browsing time allocation to infer the change in efficiency as a result of generative AI technology. Observed shifts in time allocation and the leisure Engel curves identify changes in technology up to the scaling parameter $\bar{\eta}$. Recall the assumption that generative AI adoption raises only the efficiency of productive digital tasks, and that we can therefore treat leisure browsing as our reference activity.

 To estimate the implied scaled generative AI productivity change $(1-\eta_z)\delta$ for productive digital tasks,\footnote{This is analogous to the scaled productivity shock for computer time computed by \cite{aguiar2021} under the assumption that leisure time spent on eating, sleeping, and personal care (ESP) can be treated as a reference category that does not experience a technology shock during the 2000s.} we combine our estimates of Engel elasticities with our IV estimates of the causal effect of ChatGPT adoption on browsing activity. 
As before, let $\hat{\beta}^{GPT}_a$ denote our estimate of the causal effect of ChatGPT adoption on the log of activity time $a$:
\begin{equation}
    \hat{\beta}^{GPT}_a \equiv \ln a_{\text{Adopt}} - \ln a_{\text{No adopt}},
    \qquad a \in \{z,\ell,o\},
\end{equation}
which we obtain from our IV long-difference design (Table \ref{table:2sls-bycat}).  Then, the difference in time allocation effects shown in equation \ref{eq:stylized-model-effect-difference} (using the exact version as $\delta$ may not be small) across the Adopt vs.\ No-adopt
regimes eliminates the (unobserved) shadow value of digital time. Using the Engel curve estimates to substitute $\eta_z / \eta_{\ell} = \hat{\beta}_z / \hat{\beta}_{\ell}$, then yields:
\begin{equation} \label{eq:deltascaled}
    (1- \eta_z)\ln(1 + \delta_z)
    =   (\beta_z / \beta_{\ell})\hat{\beta}^{GPT}_\ell - \hat{\beta}^{GPT}_z,
\end{equation}
if we are assuming that ChatGPT only impacts the efficiency of productive digital tasks and we are using leisure digital tasks as the unaffected reference category.
Here, we have also multiplied both sides by $-1$, in order to ensure the technology shock on the left-hand side has a positive sign when $\eta_z<1$, which we expect to be the case if productive browsing activity behaves as a ``digital time necessity''.
Intuitively, this equation shows that the implied (scaled) efficiency shock due to ChatGPT use  $(1- \eta_z)\ln(1 + \delta_z)$ can be inferred from the \textit{relative} change in browsing time. As reported in column (2) of Table \ref{table:2sls-bycat}, leisure digital browsing time increased by 151 log points over the 2022-2024 period as a result of ChatGPT use. The estimates in Table \ref{tab:engel_regs_main} give $\frac{\hat{\beta}^z}{\hat{\beta}^{\ell}} = 0.68$. This
implies that, absent any technological change in productive digital tasks, the change in leisure browsing implies that  productive browsing time should have increased by 102 log points as well. This is the term on the left-hand side of equation \ref{eq:deltascaled}, which corresponds to the predicted movement along the Engel curve for productive digital tasks. However, productive browsing time actually rose by only 1.1 log points. The model interprets this gap relative to the expected productive digital time allocation as having to result from greater efficiency.  We therefore estimate the change in $(1-\eta_z)\ln(1 + \delta_z)$ to be a 101 log point, or 175\%, increase in scaled efficiency. 

\paragraph{Allowing for leisure efficiency gains.} This calculation assumed that leisure digital tasks experience no change in efficiency due to ChatGPT use. This is likely too strong an assumption. While it seems plausible that leisure efficiency gains are weakly smaller than productive task efficiency gains---which is also consistent with ChatGPT being mainly used in the context of productive digital tasks---we can relax this assumption and allow leisure task efficiency gains to lie in a range between zero and being the same magnitude as productive task efficiency gains, i.e., we define a parameter $\psi$, the ``efficiency gain ratio'', defined by $\delta_{\ell} = \psi \; \delta_z$, where $\psi \in [0,1]$.

The relative impact on the observed changes in log browsing time for the two types of tasks is then given by the following modification of equation \ref{eq:stylized-model-effect-difference}:
\begin{equation}
\frac{\beta^{GPT}_{z}}{\eta^z} - \frac{\beta^{GPT}_{\ell} }{\eta^{\ell}} = \frac{\eta^z - 1}{\eta^z}  \ln(1+ \delta_z )  - \frac{\eta^{\ell} - 1}{\eta^{\ell}} \ln(1+ \delta_{\ell} ).
\end{equation}
and equation \ref{eq:deltascaled} becomes
\[
(1 - \eta_z)\ln(1 + \delta_z)
\;-\;
\frac{\beta_z}{\beta_\ell} (1 - \eta_\ell)\ln(1 + \psi \delta_z)
\;=\;
 \frac{\beta_z} {\beta_{\ell}}\hat{\beta}^{GPT}_\ell - \hat{\beta}^{GPT}_z.
\]
In the case where $\psi = 0$, this is identical to equation \ref{eq:deltascaled}, but if $\psi>0$, it shows that the size of the measured shock on the right-hand side can now be explained not just by a scaled efficiency shock to productive tasks (the first term on the LHS), but also potentially by a scaled efficiency shock to leisure tasks, as long as $\eta_{\ell}>1$, which means that leisure tasks behave as a ``time luxury'', such that the second term on the LHS is positive. Intuitively, as long as  this will lead to smaller estimates of the scaled efficiency gain, because any gap between increases in leisure task time and increases in productive task time can  no longer  be explained only by a productivity shock lowering time spent on the (time necessity) productive tasks. Instead, part of the increase in leisure task time can now be due to its own efficiency gain and the fact that it behaves as a time luxury, such that the inferred efficiency gain for productive tasks is smaller.

As we do not know the \textit{levels} of $\eta_z$ and $\eta_{\ell}$, we cannot generally solve this for the scaled efficiency gain. However, if we make assumptions about $\bar{\eta}$ in addition to assuming values of $\psi$, we can compute the scaled efficiency gain for productive tasks under these different scenarios.  \cite{aguiar2021} find $\bar{\eta}\approx 0.73$ and show robustness for a range $\bar{\eta} \in [0.73, 1.0, 1.19]$. In our setting, we estimated $\hat{\beta}_z = 0.931$ as the elasticity of productive browsing time
  with respect to total browsing time (see Table \ref{tab:engel_regs_main}). \cite{aguiar2021} show that  $\beta_i=\eta_i/\bar{\eta}$,  and we find in Table \ref{table:2sls-bycat} that, following adoption, total browsing rises, implying a proportional increase in productive browsing time,  but our estimates in Table \ref{table:2sls-bycat} show no economically significant increase. This implies that $\eta_z < 1$, which is consistent with productive browsing behaving as a time necessity. As a result, we can infer that $\eta_z = \beta_z \bar{\eta} < 1$, which implies that $\bar{\eta} < 1/\beta_z \approx 1.07$.
That is, $\bar{\eta}$ cannot be much above one if productive tasks are behaving as a time necessity. Coincidentally, the estimate of $\bar{\eta}\approx 0.73$ from \cite{aguiar2021} also corresponds to the lowest $\bar{\eta}$ in our setting that is consistent with $\eta_{\ell}>1$, which reflects the sensible assumption that leisure digital tasks behave as a time luxury good.
We therefore consider values $\bar{\eta} \in [0.73, 0.9, 1.0, 1.07]$ as possible scenarios (where 1 is a salient benchmark and $\bar{\eta} = 0.9$ is the midpoint between the upper and lower bounds). Given the large observed increases in total browsing, without a change in productive browsing time, $\eta_z$ seems more likely to be well below 1, which implies that $\bar{\eta}$ is plausibly substantially smaller than the upper bound considered here.

\paragraph{Calibrating the efficiency gain ratio.} We can determine plausible values for $\psi$ by assuming it ranges from leisure not being affected by ChatGPT ($\psi = 0$) to assuming that leisure and productive digital tasks see the same efficiency gains ($\psi = 1$), which provides plausible outer bounds for the relative effects. 
We can further narrow this range, by using our data to construct empirical proxies for the share of activities on leisure sites that are exposed to ChatGPT relative to the share of productive site activities that are exposed. In Appendix Table \ref{table:exp-bycat}, we show the duration-weighted average exposure during our benchmarking period in 2021 for the main usage categories. As the data show, leisure-oriented websites had an average activity exposure of 17.7\%, while productive sites average 50.7\% exposure, so the predicted efficiency gain ratio based only on this exposure proxy would be $\frac{0.117}{0.507} \approx 0.23$.  We therefore consider efficiency gain ratios $\psi\in[0,0.25,0.5, 1]$, to include a value close to our empirical proxy and also consider what doubling this proxy or assuming the same shock for both task types would mean for our estimates.

\begin{table}[!ht]
\centering \setlength{\tabcolsep}{24pt}
\caption{\\ \textbf{Scaled productive efficiency gain calibration}}
\label{tab:scaled_efficiency_gain_grid}
\begin{threeparttable}
\begin{tabular}{lcccccc}
\hline
\multicolumn{5}{c}{{Scaled productive efficiency gain (in \% increase):} } \\
 \multicolumn{5}{c}{$100\left(\exp\!\big((1-\eta_z)\ln(1+\delta_z)\big)-1\right)$} \\
\hline
$\bar{\eta}$ $\backslash$ $\psi$ & $0$ & $0.25$ & $0.5$ & $1$ \\
\hline
$0.73$ & $175.52$ & $174.46$ & $174.12$ & $173.76$ \\
$0.90$ & $175.52$ & $85.16$  & $75.59$  & $66.45$  \\
$1.00$ & $175.52$ & $33.60$  & $28.80$  & $24.22$  \\
$1.07$ & $175.52$ & $1.73$   & $1.47$   & $1.21$   \\
\hline
\end{tabular}

\begin{tablenotes}
            \item \footnotesize \emph{Note:}  $(\bar{\eta},\psi)$, $\delta_z$ solves
$(1-\eta_z)\ln(1+\delta_z)-r(1-\eta_\ell)\ln(1+\psi\,\delta_z)=A_z$, 
 with $\eta_z=0.931\,\bar{\eta}$, $\eta_\ell=1.374\,\bar{\eta}$, $\frac{\beta_z}{\beta_\ell}=0.6776$, and
$\frac{\beta_z} {\beta_{\ell}}\hat{\beta}^{GPT}_\ell - \hat{\beta}^{GPT}_z =\frac{0.931}{1.374}\cdot 1.512-0.011$. 
\end{tablenotes}
\end{threeparttable}
\end{table}

\paragraph{Discussion.} The resulting estimates of the scaled productive task efficiency shock in log points are shown in Table \ref{tab:scaled_efficiency_gain_grid}. As the table shows, the implied scaled efficiency gains in percent for the productive digital task are large  depending on the assumptions about the elasticity parameter and the efficiency gain ratio. 

We consider the midpoint $\bar{\eta} = 0.9$ between the lower and upper bounds for $\bar{\eta}$ that would be consistent with productive time  behaving as a necessity and leisure time as a luxury, as a pragmatic choice. We also assume that the actual efficiency ratio gain is no more than double our empirical proxy for the average ratio of exposure on leisure sites to that on productive sites, such that $\psi$ is plausibly between 0 and 0.5. Then, this table suggests a scaled efficiency gain range of 76\%-176\% for productive digital tasks from generative AI adoption.


Our estimated productivity gains of 76\% to 176\% from ChatGPT adoption on household productive Internet tasks are larger than, but in a similar order of magnitudes as those found in the recent literature on generative AI in workplace settings. For instance, \citet{noy2023} found that ChatGPT reduced task completion time by 40\% for professional writing tasks; \citet{brynjolfsson2023} found that AI-based conversational assistance increased customer support agent productivity by 34\% for novice workers; \citet{dell2023} found that BCG consultants using GPT-4 finished 25.1\% faster on tasks within AI's capability frontier; \citet{peng2023} found that GitHub Copilot users completed coding tasks 55.8\% faster. 

There can be several possible reasons for the effect to be larger for home setting. First, workplace experiments typically measure productivity on \emph{structured professional tasks}---writing reports, resolving customer tickets, coding---where workers already possess specialized skills and established workflows. By contrast, household productive Internet tasks---such as researching product purchases, troubleshooting home appliances, looking up recipes, seeking tax or medical information, and helping children with homework---are often performed by individuals with little domain expertise, making them closer to the ``novice'' end of the skill distribution where productivity gains are largest \citep{brynjolfsson2023,noy2023}. Second, our IV strategy identifies a local average treatment effect for compliers whose browsing patterns had the highest overlap with ChatGPT's capabilities, capturing the productivity gains among households most intensively exposed to the technology, which are known to amplify the average effect (\cite{jiang_have_2017}). Third, and perhaps most importantly, many routine household Internet tasks involve extensive browsing---visiting multiple websites, comparing information across sources, reading forums, and watching tutorial videos---that ChatGPT can substitute almost entirely with a single conversational query.\footnote{For example, a household member researching how to fix a leaking faucet might previously spend 20--30 minutes browsing plumbing forums, watching YouTube tutorials, and comparing advice across websites, whereas ChatGPT can provide a step-by-step diagnosis and repair guide in under two minutes---a time savings exceeding 1,000\%. Similarly, comparing features and reviews across consumer products, finding and adapting a recipe to dietary restrictions, understanding a health symptom before deciding whether to visit a doctor, or navigating government forms and tax questions are all tasks where a single ChatGPT prompt can replace what previously required 15--30 minutes of multi-site browsing, implying efficiency gains of 500\% to well over 1,000\% for these specific activities.} Because our model aggregates across all productive browsing tasks---including those with more modest gains---the estimated range of 76\% to 176\% represents an average that is well within the plausible range implied by the mix of near-complete task substitution for some household activities and more moderate efficiency gains for others.



\section{Conclusion} \label{sec:conclusion}

This paper provides the first comprehensive evidence on the adoption and impact of generative AI on U.S. households, utilizing detailed browsing data to explore the timing, usage patterns, and behavioral changes associated with ChatGPT. Our findings show that households predominantly use ChatGPT for productive, non-market activities, such as education and job search, and that generative AI adoption leaves time spent on productive online activities unchanged while increasing digital leisure time. This suggests that ChatGPT enhances household welfare by making productive online tasks more efficient, thus freeing up time for leisure. Moreover, we document significant demographic gaps in adoption, with younger and higher-income households adopting generative AI at higher rates. 

Our analysis also sheds light on how generative AI adoption alters household behavior. By quantifying changes in online activity before and after ChatGPT adoption, we find that households tend to shift time to leisure browsing activities, consistent with a greater efficiency of online productive tasks that allows households to reallocate the resulting time savings to leisure browsing. These results suggest that while generative AI can increase household productivity, it may also contribute to enhanced well-being by providing more time for leisure. The broader economic impact of these shifts is likely to be substantial, especially as generative AI continues to diffuse into more households across the income and age spectrum.

Finally, our findings emphasize the need for policies that address the unequal adoption of generative AI, particularly among older and lower-income households. By improving access to tools and increasing technology literacy in these groups, policymakers can help ensure that the benefits of generative AI are more evenly distributed. As generative AI continues to evolve, understanding its impact on household productivity and welfare will be critical for designing inclusive strategies that maximize its societal benefits while mitigating potential inequalities. Moreover, accurately measuring the impact of generative AI on the economy and people's lives will require making further progress in measuring the effect on non-market activities---changes in private lives---that builds on our contribution.


\clearpage
\newpage

{\small
\onehalfspacing 
  \bibliographystyle{jf}
  \bibliography{bibliogeo}

\begin{thebibliography}{34}
\expandafter\ifx\csname natexlab\endcsname\relax\def\natexlab#1{#1}\fi

\bibitem[Aguiar et~al.(2021)Aguiar, Bils, Charles, and Hurst]{aguiar2021}
Aguiar, Mark, Mark Bils, Kerwin~Kofi Charles, and Erik Hurst, 2021, Leisure luxuries and the labor supply of young men, {\em Journal of Political Economy\/} 129, 337--382.

\bibitem[Aguiar et~al.(2013)Aguiar, Hurst, and Karabarbounis]{aguiar2013}
Aguiar, Mark, Erik Hurst, and Loukas Karabarbounis, 2013, Time use during the great recession, {\em American Economic Review\/} 103, 1664--1696.

\bibitem[Bick et~al.(2024)Bick, Blandin, and Deming]{bick2024}
Bick, Alexander, Adam Blandin, and David~J Deming, 2024, The rapid adoption of generative ai, Technical report, National Bureau of Economic Research.

\bibitem[Brynjolfsson et~al.(2025)Brynjolfsson, Collis, Diewert, Eggers, and Fox]{brynjolfsson2025gdp}
Brynjolfsson, Erik, Avinash Collis, W~Erwin Diewert, Felix Eggers, and Kevin~J Fox, 2025, Gdp-b: Accounting for the value of new and free goods, {\em American Economic Journal: Macroeconomics\/} 17, 312--344.

\bibitem[Brynjolfsson et~al.(2023)Brynjolfsson, Li, and Raymond]{brynjolfsson2023}
Brynjolfsson, Erik, Danielle Li, and Lindsey~R Raymond, 2023, Generative ai at work, Technical report, National Bureau of Economic Research.

\bibitem[Burtch et~al.(2024)Burtch, Lee, and Chen]{burtch2024consequences}
Burtch, Gordon, Dokyun Lee, and Zhichen Chen, 2024, The consequences of generative ai for online knowledge communities, {\em Scientific Reports\/} 14, 10413.

\bibitem[Chatterji et~al.(2025)Chatterji, Cunningham, Deming, Hitzig, Ong, Shan, and Wadman]{chatterji_how_2025}
Chatterji, Aaron, Thomas Cunningham, David~J. Deming, Zoe Hitzig, Christopher Ong, Carl~Yan Shan, and Kevin Wadman, 2025, How {{People Use ChatGPT}}.

\bibitem[del Rio-Chanona et~al.(2024)del Rio-Chanona, Laurentsyeva, and Wachs]{del2024large}
del Rio-Chanona, R~Maria, Nadzeya Laurentsyeva, and Johannes Wachs, 2024, Large language models reduce public knowledge sharing on online q\&a platforms, {\em PNAS nexus\/} 3, pgae400.

\bibitem[Dell'Acqua et~al.(2023)Dell'Acqua, McFowland~III, Mollick, Lifshitz-Assaf, Kellogg, Rajendran, Krayer, Candelon, and Lakhani]{dell2023}
Dell'Acqua, Fabrizio, Edward McFowland~III, Ethan~R Mollick, Hila Lifshitz-Assaf, Katherine Kellogg, Saran Rajendran, Lisa Krayer, Fran{\c{c}}ois Candelon, and Karim~R Lakhani, 2023, Navigating the jagged technological frontier: Field experimental evidence of the effects of ai on knowledge worker productivity and quality, {\em Harvard Business School Technology \& Operations Mgt. Unit Working Paper\/} .

\bibitem[Eisfeldt and Papanikolaou(2013)]{eisfeldt2013}
Eisfeldt, Andrea~L., and Dimitris Papanikolaou, 2013, Organization capital and the cross-section of expected returns, {\em The Journal of Finance\/} 68, 1365 -- 1406.

\bibitem[Eisfeldt and Schubert(2024)]{eisfeldt2024}
Eisfeldt, Andrea~L, and Gregor Schubert, 2024, Ai and finance, Technical report, National Bureau of Economic Research.

\bibitem[Eisfeldt et~al.(2023)Eisfeldt, Schubert, Zhang, and Taska]{eisfeldt2023}
Eisfeldt, Andrea~L, Gregor Schubert, Miao~Ben Zhang, and Bledi Taska, 2023, The labor impact of generative ai on firm values, {\em Available at SSRN 4436627\/} .

\bibitem[Ghani et~al.(2022)Ghani, Mustafa, Hashim, Hanafi, and Bakhtiar]{ghani2022}
Ghani, Miharaini~Md, Wan Azani~Wan Mustafa, Mohd Ekram~Alhafis Hashim, Hafizul~Fahri Hanafi, and Durratul Laquesha~Shaiful Bakhtiar, 2022, Impact of generative ai on communication patterns in social media, {\em Journal of Advanced Research in Computing and Applications\/} 26, 22--34.

\bibitem[Handa et~al.(2025)Handa, Tamkin, McCain, Huang, Durmus, Heck, Mueller, Hong, Ritchie, Belonax, et~al.]{handa2025}
Handa, Kunal, Alex Tamkin, Miles McCain, Saffron Huang, Esin Durmus, Sarah Heck, Jared Mueller, Jerry Hong, Stuart Ritchie, Tim Belonax, et~al., 2025, Which economic tasks are performed with ai? evidence from millions of claude conversations, {\em arXiv preprint arXiv:2503.04761\/} .

\bibitem[Humlum and Vestergaard(2025)]{humlum2025}
Humlum, Anders, and Emilie Vestergaard, 2025, The unequal adoption of chatgpt exacerbates existing inequalities among workers, {\em Proceedings of the National Academy of Sciences\/} 122, e2414972121.

\bibitem[Jiang(2017)]{jiang_have_2017}
Jiang, Wei, 2017, Have instrumental variables brought us closer to the truth, {\em Review of Corporate Finance Studies\/} 6, 127--140.

\bibitem[Jiang et~al.(2025)Jiang, Park, Xiao, and Zhang]{jiang2025ai}
Jiang, Wei, Junyoung Park, Rachel~Jiqiu Xiao, and Shen Zhang, 2025, Ai and the extended workday: Productivity, contracting efficiency, and distribution of rents, Technical report, National Bureau of Economic Research.

\bibitem[Khan(2024)]{khan2024}
Khan, Salman, 2024, {\em Brave new words: How AI will revolutionize education (and why that's a good thing)\/} (Penguin).

\bibitem[Kosmyna et~al.(2025)Kosmyna, Hauptmann, Yuan, Situ, Liao, Beresnitzky, Braunstein, and Maes]{kosmyna2025your}
Kosmyna, Nataliya, Eugene Hauptmann, Ye~Tong Yuan, Jessica Situ, Xian-Hao Liao, Ashly~Vivian Beresnitzky, Iris Braunstein, and Pattie Maes, 2025, Your brain on chatgpt: Accumulation of cognitive debt when using an ai assistant for essay writing task, {\em arXiv preprint arXiv:2506.08872\/} .

\bibitem[Lei and Liu(2025)]{lei2025}
Lei, Ke, and Yixuan Liu, 2025, When ai becomes a shopping advisor: A study on the impact of generative ai review on consumer purchase decision, {\em SAGE Open\/} 15, 21582440251357671.

\bibitem[Ling and Imas(2025)]{ling2025}
Ling, Yier, and Alex Imas, 2025, Underreporting of ai use: The role of social desirability bias, {\em Available at SSRN\/} .

\bibitem[Lo and Ross(2024)]{lo2024}
Lo, Andrew~W, and Jillian Ross, 2024, Can chatgpt plan your retirement?: Generative ai and financial advice, {\em Harvard Data Science Review\/} .

\bibitem[Lyu et~al.(2025)Lyu, Siderius, Li, Acemoglu, Huttenlocher, and Ozdaglar]{lyu2025wikipedia}
Lyu, Liang, James Siderius, Hannah Li, Daron Acemoglu, Daniel Huttenlocher, and Asuman Ozdaglar, 2025, Wikipedia contributions in the wake of chatgpt, in {\em Companion Proceedings of the ACM on Web Conference 2025\/},  1176--1179.

\bibitem[McElheran et~al.(2024)McElheran, Li, Brynjolfsson, Kroff, Dinlersoz, Foster, and Zolas]{mcelheran2024}
McElheran, Kristina, J~Frank Li, Erik Brynjolfsson, Zachary Kroff, Emin Dinlersoz, Lucia Foster, and Nikolas Zolas, 2024, Ai adoption in america: Who, what, and where, {\em Journal of Economics \& Management Strategy\/} 33, 375--415.

\bibitem[Noy and Zhang(2023)]{noy2023}
Noy, Shakked, and Whitney Zhang, 2023, Experimental evidence on the productivity effects of generative artificial intelligence, {\em Available at SSRN 4375283\/} .

\bibitem[{OpenAI}(2026)]{openai2026}
{OpenAI}, 2026, Openai signals: Measuring {AI} adoption, protecting privacy, and empowering decisions.

\bibitem[Padilla et~al.(2025)Padilla, Lam, Lambrecht, and Hollenbeck]{padilla2025impact}
Padilla, Nicolas, H~Tai Lam, Anja Lambrecht, and Brett Hollenbeck, 2025, The impact of llm adoption on online user behavior, {\em Available at SSRN 5393256\/} .

\bibitem[Peng et~al.(2023)Peng, Kalliamvakou, Cihon, and Demirer]{peng2023}
Peng, Sida, Eirini Kalliamvakou, Peter Cihon, and Mert Demirer, 2023, The impact of ai on developer productivity: Evidence from github copilot, {\em arXiv preprint arXiv:2302.06590\/} .

\bibitem[{Pew Research Center}(2025)]{pewresearch2025}
{Pew Research Center}, 2025, 34\% of u.s. adults have used {ChatGPT}, about double the share in 2023, Survey methodology and topline results available.

\bibitem[Schubert(2025)]{schubert2025organizational}
Schubert, Gregor, 2025, Organizational technology ladders: Remote work and generative ai adoption, {\em Available at SSRN\/} .

\bibitem[Van~Dijk(2020)]{van2020digital}
Van~Dijk, Jan, 2020, {\em The digital divide\/} (John Wiley \& Sons).

\bibitem[Yotzov et~al.(2026)Yotzov, Barrero, Bloom, Bunn, Davis, Foster, Jalca, Meyer, Mizen, Navarrete, Smietanka, Thwaites, and Wang]{yotzov_firm_2026}
Yotzov, Ivan, Jose~Maria Barrero, Nicholas Bloom, Philip Bunn, Steven~J. Davis, Kevin~M. Foster, Aaron Jalca, Brent~H. Meyer, Paul Mizen, Michael~A. Navarrete, Pawel Smietanka, Gregory Thwaites, and Ben~Zhe Wang, 2026, Firm {{Data}} on {{AI}}.

\bibitem[Yu et~al.(2024)Yu, Xu, CH-Wang, and Arum]{yu2024}
Yu, Renzhe, Zhen Xu, Sky CH-Wang, and Richard Arum, 2024, Whose chatgpt? unveiling real-world educational inequalities introduced by large language models, {\em arXiv preprint arXiv:2410.22282\/} .

\bibitem[Zhang et~al.(2025)Zhang, Yu, Min, Xin, Wei, Shi, Huang, Kong, Xin, Jiang, et~al.]{zhang2025}
Zhang, Ruihan, Borou Yu, Jiajian Min, Yetong Xin, Zheng Wei, Juncheng~Nemo Shi, Mingzhen Huang, Xianghao Kong, Nix~Liu Xin, Shanshan Jiang, et~al., 2025, Generative ai for film creation: A survey of recent advances, in {\em Proceedings of the Computer Vision and Pattern Recognition Conference\/},  6267--6279.

\end{thebibliography}
}
\clearpage 






\clearpage

\clearpage

\clearpage
\newpage  


\begin{center}
\Large{\bf Internet Appendix for}

\Large{\bf {``The Household Impact of Generative AI: \\
Evidence from Internet Browsing Behavior''}} \end{center} \centerline{\large  } \centerline{\large  {Michael Blank} \ \ \ \ \  {Gregor Schubert } \ \ \ \ \ \     {Miao Ben Zhang}  } \setcounter{table}{0} \setcounter{figure}{0} \setcounter{section}{0} \renewcommand{\thetable}{IA.\arabic{table}} \renewcommand{\thefigure}{IA.\arabic{figure}}  

\clearpage 
\renewcommand{\thetable}{IA.\arabic{table}}
\renewcommand{\thefigure}{IA.\arabic{figure}}
\setcounter{equation}{0}
\renewcommand{\thesection}{IA.\arabic{section}}
\renewcommand{\theHsection}{IA.\arabic{section}}
\setcounter{section}{0}
\onehalfspacing
\section{Methodology Appendix} \label{ia.sec:appx_methodology}

\subsection{Website classification}

We sort websites in the public use version of Comscore data by total  page views in 2023 and focus our labeling on the top 160K websites, which correspond to $>95\%$ of all Internet use.

The website classification pipeline then processes this domain-level data through multiple stages to create a dataset of websites with their usage classification---whether the website use is most likely "Productive", "Leisure", or "Mixed-use":
\begin{enumerate}
    \item \textbf{Domain Selection}: We start with a ranked list of domains.  Duplicates are removed to ensure each domain appears only once in the dataset.
    
    \item \textbf{Metadata Extraction}: For each selected domain, we use the Beautiful Soup package in Python to access the domain and extract metadata from the website's source code, focusing on information contained within `title', `description' and `keyword' tags provided by the website owner.

    \item \textbf{Website description and activity extraction}: For each domain, we use the GPT-4.1 Mini model from OpenAI  (accessed in the May 28, 2025 version of the model via Azure OpenAI) to generate:
    \begin{itemize}
        \item A concise description (2-3 sentences) of the website's nature and primary function
        \item A list of the top 5 meaningful user activities typically performed on the website
        \item The full prompt used is the following: \tiny
\begin{verbatim} 
I need a detailed analysis of the activities one would expect on the website with domain "{domain}".

Here is the available metadata about this website:
{available_data if metadata else "No additional metadata available."}

Based on this information, please:
1. Provide a concise description (2-3 sentences) of the nature of the website and its primary function.
2. List the top 5 meaningful activities that users would likely perform on this website.

Omit auxiliary activities, such as "set up an account with the website", "provide their payment details to pay for the website" etc.
that are not core to the purpose of the website.

Return your response in this exact JSON format:
{{
    "website_description": "Description of the website's nature and primary function",
    "top_activities": [
        "Activity 1",
        "Activity 2",
        "Activity 3",
        "Activity 4",
        "Activity 5"
    ]
}}

If the domain name alone doesn't provide enough information for a confident assessment, make reasonable inferences based on similar websites.
\end{verbatim}

\item Note that this approach relies on the website's metadata if it's available as grounding for the model's response, but if there is no further description also allows the model to rely on the knowledge obtained from its training data (which contains large amounts of Internet text) in judging what the likely use of a website is.
  \item Websites triggering content management policy violation errors from the LLM (because of adult content) are classified as ``adult content, likely of a sexual nature'' if the LLM refuses to respond.
    \end{itemize}

\end{enumerate}

\subsection*{Usage Classification: productive vs. leisure use websites}

 Each website is classified along two dimensions using GPT-4.1 mini:
    \begin{itemize}
        \item \textbf{Usage Type}: Categorized as ``Productive'' (market work, education, childcare, etc.), ``Leisure'' (gaming, social activity, TV, etc.), or ``Ad platform/CDN'' (advertising networks, content delivery networks)
        \item \textbf{Usage Diversity}: Labeled as either ``Single-use'' (primarily productive OR leisure) or ``Mixed-use'' (has important productive AND leisure uses)
    \end{itemize}

The prompt used in the classification is shown below: 
\tiny{
\begin{verbatim}
I need to classify a website based on its domain name and description.

Website Domain: {domain}
Website Description: {website_description}

Please classify this website in two ways:

1. Usage Type: Determine if the website is primarily:
"Productive" (Market Work, Other Income-Generating Work, Job Search, Childcare, Non-market work, Education, Civic Activities, Own Medical Care) or
"Leisure" (TV, Social Activity, Sleep, Eating and Personal Care, Gaming, Other Leisure) or
"Ad platform/CDN" (Advertising networks, ad servers, analytics platforms, tracking services, content delivery networks etc. that primarily serve ads or video for other sites, or track user behavior)

2. Usage Diversity: Determine if the website is
"Single-use" (only has important Productive OR Leisure uses) or
"Mixed-use" (has important Productive uses AND Leisure uses).

Consider:
- The website's primary purpose
- The typical user activities and motivations
- Whether users would engage in both leisure and productive uses on the website
- For ad platforms/CDNs, look for domains related to ad serving, tracking, analytics, content delivery infrastructure, or marketing infrastructure

Return your response in this exact JSON format:
{{
    "usage_type_reasoning": "Brief explanation of why this is classified as productive, leisure, or ad platform/CDN",
    "usage_type": "Productive" OR "Leisure" OR "Ad platform/CDN",
    "usage_diversity_reasoning": "Brief explanation of why this is single-use or mixed-use",
    "usage_diversity": "Single-use" OR "Mixed-use"
}}
\end{verbatim}}

To validate the robustness of this overall procedure, we also used a separate methodology that first converts the website activities resulting from the first LLM call into semantic embeddings. Then, each website activity is matched to similar activities from the American Time Use Survey (ATUS) that could conceivably be done on a computer, by also embedding the ATUS activities and computing cosine similarities. Next, an LLM is asked to determine which ATUS activity is the main one for this website by choosing  one from the list of matches. Finally, this ATUS activity is mapped to the categories of activities that reflect productive or leisure uses based on on the categories of activities in \cite{aguiar2013}. This procedure resulted in labels of productive or leisure uses of websites that were very similar to the more direct procedure described above, but substantially reduced the coverage of domains because for many website activities no good ATUS activity matches could be found.

\subsection{Website exposure classification} \label{sec:exposure_prompt}

Here is the prompt used for labeling websites as exposed:

\tiny{
\begin{verbatim}
    I need to classify a website based on its domain name and description to determine if it's likely to be exposed to competition from ChatGPT chatbot use.

Website Domain: {domain}
Website Description: {website_description}

Please classify this website based on whether it provides services or content that users might partially be able to substitute with ChatGPT adoption.

Here are the website capabilities that ChatGPT might replace:

### 1. Language & Translation Sites
- **Translation Platforms:** Websites offering language conversion services
- **Grammar & Writing Tools:** Sites providing spelling, grammar, and style checking
- **Language Learning Resources:** Educational platforms teaching languages
- **Multilingual Customer Service:** Sites whose primary value is language bridging

### 2. Content Creation Websites
- **Creative Writing Platforms:** Sites offering story, poem, or script generation
- **Business Document Services:** Websites specializing in professional document templates
- **Marketing Copy Generators:** Sites creating advertising text and slogans
- **Email Writing Assistants:** Platforms helping draft professional communications
- **Content Mills:** Websites producing basic informational articles

### 3. Information & Research Sites
- **General Knowledge Platforms:** Encyclopedia and fact-aggregation websites
- **Simple Q&A Forums:** Sites answering straightforward informational questions
- **Educational Summaries:** Websites offering simplified explanations of concepts
- **Product Comparison Sites:** Basic review aggregators without detailed testing
- **Academic Resource Centers:** Sites providing standard explanations of subjects

### 4. Basic Programming Help
- **Code Snippet Libraries:** Websites offering common programming solutions
- **Beginner Programming Tutorials:** Sites explaining fundamental coding concepts
- **Simple Debugging Platforms:** Services that identify basic code errors
- **Standard Documentation Sites:** Websites with common technical explanations

### 5. Data Analysis Services
- **Basic Visualization Tools:** Sites generating simple charts and graphs
- **Data Interpretation Services:** Platforms explaining statistical information
- **Simple Reporting Tools:** Websites creating standard data reports
- **Algorithm Explanation Sites:** Services describing common computational approaches

### 6. Technical Support Forums
- **Common Troubleshooting Platforms:** Sites offering solutions to frequent issues
- **Software Usage Guides:** Websites explaining how to use popular programs
- **IT Knowledge Bases:** Repositories of standard technical solutions
- **Command Reference Sites:** Platforms listing command structures and usage

### 7. Basic Professional Assistance
- **Resume Builders:** Websites generating standard resume formats
- **Cover Letter Templates:** Platforms offering professional letter frameworks
- **Business Plan Generators:** Sites creating basic business documentation
- **Simple Legal Document Services:** Platforms for standard legal forms

### 8. Advisory Content Sites
- **General Financial Advice:** Websites offering basic investment information
- **Career Guidance Platforms:** Sites providing common professional development tips
- **Business Strategy Blogs:** Publications sharing standard business insights
- **Health Information Resources:** General wellness and nutrition advice sites

### 9. Lifestyle Support Services
- **Relationship Advice Columns:** Sites offering general interpersonal guidance
- **Basic Trip Planning Tools:** Platforms generating standard travel itineraries
- **Fitness Program Generators:** Websites creating common exercise routines
- **Study Aid Resources:** Services offering standard learning assistance

### 10. Creative Planning Tools
- **Event Planning Templates:** Sites offering standard organizational frameworks
- **Project Management Guides:** Platforms providing basic structural approaches
- **Brainstorming Tools:** Websites facilitating ideation and creative processes
- **Organizational Systems:** Services creating standard management frameworks

### 11. Visual Content Services
- **Image Description Sites:** Platforms that caption or explain visual content
- **Basic Visual Analysis Tools:** Services identifying elements in images
- **Chart Interpretation Services:** Websites explaining graphs and visualizations
- **Document Reading Services:** Platforms extracting text from images

### 12. Audio Processing Sites
- **Transcription Services:** Websites converting speech to text
- **Basic Audio Analysis:** Platforms identifying elements in audio content
- **Podcast Summarization:** Services creating text summaries of audio content
- **Voice-to-Text Applications:** Sites offering speech recognition functionality

### 13. Multimedia Integration Platforms
- **Basic Media Conversion:** Services translating between different content formats
- **Simple Virtual Assistants:** Websites offering basic multimodal interactions
- **Content Accessibility Tools:** Platforms making content available across formats
- **Media Description Services:** Sites generating text descriptions of visual/audio content

### 14. Beginner Tutorial Sites
- **Introductory Learning Platforms:** Websites offering fundamental knowledge
- **How-To Guides:** Services providing step-by-step instructions
- **Basic Skill Development:** Sites teaching entry-level capabilities
- **Simplified Explanations:** Platforms reducing complex topics to basics

### 15. Design Template Services
- **Basic UI/UX Resources:** Websites offering standard design patterns
- **Template Libraries:** Platforms providing pre-made design frameworks
- **Simple Logo Generators:** Services creating basic brand elements
- **Design Guidelines:** Sites explaining fundamental aesthetic principles

Determine if ANY activities on the given website is exposed to ChatGPT competition based on the above categories. Note that exposure may come from only some activities being done by ChatGPT, even if the website has other non-exposed activities.

Return your response in this exact JSON format:
{{
    "exposure_reasoning": "Brief explanation of why this website is or is not exposed to ChatGPT competition",
    "substitution_examples": "If exposed, provide 1-2 specific examples of how users might substitute this website's services with ChatGPT",
    "is_exposed": "Yes" OR "No",
    "exposure_level": "High" OR "Medium" OR "Low" OR "None",
}}
\end{verbatim}

}

\subsection{Website activity exposure classification} \label{sec:exposure_prompt_activity}

Here is the prompt used for labeling website activities as exposed:

\tiny
\begin{verbatim}
I need to classify a specific activity on a website based on whether users might be able to substitute this activity with ChatGPT adoption.

Website Domain: {domain}
Website Description: {website_description}
Specific Website Activity: {activity}

Please classify this specific website activity based on whether it could be partially or completely substituted by using ChatGPT.

Here are the website capabilities that ChatGPT might replace:

### 1. Language & Translation Services
- **Translation Platforms:** Converting text between languages
- **Grammar & Writing Tools:** Spell checking, grammar correction, style suggestions
- **Language Learning Resources:** Language explanations, practice exercises
- **Multilingual Customer Service:** Bridging language gaps in customer communications

### 2. Content Creation
- **Creative Writing:** Story, poem, or script generation
- **Business Documents:** Professional document creation and templates
- **Marketing Copy:** Advertising text, slogans, product descriptions
- **Email Writing:** Professional communications drafting
- **Content Generation:** Basic informational articles

### 3. Information & Research
- **General Knowledge:** Facts, explanations, encyclopedia-like information
- **Simple Q&A:** Straightforward informational questions and answers
- **Educational Summaries:** Simplified explanations of concepts
- **Product Comparisons:** Basic review information
- **Academic Resources:** Standard subject explanations

### 4. Basic Programming Help
- **Code Snippets:** Common programming solutions
- **Programming Tutorials:** Fundamental coding concept explanations
- **Simple Debugging:** Basic code error identification
- **Standard Documentation:** Common technical explanations

### 5. Data Analysis
- **Basic Visualization:** Simple chart and graph generation
- **Data Interpretation:** Statistical information explanation
- **Simple Reporting:** Standard data report creation
- **Algorithm Explanation:** Common computational approach descriptions

### 6. Technical Support
- **Common Troubleshooting:** Solutions to frequent issues
- **Software Usage Guides:** Popular program usage explanations
- **IT Knowledge Base:** Standard technical solutions
- **Command References:** Command structures and usage information

### 7. Basic Professional Assistance
- **Resume Building:** Standard resume format generation
- **Cover Letters:** Professional letter framework creation
- **Business Plans:** Basic business documentation
- **Simple Legal Documents:** Standard legal form assistance

### 8. Advisory Content
- **General Financial Advice:** Basic investment information
- **Career Guidance:** Common professional development tips
- **Business Strategy:** Standard business insights
- **Health Information:** General wellness and nutrition advice

### 9. Lifestyle Support
- **Relationship Advice:** General interpersonal guidance
- **Trip Planning:** Standard travel itineraries
- **Fitness Programs:** Common exercise routines
- **Study Aids:** Standard learning assistance

### 10. Creative Planning
- **Event Planning:** Standard organizational frameworks
- **Project Management:** Basic structural approaches
- **Brainstorming:** Ideation and creative process facilitation
- **Organizational Systems:** Standard management frameworks

### 11. Visual Content Services
- **Image Description:** Caption or explanation of visual content
- **Basic Visual Analysis:** Element identification in images
- **Chart Interpretation:** Graph and visualization explanation
- **Document Reading:** Text extraction from images

### 12. Audio Processing
- **Transcription:** Speech-to-text conversion
- **Basic Audio Analysis:** Element identification in audio
- **Content Summarization:** Text summaries of audio content
- **Voice-to-Text:** Speech recognition functionality

### 13. Multimedia Integration
- **Basic Media Conversion:** Translation between content formats
- **Simple Virtual Assistance:** Basic multimodal interactions
- **Content Accessibility:** Making content available across formats
- **Media Description:** Text descriptions of visual/audio content

### 14. Beginner Tutorials
- **Introductory Learning:** Fundamental knowledge provision
- **How-To Guides:** Step-by-step instructions
- **Basic Skill Development:** Entry-level capability teaching
- **Simplified Explanations:** Complex topic reduction to basics

### 15. Design Template Services
- **Basic UI/UX Resources:** Standard design patterns
- **Template Libraries:** Pre-made design frameworks
- **Simple Logo Generation:** Basic brand element creation
- **Design Guidelines:** Fundamental aesthetic principle explanation

Determine if this specific activity is exposed to ChatGPT competition based on the above categories.

Return your response in this exact JSON format:
{{
    "activity_type": "Categorize this activity into one of the 15 categories above, or 'Other' if none apply",
    "exposure_reasoning": "Brief explanation of why this activity is or is not exposed to ChatGPT competition",
    "substitution_examples": "If exposed, provide 1-2 specific examples of how users might substitute this activity with ChatGPT",
    "is_exposed": "Yes" OR "No",
    "exposure_level": "High" OR "Medium" OR "Low" OR "None",
}}
\end{verbatim}

\subsection*{Output Data}
\normalsize
The final dataset includes the following key fields for each domain:
\begin{itemize}
    \item Domain name
    \item Website description
    \item Top user activities (up to 5)
    \item Usage type (Productive/Leisure/Ad platform)
    \item Usage type reasoning
    \item Usage diversity (Single-use/Mixed-use)
    \item Usage diversity reasoning
\end{itemize}

This processed dataset enables analysis of Internet usage patterns, classification of time spent online, and economic analysis of productivity and leisure trade-offs in digital environments.

\subsection{Rainfall instrument construction} \label{sec:rain-instrument}

This section shows how we construct the measure of local precipitation used to instrument for total online browsing time in a quarter.

\paragraph{Data Sources} We construct a metro region-level monthly weather panel from the grid-level daily weather
dataset maintained by Wolfram Schlenker.\footnote{Available at
\url{https://wolfram-schlenker.info/links.html}.} The raw data provide daily observations of precipitation (\texttt{prec}) for a fine grid of cells covering the contiguous United States, Alaska, and Hawaii. Each grid cell is
pre-assigned to a county FIPS code via a grid-to-county crosswalk distributed  with the data. Our sample covers the years 2019--2023. For each year, the data contain
3,105 county-level files, with each file containing daily observations for all grid cells assigned to that county. The typical county contains between 85 and
150 grid cells, depending on the geographic area.

\paragraph{Geographic Mapping} Aggregation from grid cells to our target regions proceeds through a three-step
geographic crosswalk:
\begin{enumerate}
    \item \textbf{Grid cell $\to$ County.} This mapping is pre-assigned in the raw data. Each grid cell belongs to exactly one county, identified by its
    5-digit FIPS code.
    \item \textbf{County $\to$ MSA.} We use the Bureau of Labor Statistics Quarterly Census of Employment and Wages (QCEW) county-to-MSA/CBSA
    crosswalk (July 2023 vintage) to map county FIPS codes to Metropolitan and Micropolitan Statistical Area titles. This step successfully maps 1,843 counties to 924 unique CBSA titles. Counties not assigned to any
    MSA---primarily rural counties---are excluded from the regional aggregation.


    \item \textbf{MSA $\to$ Region.} We map MSA titles to the 210 metro regions included in the Comscore data
    using a semi-manual approach. Because MSA titles (e.g., ``Dallas-Fort
    Worth-Arlington, TX MSA'') differ systematically from Comscore region names (e.g.,
    ``Dallas-Ft.\ Worth, TX''), simple fuzzy matching approaches alone are likely unreliable.
    Instead, we automatically match MSAs where (i) the MSA's primary state matches the Comscore region's primary state and (ii) the MSA's primary city name matches a city component in the Comscore region name. This automatic process matches 178 MSAs. We manually match an additional 95 of the Comscore regions to the QCEW MSA titles. In total, 273 of 924 MSA titles are mapped to one of 210 target Comscore regions. The remaining 651 MSAs correspond to areas outside our Comscore target regions. Three counties that lack QCEW MSA assignments but map to Comscore target regions are
    assigned directly  (Aroostook County, ME
    $\to$ Presque Isle-Caribou, ME; Dawson County, MT $\to$ Glendive, MT).
\end{enumerate}
The final crosswalk maps 1,013 unique counties to 210 Comscore regions.

\paragraph{Aggregation Methodology.}
Aggregation from the grid-level rainfall data then proceeds in two stages:
\begin{enumerate}
    \item {Grid cells $\to$ County-month:} For each county and calendar day, we compute simple (unweighted) arithmetic
means of precipitation across all grid cells
assigned to that county. We then aggregate county-day observations to county-month observations by computing the mean daily precipitation rate in the month.
\item {County-month $\to$ Region-month:}
For each region and month, we compute simple (unweighted) arithmetic means of weather variables across the counties assigned to that region. No population or area weighting is applied; each county within a region receives equal weight.

\end{enumerate}

\paragraph{Geographic coverage.}
Of 210 target regions, 206 appear in the final panel with complete 60-month
time series (12 months $\times$ 5 years). The four missing regions---Anchorage,
AK; Fairbanks, AK; Juneau, AK; and Honolulu, HI---are absent because the
weather data files for Alaska and Hawaii counties, while present in the raw
data, do not match to MSA titles in the QCEW crosswalk for these regions.

\section{Additional Tables}


\begin{table}[htbp]
\centering
\caption{\\ \textbf{Sample Restriction Criteria}}
\label{tab:sample_restrictions}
\begin{threeparttable}
\begin{tabularx}{\textwidth}{>{\raggedright\arraybackslash}X S[table-format=8.0] S[table-format=3.0] S[table-format=8.0] S[table-format=3.0]}
\toprule
\multicolumn{1}{c}{Exclusion Criteria} &
\multicolumn{2}{c}{Number of Machines} &
\multicolumn{2}{c}{Number of Machine - Quarters} \\
\cmidrule(lr){2-3}\cmidrule(lr){4-5}
& \multicolumn{1}{c}{Number} & \multicolumn{1}{c}{Percent}
& \multicolumn{1}{c}{Number} & \multicolumn{1}{c}{Percent} \\
\midrule
None  & 303244 & 100 & 2439710 & 100   \\
- Machines for work & 291649 & 96 & 2333246 & 96  \\
- Machines w/o demographic data 
                                         & 243650 & 80 & 1953678 & 80  \\
- Machines w/ $<1$ hour of \\ browsing from 7/2021 to 6/2022 
                                         &  199783 & 66 & 1741563 & 71 \\
\bottomrule
\end{tabularx}
\end{threeparttable}
\end{table}

\clearpage


\begin{table}[htbp]
\caption{\\ \centering \textbf{OLS estimates of ChatGPT adoption on browsing activity by category}}  

\setlength{\abovedisplayskip}{0pt}
\setlength{\belowdisplayskip}{0.5pt}
\setlength{\abovedisplayshortskip}{0pt}
\setlength{\belowdisplayshortskip}{0.5pt}
\vspace{-0.1cm} \label{table:ols-bycat} \footnotesize  This table shows estimates of long difference OLS regressions comparing 2024 browsing characteristics to 2022 browsing characteristics, in an estimatin of the form
$$\Delta\text{BrowsingOutcome}_{i} = \gamma\; \text{ChatGPTUse}_{i}  + \lambda_{FEs}  + \varepsilon_{i}.$$
The only RHS variable is a dummy for whether household $j$ has ever used ChatGPT by the end of 2024. The dependent variable in each regression is a measure of the household's browsing activity in different categories. In Column (1) it is the quarterly log of browsing duration across all sites; in Columns (2),(3) and (4) it is the quarterly log of browsing duration on  leisure, productive, and mixed sites; and in Columns (5), (6), and (7) the change in the share of the household's total browsing that is  on  leisure, productive, and mixed sites. All regressions include fixed effects for income bin X age bin X MSA. T-statistics based on heteroskedasticity-robust standard errors clustered at the income bin X age bin X MSA level are in parentheses: * p$<$0.10, ** p$<$0.05, *** p$<$0.01.  \vspace{0.1cm}

\footnotesize
 \centering \setlength{\tabcolsep}{2pt}
\begin{adjustbox}{max width=\textwidth}
\begin{tabular}{@{}l*{7}{c}@{}} 
\toprule
\textit{Specification:} &  \multicolumn{7}{c}{\textit{Long differences OLS: 2024 vs. 2022}}\\
\cmidrule(lr){2-8} 
\addlinespace
\textit{Dep. var.} &  $\substack{\Delta \text{ Log Duration} \\ \text{All}}$  &   $\substack{\Delta \text{ Log Duration} \\ \text{Leisure}}$ &  $\substack{\Delta \text{ Log Duration} \\ \text{Productive}}$   &  $\substack{\Delta \text{ Log Duration} \\ \text{Mixed}}$   &  $\substack{\Delta \text{ Share} \\ \text{Leisure}}$  &  $\substack{\Delta \text{ Share} \\ \text{Productive}}$  &  $\substack{\Delta \text{ Share} \\ \text{Mixed}}$ \\
   \cmidrule(lr){2-2}   \cmidrule(lr){3-3}   \cmidrule(lr){4-4}   \cmidrule(lr){5-5}   \cmidrule(lr){6-6}   \cmidrule(lr){7-7}   \cmidrule(lr){8-8} 
 &  (1) & (2) & (3) & (4) & (5) & (6) & (7) \\
   \addlinespace
   \midrule
\addlinespace ChatGPT Use&       0.508***&       0.392***&       0.539***&       0.407***&      -0.009** &       0.011** &      -0.002   \\
            &    (14.566)   &     (8.480)   &    (14.836)   &     (9.443)   &    (-2.095)   &     (2.281)   &    (-0.608)   \\
\addlinespace Observations&      43,874   &      43,874   &      43,874   &      43,874   &      43,874   &      43,874   &      43,874   \\

\midrule
 	Estimation & OLS & OLS & OLS & OLS & OLS & OLS & OLS   \\ 
Inc. X Age X MSA FEs     & X & X	& X & X	& X & X	& X      \\	
    \bottomrule
\end{tabular}
\end{adjustbox}
\end{table}

\clearpage

\begin{table}[htbp]
\caption{\\ \centering \textbf{Effect of ChatGPT adoption on browsing activity by Age}}  

\setlength{\abovedisplayskip}{0pt}
\setlength{\belowdisplayskip}{0.5pt}
\setlength{\abovedisplayshortskip}{0pt}
\setlength{\belowdisplayshortskip}{0.5pt}
\vspace{-0.1cm} \label{table:2sls-bycat-byage} \footnotesize  This table shows estimates of long difference IV regressions comparing 2024 browsing characteristics to 2022 browsing characteristics, in an estimatin of the form
$$\Delta\text{BrowsingOutcome}_{i} = \gamma\; \text{ChatGPTUse}_{i}  + \lambda_{FEs} + X_i'\xi + \varepsilon_{i}.$$
The endogenous variable is a dummy for whether household $j$ has ever used ChatGPT by the end of 2024. The instrument is the panelist household's predicted exposure to ChatGPT in 2021, based on the household's share of browsing on highly substitutable sites. The dependent variable in each regression is a measure of the household's browsing activity in different categories. In Column (1) it is the quarterly log of browsing duration across all sites; in Columns (2),(3) and (4) it is the quarterly log of browsing duration on  leisure, productive, and mixed sites; and in Columns (5), (6), and (7) the change in the share of the household's total browsing that is  on  leisure, productive, and mixed sites. All regressions include fixed effects for income bin X age bin X MSA. All regressions also include control variables for the generative AI exposure predicted by the households 2021 browsing composition by Comscore content category, and for the sum of the browsing share of websites with a GenAI exposure label in 2021 - the 'Bartik share control'. T-statistics based on heteroskedasticity-robust standard errors clustered at the income bin X age bin X MSA level are in parentheses: * p$<$0.10, ** p$<$0.05, *** p$<$0.01.  \vspace{0.1cm}

\footnotesize
 \centering \setlength{\tabcolsep}{2pt}
\begin{adjustbox}{max width=\textwidth}
\begin{tabular}{@{}l*{7}{c}@{}} 
\toprule
\textit{Specification:} &  \multicolumn{7}{c}{\textit{Long differences: 2024 vs. 2022}}\\
\cmidrule(lr){2-8} 
\addlinespace
\textit{Dep. var.} &  $\substack{\Delta \text{ Log Duration} \\ \text{All}}$  &   $\substack{\Delta \text{ Log Duration} \\ \text{Leisure}}$ &  $\substack{\Delta \text{ Log Duration} \\ \text{Productive}}$   &  $\substack{\Delta \text{ Log Duration} \\ \text{Mixed}}$   &  $\substack{\Delta \text{ Share} \\ \text{Leisure}}$  &  $\substack{\Delta \text{ Share} \\ \text{Productive}}$  &  $\substack{\Delta \text{ Share} \\ \text{Mixed}}$ \\
   \cmidrule(lr){2-2}   \cmidrule(lr){3-3}   \cmidrule(lr){4-4}   \cmidrule(lr){5-5}   \cmidrule(lr){6-6}   \cmidrule(lr){7-7}   \cmidrule(lr){8-8} 
 &  (1) & (2) & (3) & (4) & (5) & (6) & (7) \\
   \addlinespace
   \midrule
\multicolumn{8}{l}{\textit{Panel A: Young Households}} \\
\addlinespace ChatGPT Use&       0.314   &       2.375   &      -1.590   &      -1.023   &       0.366*  &      -0.514** &       0.147   \\
            &     (0.248)   &     (1.270)   &    (-1.130)   &    (-0.588)   &     (1.798)   &    (-2.183)   &     (0.967)   \\
\addlinespace Observations&       6,924   &       6,924   &       6,924   &       6,924   &       6,924   &       6,924   &       6,924   \\
1st-stage KP F-stat.&          11   &          11   &          11   &          11   &          11   &          11   &          11   \\

\midrule
\multicolumn{8}{l}{\textit{Panel B: Middle Age Households}} \\
\addlinespace ChatGPT Use&       0.504   &       1.238   &      -0.540   &       0.030   &       0.334***&      -0.305***&      -0.029   \\
            &     (0.873)   &     (1.416)   &    (-0.872)   &     (0.041)   &     (3.838)   &    (-3.124)   &    (-0.444)   \\
\addlinespace Observations&      20,260   &      20,260   &      20,260   &      20,260   &      20,260   &      20,260   &      20,260   \\
1st-stage KP F-stat.&          88   &          88   &          88   &          88   &          88   &          88   &          88   \\

\midrule
\multicolumn{8}{l}{\textit{Panel C: Old Age Households}} \\
\addlinespace ChatGPT Use&       3.356***&       1.351   &       2.302*  &      -0.277   &       0.206   &       0.190   &      -0.395***\\
            &     (2.692)   &     (0.922)   &     (1.830)   &    (-0.207)   &     (1.342)   &     (1.207)   &    (-2.972)   \\
\addlinespace Observations&      15,702   &      15,702   &      15,702   &      15,702   &      15,702   &      15,702   &      15,702   \\
1st-stage KP F-stat.&          34   &          34   &          34   &          34   &          34   &          34   &          34   \\

\midrule

 	Estimation & IV & IV & IV & IV & IV & IV & IV   \\ 
Inc. X Age X MSA FEs     & X & X	& X & X	& X & X	& X      \\	
Bartik share control     & X & X	& X & X	& X & X	& X      \\	
Website category exposure    & X & X	& X & X	& X & X	& X      \\	
    \bottomrule
\end{tabular}
\end{adjustbox}
\end{table}

\clearpage

\begin{table}[htbp]
\caption{\\ \centering \textbf{Effect of ChatGPT adoption on browsing activity by Income}}  

\setlength{\abovedisplayskip}{0pt}
\setlength{\belowdisplayskip}{0.5pt}
\setlength{\abovedisplayshortskip}{0pt}
\setlength{\belowdisplayshortskip}{0.5pt}
\vspace{-0.1cm} \label{table:2sls-bycat-byinc} \footnotesize  This table shows estimates of long difference IV regressions comparing 2024 browsing characteristics to 2022 browsing characteristics, in an estimatin of the form
$$\Delta\text{BrowsingOutcome}_{i} = \gamma\; \text{ChatGPTUse}_{i}  + \lambda_{FEs} + X_i'\xi + \varepsilon_{i}.$$
The endogenous variable is a dummy for whether household $j$ has ever used ChatGPT by the end of 2024. The instrument is the panelist household's predicted exposure to ChatGPT in 2021, based on the household's share of browsing on highly substitutable sites. The dependent variable in each regression is a measure of the household's browsing activity in different categories. In Column (1) it is the quarterly log of browsing duration across all sites; in Columns (2),(3) and (4) it is the quarterly log of browsing duration on  leisure, productive, and mixed sites; and in Columns (5), (6), and (7) the change in the share of the household's total browsing that is  on  leisure, productive, and mixed sites. All regressions include fixed effects for income bin X age bin X MSA. All regressions also include control variables for the generative AI exposure predicted by the households 2021 browsing composition by Comscore content category, and for the sum of the browsing share of websites with a GenAI exposure label in 2021 - the 'Bartik share control'. T-statistics based on heteroskedasticity-robust standard errors clustered at the income bin X age bin X MSA level are in parentheses: * p$<$0.10, ** p$<$0.05, *** p$<$0.01.  \vspace{0.1cm}

\footnotesize
 \centering \setlength{\tabcolsep}{2pt}
\begin{adjustbox}{max width=\textwidth}
\begin{tabular}{@{}l*{7}{c}@{}} 
\toprule
\textit{Specification:} &  \multicolumn{7}{c}{\textit{Long differences: 2024 vs. 2022}}\\
\cmidrule(lr){2-8} 
\addlinespace
\textit{Dep. var.} &  $\substack{\Delta \text{ Log Duration} \\ \text{All}}$  &   $\substack{\Delta \text{ Log Duration} \\ \text{Leisure}}$ &  $\substack{\Delta \text{ Log Duration} \\ \text{Productive}}$   &  $\substack{\Delta \text{ Log Duration} \\ \text{Mixed}}$   &  $\substack{\Delta \text{ Share} \\ \text{Leisure}}$  &  $\substack{\Delta \text{ Share} \\ \text{Productive}}$  &  $\substack{\Delta \text{ Share} \\ \text{Mixed}}$ \\
   \cmidrule(lr){2-2}   \cmidrule(lr){3-3}   \cmidrule(lr){4-4}   \cmidrule(lr){5-5}   \cmidrule(lr){6-6}   \cmidrule(lr){7-7}   \cmidrule(lr){8-8} 
 &  (1) & (2) & (3) & (4) & (5) & (6) & (7) \\
   \addlinespace
   \midrule
\multicolumn{8}{l}{\textit{Panel A: Low-Income Households}} \\
\addlinespace ChatGPT Use&       3.132***&       4.530***&       0.887   &       1.624   &       0.407***&      -0.355** &      -0.052   \\
            &     (2.996)   &     (3.459)   &     (0.853)   &     (1.310)   &     (2.873)   &    (-2.469)   &    (-0.471)   \\
\addlinespace Observations&      16,489   &      16,489   &      16,489   &      16,489   &      16,489   &      16,489   &      16,489   \\
1st-stage KP F-stat.&          55   &          55   &          55   &          55   &          55   &          55   &          55   \\

\midrule
\multicolumn{8}{l}{\textit{Panel B: Middle-Income Households}} \\
\addlinespace ChatGPT Use&       0.992   &       0.716   &      -0.190   &      -0.123   &       0.257** &      -0.227*  &      -0.030   \\
            &     (1.117)   &     (0.639)   &    (-0.195)   &    (-0.117)   &     (2.102)   &    (-1.685)   &    (-0.328)   \\
\addlinespace Observations&      12,941   &      12,941   &      12,941   &      12,941   &      12,941   &      12,941   &      12,941   \\
1st-stage KP F-stat.&          30   &          30   &          30   &          30   &          30   &          30   &          30   \\

\midrule
\multicolumn{8}{l}{\textit{Panel C: High-Income Households}} \\
\addlinespace ChatGPT Use&       0.443   &       0.614   &      -0.272   &      -1.433   &       0.296** &      -0.128   &      -0.168*  \\
            &     (0.549)   &     (0.488)   &    (-0.336)   &    (-1.335)   &     (2.407)   &    (-1.014)   &    (-1.752)   \\
\addlinespace Observations&      13,456   &      13,456   &      13,456   &      13,456   &      13,456   &      13,456   &      13,456   \\
1st-stage KP F-stat.&          39   &          39   &          39   &          39   &          39   &          39   &          39   \\

\midrule

 	Estimation & IV & IV & IV & IV & IV & IV & IV   \\ 
Inc. X Age X MSA FEs     & X & X	& X & X	& X & X	& X      \\	
Bartik share control     & X & X	& X & X	& X & X	& X      \\	
Website category exposure    & X & X	& X & X	& X & X	& X      \\	
    \bottomrule
\end{tabular}
\end{adjustbox}
\end{table}


\clearpage

\begin{table}[htbp]
\caption{\\ \centering \textbf{Full Engel Curve Estimation Results}}

\vspace{-0.1cm}  \footnotesize  This table shows estimates of age-income-region-level panel regressions of the form
\begin{center}$
s_{it}^{a}
=
\delta_{t}^{a}
+
\gamma^{a}\,\ln H_{it}
+
u_{it}^{a},
$
\end{center}
where $\ln H_{it}$ is the log of total browsing time, and $s_{it}^{a}$ is the browsing share of activity $a$ (e.g., leisure or productive activity). In columns 3 and 4, the dependent variable is instead the log of the browsing time for each activity category - the estimates are identical to those in Table \ref{tab:engel_regs_main}. Implied Engel elasticities are computed in columns 1 and 2 as $\hat{\beta}^a_{\text{Engel}} = 1+ \frac{\hat{\gamma}^a}{\bar{s}^a}$. Columns (2) and (4) show IV estimates where the instrument is the log of total precipitation in the region in the year-quarter.
All regressions include fixed effects that capture the interaction between income and age categories, and also fixed effects in the time dimension (year-quarter level). Heteroskedasticity-robust standard errors double clustered at the age-income-region level and at the year-quarter level in parentheses: * p$<$0.10, ** p$<$0.05, *** p$<$0.01. \label{tab:engel_regs}  \vspace{.2cm}
 
 \centering
\begin{adjustbox}{max width=\textwidth}
\begin{tabular}{@{}l*{6}{c}@{}}
\toprule
\textit{Dep. var} & \multicolumn{2}{c}{Browsing share} &  \multicolumn{2}{c}{Log Activity Time} \\
   \cmidrule(lr){2-3} \cmidrule(lr){4-5}
   \addlinespace
   	Estimation & OLS & IV & OLS & IV \\ 
      \addlinespace
      \cmidrule(lr){2-2} \cmidrule(lr){5-5}  \cmidrule(lr){3-3} \cmidrule(lr){4-4}
                                        &\multicolumn{1}{c}{(1)}   &\multicolumn{1}{c}{(2)}   &\multicolumn{1}{c}{(3)}   &\multicolumn{1}{c}{(4)}     \\ 
                   \midrule
\emph{Panel A: Productive digital activities}  &&&& \\
\addlinespace
                   
                   \addlinespace Log(Total Browsing Time) &      -0.015***&      -0.045***&       0.991***&       0.931***\\
            &   (-20.871)   &    (-3.520)   &   (530.879)   &    (29.244)   \\
Observations&     129,498   &      99,608   &     129,146   &      99,430   \\
1st-stage KP F-stat.&               &          55   &               &          54   \\
Implied Engel Elasticity&        0.98   &        0.93   &        0.99   &        0.93   \\
  
\midrule
\emph{Panel B: Leisure digital activities}  &&&& \\
\addlinespace
                   
                   \addlinespace Log(Total Browsing Time) &       0.017***&       0.038***&       1.130***&       1.374***\\
            &    (28.441)   &     (3.663)   &   (241.949)   &    (15.823)   \\
Observations&     129,498   &      99,608   &     120,785   &      94,691   \\
1st-stage KP F-stat.&               &          55   &               &          48   \\
Implied Engel Elasticity&        1.09   &        1.20   &        1.13   &        1.37   \\
  
\midrule
\emph{Panel C: Other digital activities}  &&&& \\
\addlinespace
                   
                   \addlinespace Log(Total Browsing Time) &      -0.002***&       0.008   &       1.026***&       1.110***\\
            &    (-3.749)   &     (0.857)   &   (276.169)   &    (17.912)   \\
Observations&     129,498   &      99,608   &     125,862   &      97,739   \\
1st-stage KP F-stat.&               &          55   &               &          51   \\
Implied Engel Elasticity&        0.99   &        1.04   &        1.03   &        1.11   \\
  
\midrule
	Age x Income Group FEs & X & X	 & X & X    \\		
	Time FEs & X & X	 & X & X    \\		
    \bottomrule
\end{tabular}
\end{adjustbox}
\end{table}

\begin{table}[htbp]
\caption{\\ \centering \textbf{Websites with lowest relative prevalence around ChatGPT use.}}  

\vspace{-0.1cm} \label{table:bottomsites-gptwindow} \footnotesize  This table shows websites with the {lowest} prevalence in 30 minute intervals with ChatGPT use compared to non-user browsing matched on time of day and age bin $\times$ income bin demographics.

\vspace{0.2cm}

\centering
\begin{adjustbox}{max width=\textwidth}   
\begin{tabular}{@{}l*{4}{c}@{}}
\toprule
Rank & Domain & $\Delta$ Browsing \% \\
\midrule
      \addlinespace
                   1&youtube.com&-5\\
2&msn.com&-1.8\\
3&yahoo.com&-1.7\\
4&facebook.com&-1.7\\
5&aol.com&-1.2\\
6&amazon.com&-1.1\\
7&capitaloneshopping.com&-1.1\\
8&pornhub.com&-1\\
9&chaturbate.com&-.9\\
10&ebay.com&-.7\\
11&live.com&-.65\\
12&sprig.com&-.56\\
13&xvideos.com&-.51\\
14&popin.cc&-.5\\
15&deviantart.com&-.48\\
  
                   \bottomrule
                   \end{tabular}
\end{adjustbox}
\begin{adjustbox}{max width=\textwidth}
\begin{tabular}{@{}l*{4}{c}@{}}
\toprule
Rank & Domain & $\Delta$ Browsing \% \\
\midrule
      \addlinespace
                   16&fanfiction.net&-.46\\
17&xhamster.com&-.44\\
18&espn.com&-.38\\
19&crunchyroll.com&-.33\\
20&bing.com&-.33\\
21&centurygames.com&-.31\\
22&walmart.com&-.28\\
23&tubitv.com&-.27\\
24&intuit.com&-.27\\
25&xnxx.com&-.26\\
26&weather.com&-.25\\
27&searchmask.com&-.24\\
28&instagram.com&-.23\\
29&wikipedia.org&-.23\\
30&kick.com&-.23\\
  
                   \bottomrule
                   \end{tabular}
\end{adjustbox}
\end{table}

\begin{table}[!htbp]
\caption{\\ \centering \textbf{Generative AI exposure by website category.}}  

\vspace{-0.1cm} \label{table:exp-bycat} \footnotesize  This table shows the average 2021 duration-weighted mean exposure score (based on website activities) for websites in each usage category.

\vspace{0.2cm}
\centering
\begin{tabular}{l r}
\hline
{Usage type} & Activity exposure score \\
\hline
Leisure & 0.177 \\
Mixed-use & 0.299 \\
Productive & 0.507 \\
\hline
\end{tabular}
\end{table}

\FloatBarrier
\section{Additional Figures}

\begin{figure}[!hbt]
\caption[.]{\\\textbf{Website classification methodology}}\label{fig:llm}

\vspace{-0.1cm} \small  This figure illustrates the data processing pipeline used for generating labels that classify websites into having activities exposed to generative AI and into mainly having \textit{Productive} or \textit{Leisure} uses, with separate labels for \textit{Ad platforms \& Content distribution networks}, as well as \textit{Mixed-use} sites that cannot be labeled definitively.

\centering
\includegraphics[width=0.75\textwidth]{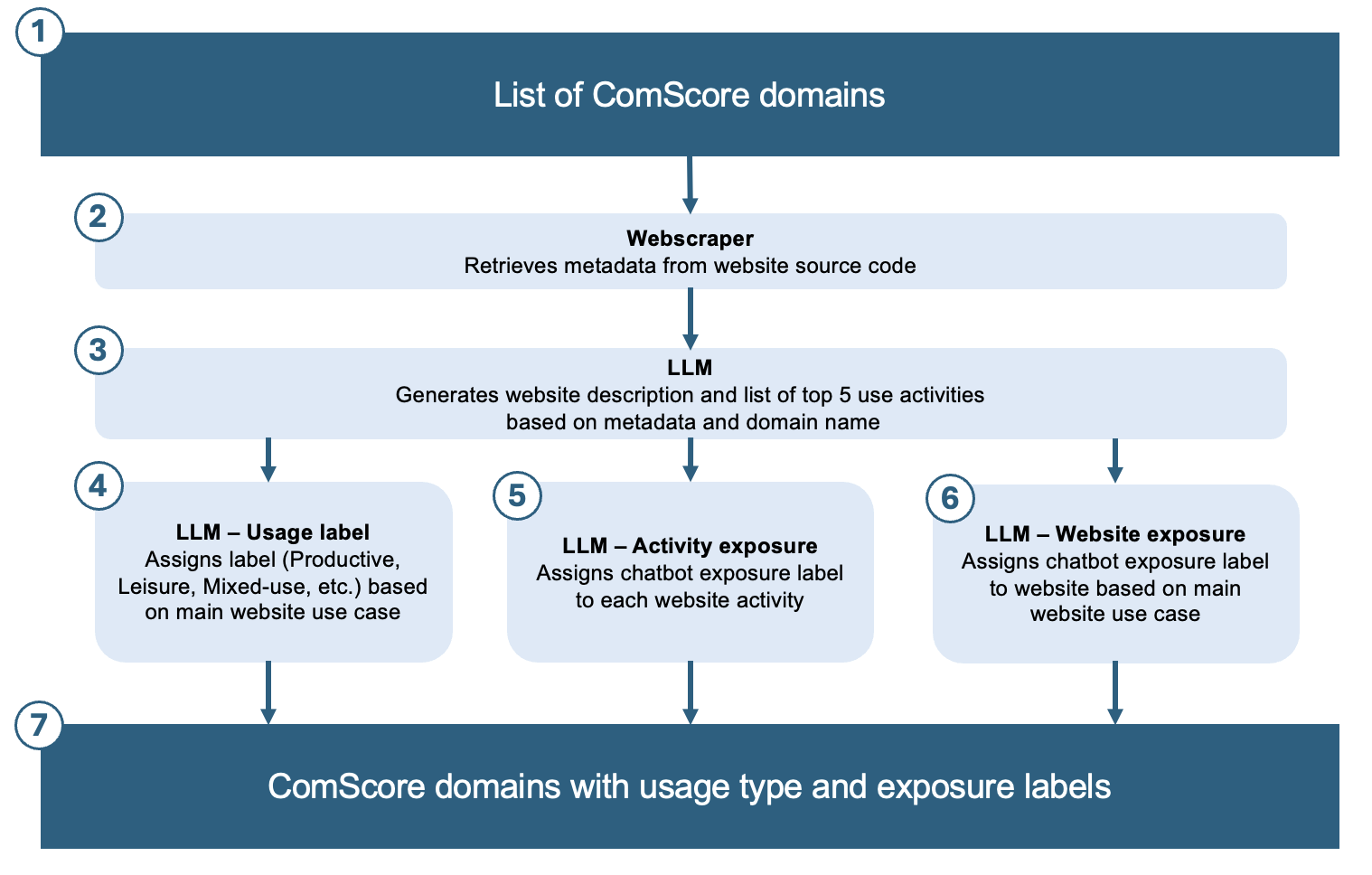} 
\end{figure}

\begin{figure}[!hbt]
\caption{Duration-weighted double sorts of website overlap fractions and Comscore activity categories} \label{fig:exposure_by_activity}
\centering
\begin{subfigure}{.5\textwidth}
  \centering
  \includegraphics[width=\linewidth]{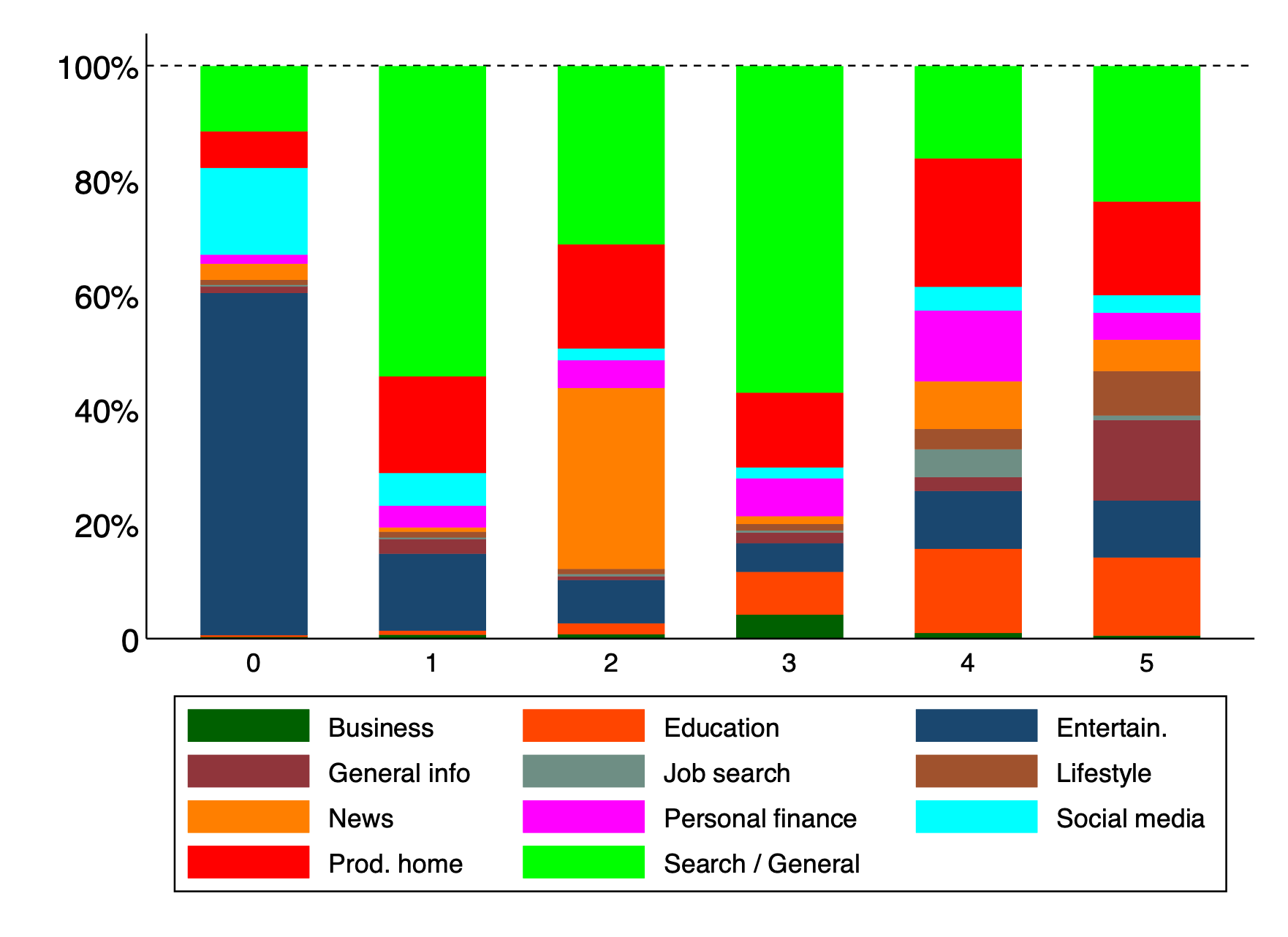}
  \caption{Duration shares of overlap fractions}
\end{subfigure}%
\begin{subfigure}{.5\textwidth}
  \centering
  \includegraphics[width=\linewidth]{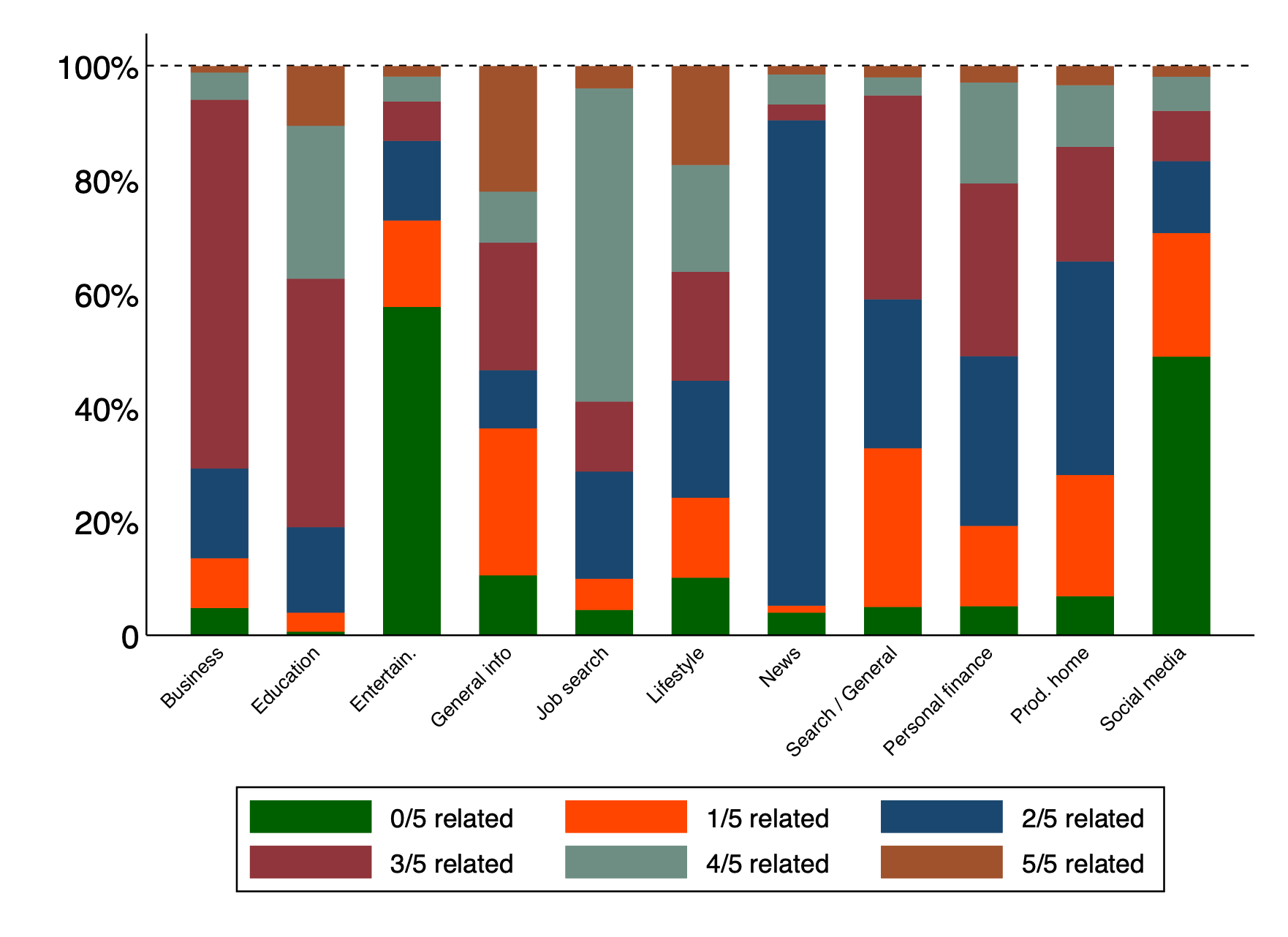}
  \caption{Duration shares of site activity categories}
  \label{fig:sub2}
\end{subfigure}
\end{figure}

\begin{figure}
\caption[.]{\\\textbf{Exposure of website activity to ChatGPT substitution.}}\label{fig:exposureshares}

\vspace{-0.1cm} \small  This figure shows the share of online household activity duration that consists of website activities that ChatGPT can be useful for. Panel A shows the share of all activity that is labeled as exposed, while Panel B shows the distribution of browsing duratoin over websites with different levels of exposure, proxied by the share of activities associated with the website that are labeled as exposed.

\centering
\begin{subfigure}{.34\textwidth}
  \centering
  \includegraphics[width=\linewidth]{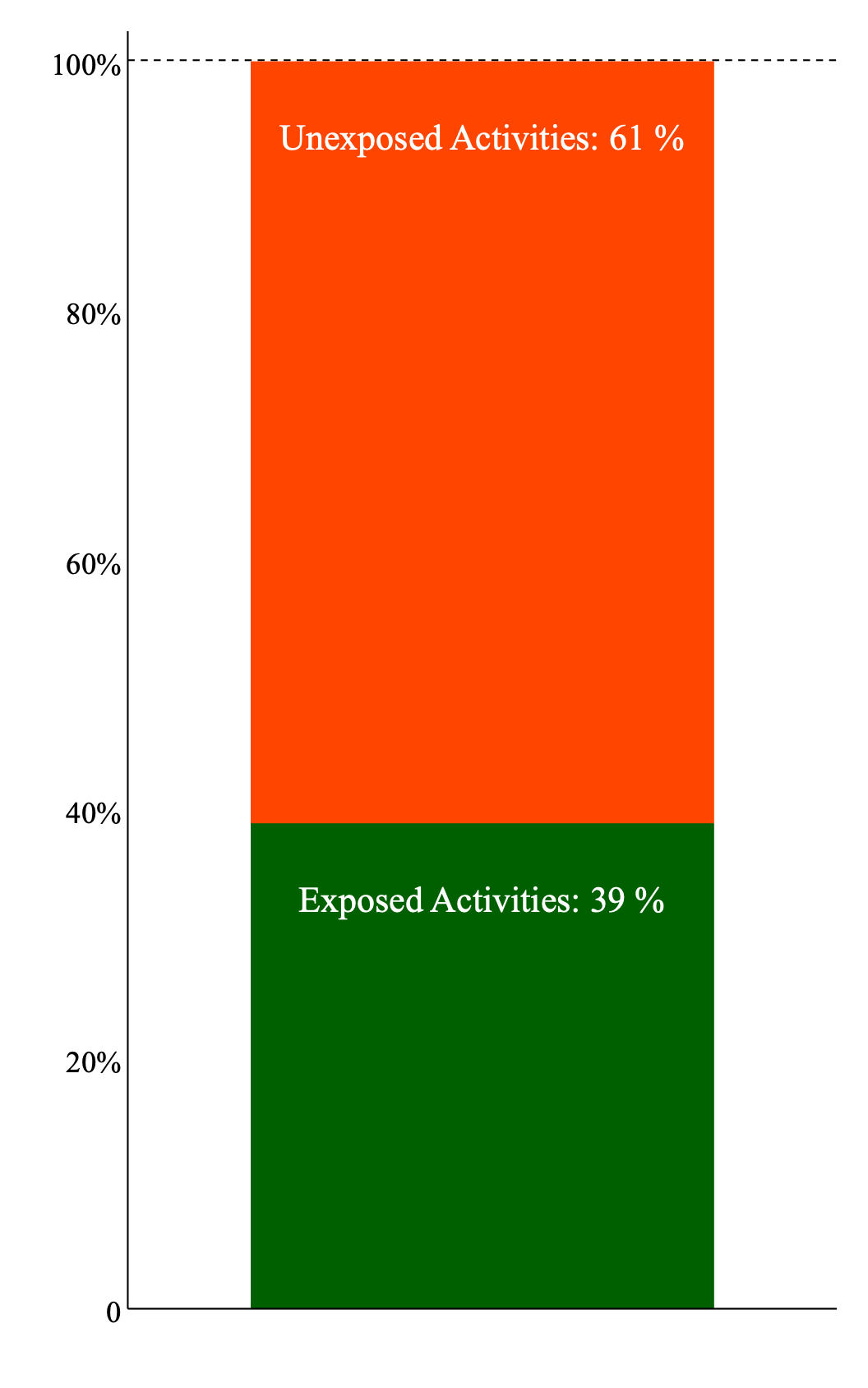}
  \caption{Duration share of exposed activities}
\end{subfigure}%
\begin{subfigure}{.66\textwidth}
  \centering
  \includegraphics[width=\linewidth]{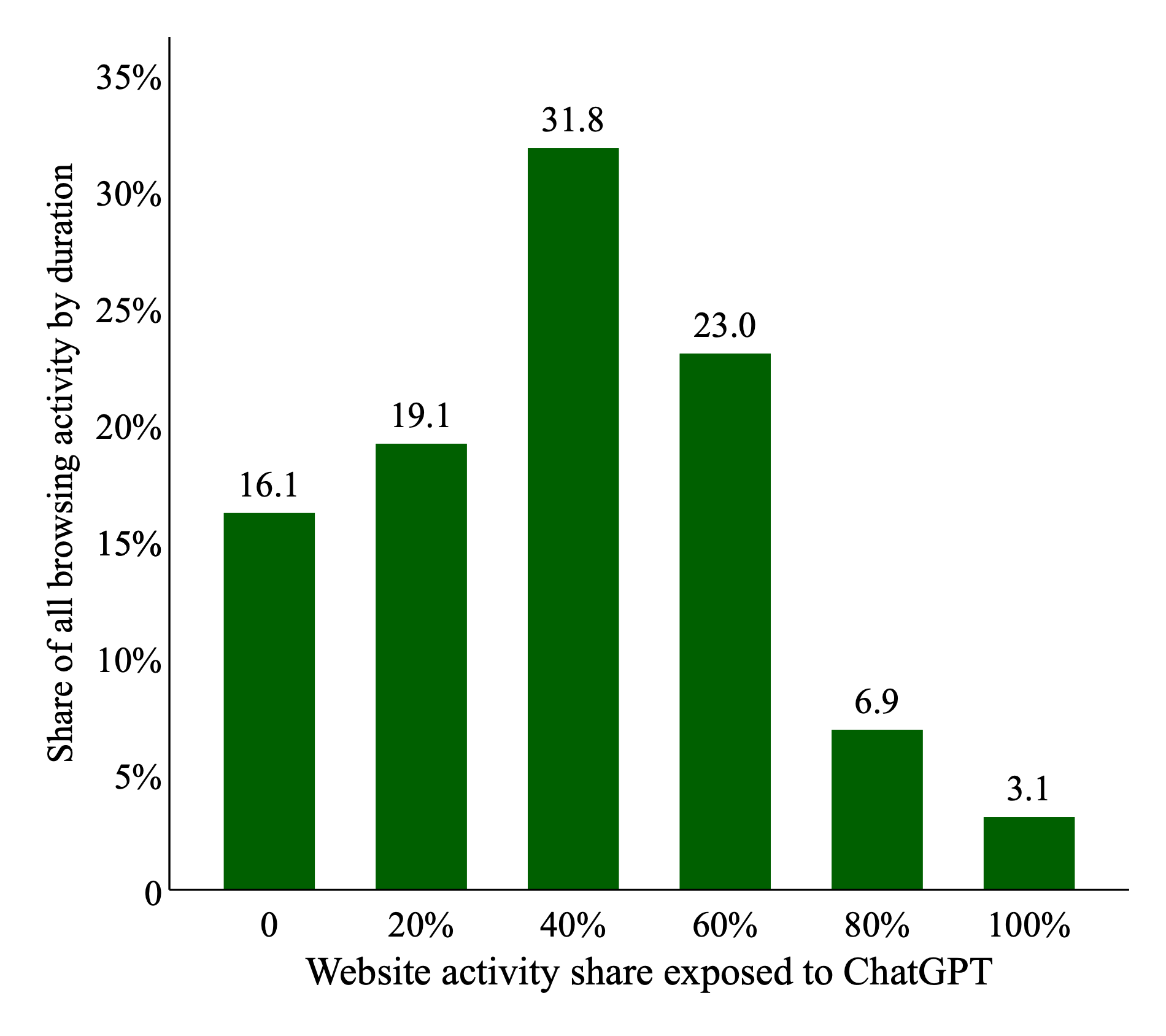}
  \caption{Duration distribution over website exposure levels}
  \label{fig:sub2}
\end{subfigure}
\end{figure}

\clearpage


\begin{figure}[!hbt]
\caption{Pre-ChatGPT release duration shares for different site categories, by household income bucket}
\label{fig:ts-first=stage}
    \centering 
     \includegraphics[width=1\linewidth]{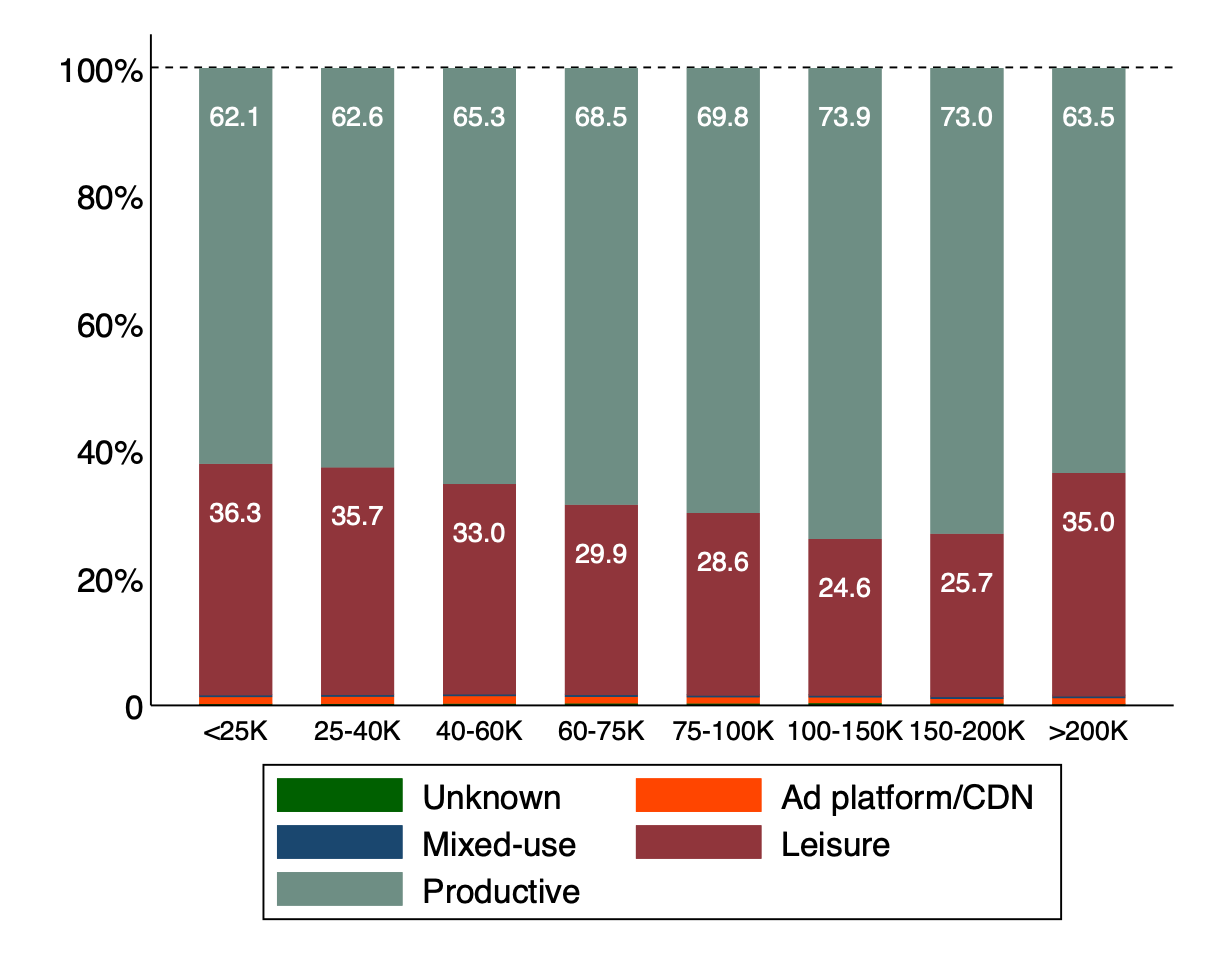}  
\end{figure}

\clearpage

\clearpage


\begin{figure}[!hbt]
\caption{\textbf{Usage of sites associated with different activities in vs. outside ChatGPT adoption windows}}
\label{fig:hf-activity}

\vspace{-0.1cm} \small 

\centering
\begin{subfigure}{.3\textwidth}
  \centering
  \includegraphics[width=\linewidth]{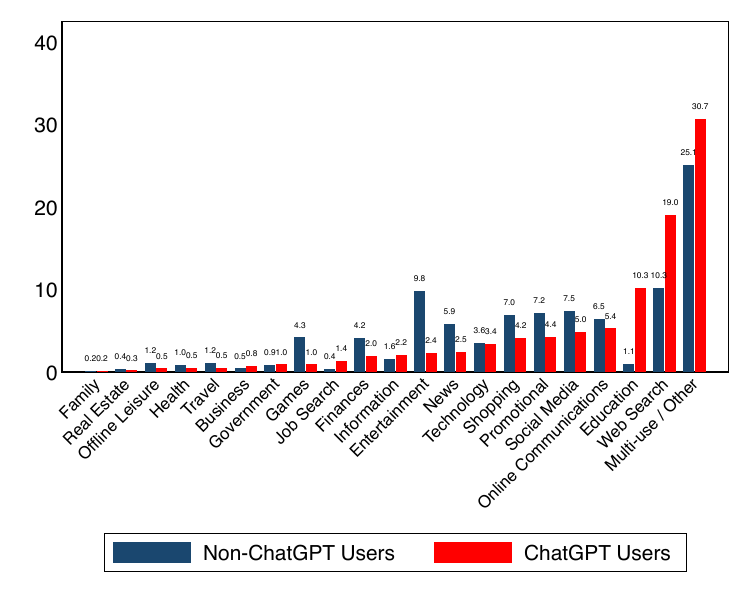}
  \caption{GPT window: \\ browsing shares}
\end{subfigure}
\begin{subfigure}{.3\textwidth}
  \centering
  \includegraphics[width=\linewidth]{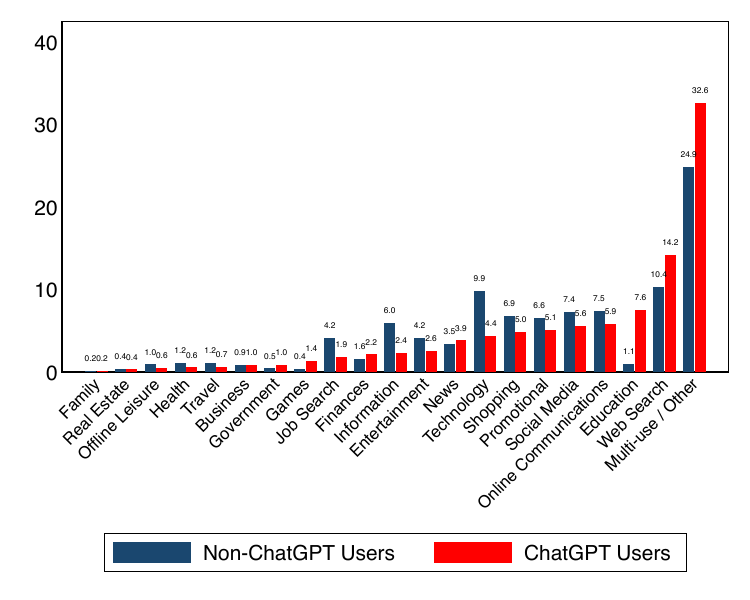}
  \caption{GPT day, outside window: \\ browsing shares}
\end{subfigure} 
\begin{subfigure}{.3\textwidth}
  \centering
  \includegraphics[width=\linewidth]{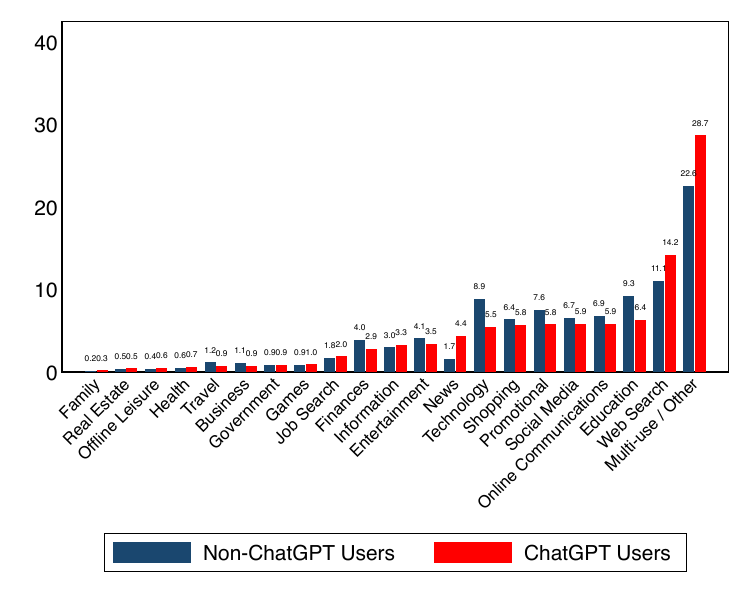}
  \caption{Pre-GPT use: \\ browsing shares}
\end{subfigure} 
\begin{subfigure}{.3\textwidth}
  \centering
  \includegraphics[width=\linewidth]{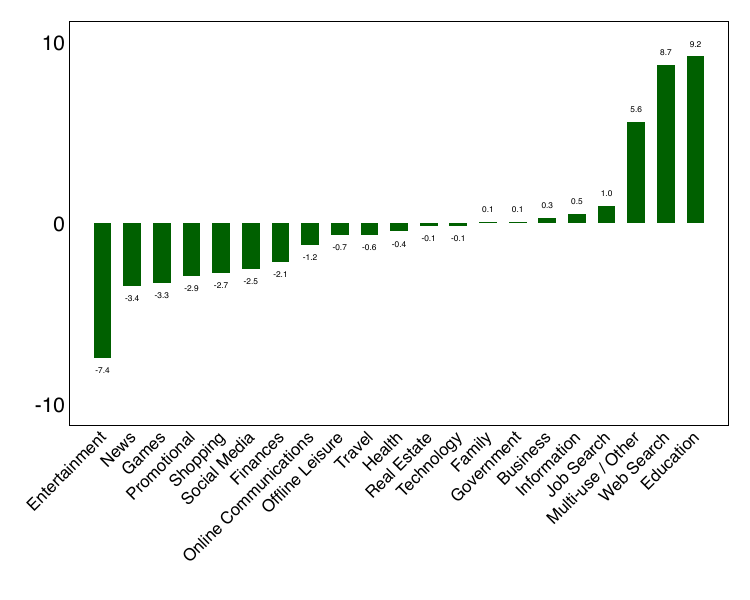}
  \caption{GPT window: \\ differences in shares}
\end{subfigure}
\begin{subfigure}{.3\textwidth}
  \centering
  \includegraphics[width=\linewidth]{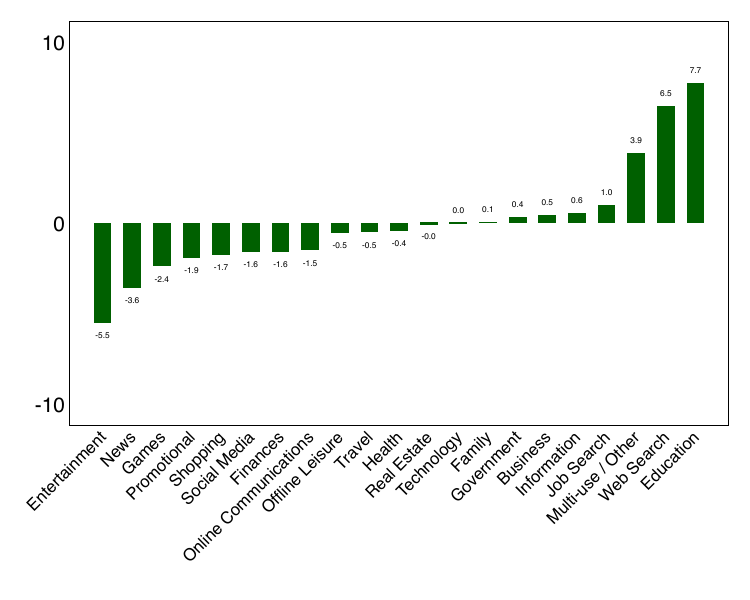}
  \caption{GPT day, outside window: \\ differences in shares}
\end{subfigure} 
\begin{subfigure}{.3\textwidth}
  \centering
  \includegraphics[width=\linewidth]{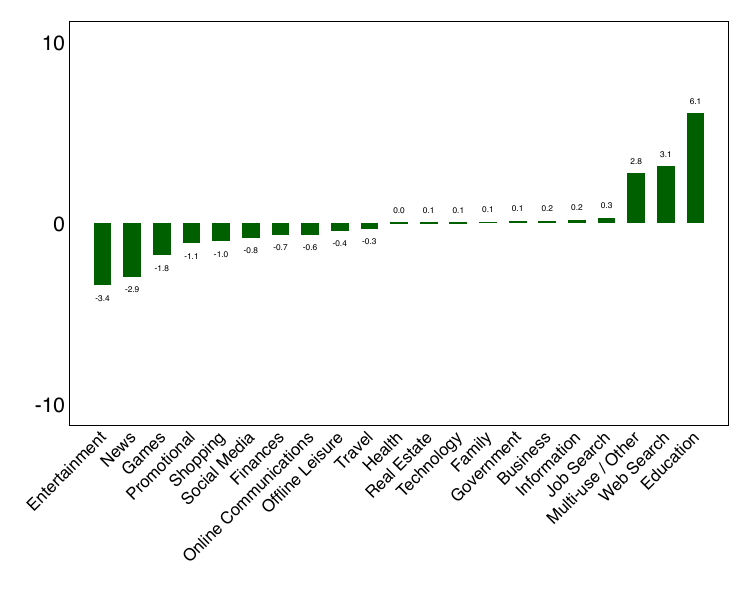}
  \caption{Pre-GPT use: \\ differences in shares}
\end{subfigure} 
\end{figure}


\clearpage


\begin{figure}[!hbt]
\caption{\textbf{Usage of sites with different GenAI overlap in vs. outside ChatGPT adoption windows}}
\label{fig:hf-exp}

\vspace{-0.1cm} \small 

\centering
\begin{subfigure}{.3\textwidth}
  \centering
  \includegraphics[width=\linewidth]{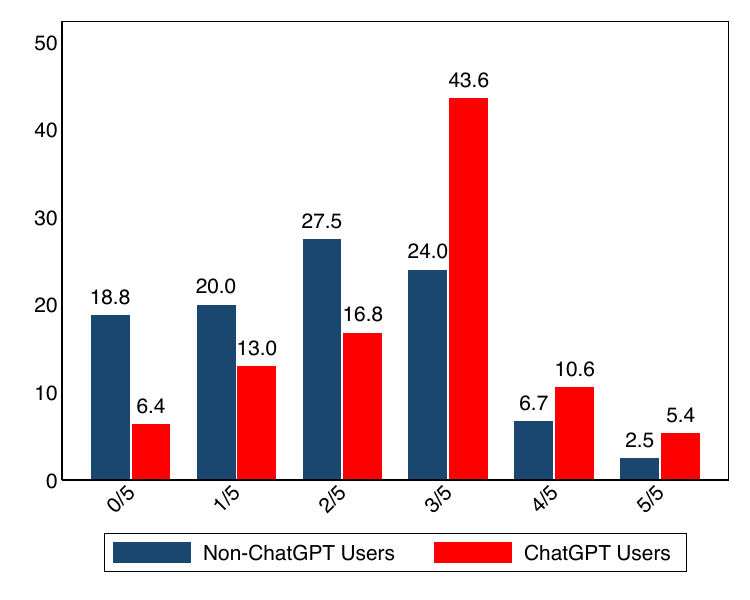}
  \caption{GPT window: \\ browsing shares}
\end{subfigure}
\begin{subfigure}{.3\textwidth}
  \centering
  \includegraphics[width=\linewidth]{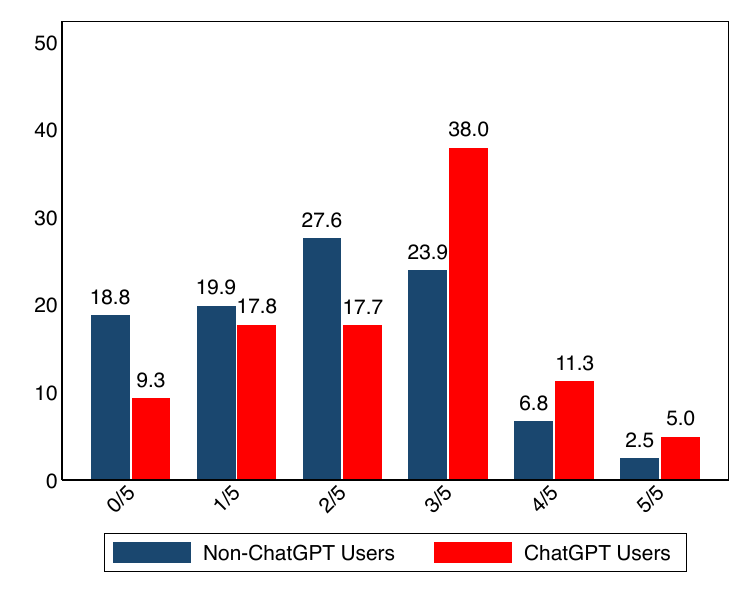}
  \caption{GPT day, outside window: \\ browsing shares}
\end{subfigure} 
\begin{subfigure}{.3\textwidth}
  \centering
  \includegraphics[width=\linewidth]{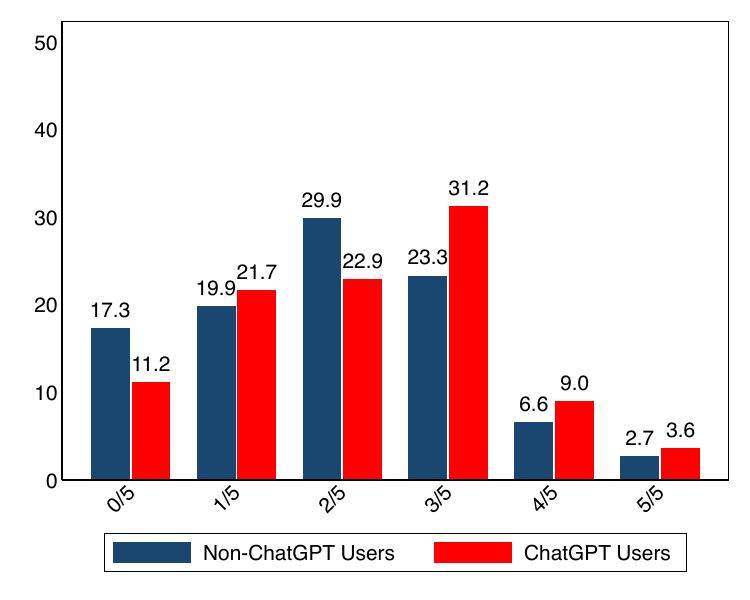}
  \caption{Pre-GPT use: \\ browsing shares}
\end{subfigure} 
\begin{subfigure}{.3\textwidth}
  \centering
  \includegraphics[width=\linewidth]{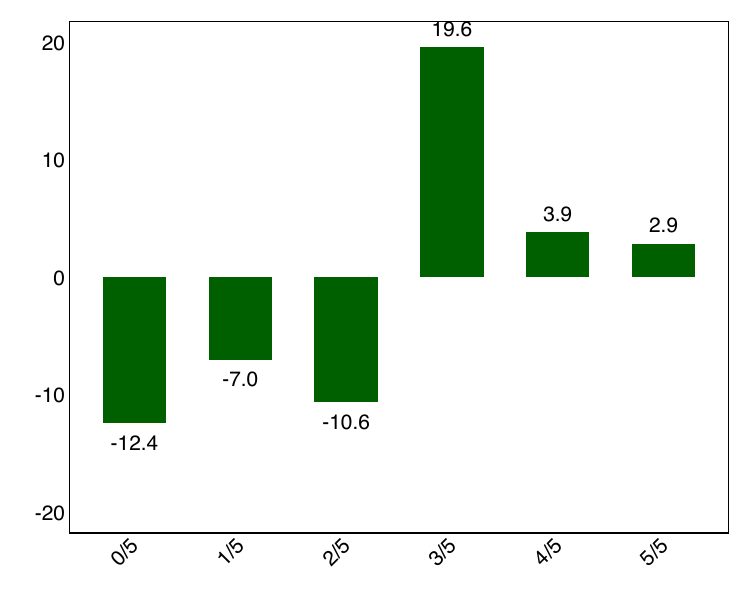}
  \caption{GPT window: \\ differences in shares}
\end{subfigure}
\begin{subfigure}{.3\textwidth}
  \centering
  \includegraphics[width=\linewidth]{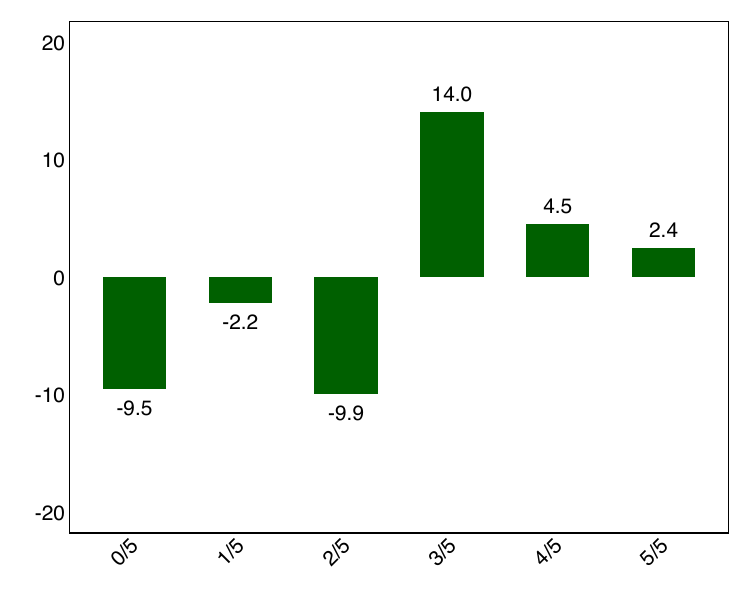}
  \caption{GPT day, outside window: \\ differences in shares}
\end{subfigure} 
\begin{subfigure}{.3\textwidth}
  \centering
  \includegraphics[width=\linewidth]{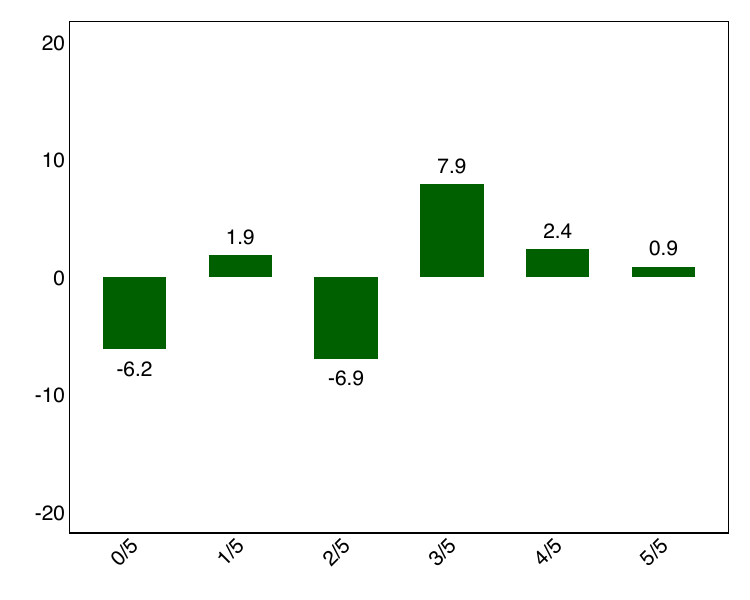}
  \caption{Pre-GPT use: \\ differences in shares}
\end{subfigure} 
\end{figure}

\section{Derivations} \normalsize
\label{app:derivations}

This appendix provides the step-by-step derivations underlying the conceptual framework in Section~\ref{sec:conceptual-framework}.

\subsection{Browsing activity demand system derivation}

\paragraph{Connection to \cite{aguiar2021} demand system}
\cite{aguiar2021} introduce an additively separable aggregator over time allocated to activities,
\[
v(h;\theta,\xi)=\sum_{j=1}^J \frac{\left(\theta_j \xi_j h_j\right)^{1-\frac{1}{\eta_j}}}{1-\frac{1}{\eta_j}},
\]
embedded in an outer utility $U(c,v)$. The key static subproblem is: allocate a fixed time endowment across activities to maximize $v$. Our framework is the special case with $J=2$ activities, $(\ell,z)$, and we work directly with $v$ (i.e., no outer $U(\cdot)$). This produces the same demand system mapping activity ``quality'' $(\theta\xi)$ and the shadow value of time into optimal time allocations.

\paragraph{Household problem.} \label{app:focs}
At time $t$, household $i$ chooses $(z_{it},\ell_{it})$ to maximize
\begin{equation}
\label{eq:app_util}
v_{it}(\ell_{it},z_{it})
= \frac{\left(\theta_{it}^{\ell}\xi_{it}^{\ell}\ell_{it}\right)^{1-\frac{1}{\eta^{\ell}}}}{1-\frac{1}{\eta^{\ell}}}
+ \frac{\left(\theta_{it}^{z}\xi_{it}^{z}z_{it}\right)^{1-\frac{1}{\eta^{z}}}}{1-\frac{1}{\eta^{z}}}
\end{equation}
subject to the time budget
\begin{equation}
\label{eq:app_time}
z_{it}+\ell_{it}\le 1.
\end{equation}
For $\eta^{\ell},\eta^{z}>0$, each term is increasing in its time argument, so the constraint binds at the optimum:
\[
z_{it}+\ell_{it}=1, \qquad \omega_{it}>0.
\]

\paragraph{First-order conditions}
Suppressing $(i,t)$ subscripts for readability, the Lagrangian is
\[
\mathcal{L}
=
\frac{\left(\theta^{\ell}\xi^{\ell}\ell\right)^{1-\frac{1}{\eta^{\ell}}}}{1-\frac{1}{\eta^{\ell}}}
+
\frac{\left(\theta^{z}\xi^{z}z\right)^{1-\frac{1}{\eta^{z}}}}{1-\frac{1}{\eta^{z}}}
+
\omega(1-z-\ell),
\]
where $\omega$ is the Lagrange multiplier (shadow value of one unit of digital time).

Define the exponents $\alpha_z \equiv 1-\frac{1}{\eta^z}$ and $\alpha_\ell\equiv 1-\frac{1}{\eta^\ell}$. Differentiating the $z$-term and using the chain rule:
\[
\frac{d}{dz}\left[\frac{(\theta^z\xi^z z)^{\alpha_z}}{\alpha_z}\right]
=(\theta^z\xi^z z)^{\alpha_z-1}(\theta^z\xi^z).
\]
Since $\alpha_z-1=-\frac{1}{\eta^z}$, we have
\[
(\theta^z\xi^z z)^{\alpha_z-1}=(\theta^z\xi^z)^{-\frac{1}{\eta^z}}z^{-\frac{1}{\eta^z}},
\]
and the first-order condition for $z$ is
\[
(\theta^z\xi^z)^{1-\frac{1}{\eta^z}}\,z^{-\frac{1}{\eta^z}}=\omega.
\]
Taking logs and solving for $\ln z$:
\begin{align*}
\left(1-\frac{1}{\eta^z}\right)(\ln\theta^z+\ln\xi^z)-\frac{1}{\eta^z}\ln z &= \ln\omega,\\
\ln z &= (\eta^z-1)(\ln\theta^z+\ln\xi^z)-\eta^z\ln\omega.
\end{align*}
Thus,
\begin{equation}
\label{eq:app_logz}
\ln(z_{it})
=
(\eta^z-1)\left(\ln\theta_{it}^z+\ln\xi_{it}^z\right)-\eta^z\ln\omega_{it}.
\end{equation}
Analogously, the FOC for $\ell$ yields
\begin{equation}
\label{eq:app_logl}
\ln(\ell_{it})
=
(\eta^\ell-1)\left(\ln\theta_{it}^\ell+\ln\xi_{it}^\ell\right)-\eta^\ell\ln\omega_{it}.
\end{equation}

\paragraph{Activity gap condition.} \label{app:gap}
Dividing~\eqref{eq:app_logz} by $\eta^z$ and~\eqref{eq:app_logl} by $\eta^\ell$, then subtracting the second from the first to eliminate $\ln\omega_{it}$:
\begin{equation}
\label{eq:app_gap}
\frac{\ln(z_{it})}{\eta^z}-\frac{\ln(\ell_{it})}{\eta^\ell}
=
\frac{\eta^z-1}{\eta^z}\left(\ln\theta_{it}^z+\ln\xi_{it}^z\right)
-
\frac{\eta^\ell-1}{\eta^\ell}\left(\ln\theta_{it}^\ell+\ln\xi_{it}^\ell\right).
\end{equation}

\paragraph{Interpretation:}
Holding $\omega_{it}$ fixed, a rise in an activity's effective ``quality'' $(\theta\xi)$ increases time in that activity if $\eta>1$ (a ``time luxury'') and decreases time if $0<\eta<1$ (a ``time necessity''). The knife-edge case $\eta=1$ corresponds to log utility in that activity.

\subsection{First-order approximation of the benefit of ChatGPT adoption} \label{app:adoption}

\paragraph{Technology shift.}
Let $\xi_{i,pre}^z=\underline{\xi}$ and, in $post$, if the household adopts,
\[
\xi_{i,post}^z=\underline{\xi}(1+\delta),
\]
while if it does not adopt, $\xi_{i,post}^z=\underline{\xi}$. All other objects are held fixed. Let $(z^N,\ell^N)$ denote the post-period optimum under No-adopt, and $(z^A,\ell^A)$ under Adopt.

\paragraph{Envelope approximation.}
For small $\delta$, approximate the utility gain from adoption by evaluating the direct effect of changing $\xi^z$ at the no-adopt choice $(z^N,\ell^N)$. The $z$-component of utility is
\[
u_z(\xi^z)=\frac{(\theta^z\xi^z z^N)^{\alpha_z}}{\alpha_z},
\qquad \alpha_z=1-\frac{1}{\eta^z}.
\]
Under adoption, $\xi^z$ becomes $\xi^z(1+\delta)$, giving a utility gain
\[
\Delta u_z
=
\frac{\left((1+\delta)^{\alpha_z}-1\right)(\theta^z\xi^z z^N)^{\alpha_z}}{\alpha_z}.
\]
Using the first-order expansion $(1+\delta)^{\alpha_z}\approx 1+\alpha_z\delta$ for small $\delta$:
\[
\Delta u_z
\approx \delta(\theta^z\xi^z z^N)^{\alpha_z}
=\delta\cdot\Big(\theta^z\underline{\xi}\,z^N\Big)^{\frac{\eta^z-1}{\eta^z}}.
\]
From the $z$ FOC, we have $\omega^N=(\theta^z\xi^z)^{1-\frac{1}{\eta^z}}(z^N)^{-\frac{1}{\eta^z}}$. Multiplying both sides by $z^N$:
\[
z^N\omega^N
=(\theta^z\xi^z z^N)^{1-\frac{1}{\eta^z}}
=(\theta^z\xi^z z^N)^{\frac{\eta^z-1}{\eta^z}}.
\]
Therefore the adoption gain can be written as
\begin{equation}
\label{eq:app_gain}
v^A-v^N \approx \delta\cdot z^N\omega^N.
\end{equation}

\paragraph{Adoption condition and time-cost interpretation.}
If the (utility) cost of adopting is $c$, the adoption rule is
\[
v^A-v^N\ge c
\quad\Rightarrow\quad
\delta\cdot z^N\omega^N \ge c.
\]
If instead the cost is a time cost $c^{time}$ (which has utility cost $c^{time}\omega^N$), the adoption condition becomes
\[
\delta z^N \ge c^{time.}
\]
Thus, to first order, adoption compares ``time saved'' $\delta z^N$ to time required to learn/use the technology.

\subsection{Time-allocation treatment effects of ChatGPT adoption} \label{app:effects}

\paragraph{Closed-form demand system}
From the FOCs, each activity's time choice can be expressed as
\begin{equation}
\label{eq:app_demand}
z=(\theta^z\xi^z)^{\eta^z-1}\omega^{-\eta^z},
\qquad
\ell=(\theta^\ell\xi^\ell)^{\eta^\ell-1}\omega^{-\eta^\ell}.
\end{equation}

\paragraph{Log-differentials.}
Let No-adopt be $N$ and Adopt be $A$. Define the log treatment effects:
\[
\beta_z^{GPT}\equiv \ln z^A-\ln z^N,
\qquad
\beta_\ell^{GPT}\equiv \ln \ell^A-\ln \ell^N.
\]
Adoption changes $\xi^z$ by factor $(1+\delta)$, so $d\ln\xi^z=\ln(1+\delta)\approx \delta$, while $\xi^\ell$ is unchanged.

Log-differentiating~\eqref{eq:app_logz} and~\eqref{eq:app_logl} with respect to $\xi^z$:
\begin{align*}
d\ln z &= (\eta^z-1)d\ln\xi^z-\eta^z d\ln\omega,\\
d\ln \ell &= -\eta^\ell d\ln\omega.
\end{align*}

The binding time constraint $z+\ell=1$ implies $dz+d\ell=0$, or $z\,d\ln z+\ell\,d\ln \ell=0$. Substituting the differentials above:
\[
z\big[(\eta^z-1)d\ln\xi^z-\eta^z d\ln\omega\big]+\ell\big[-\eta^\ell d\ln\omega\big]=0,
\]
which yields
\[
d\ln\omega=\frac{z(\eta^z-1)}{z\eta^z+\ell\eta^\ell}\,d\ln\xi^z.
\]

\paragraph{Treatment effects.}
Substituting back into the differential for $z$ and simplifying:
\begin{align*}
d\ln z
&=(\eta^z-1)d\ln\xi^z\left[1-\frac{\eta^zz}{z\eta^z+\ell\eta^\ell}\right]\\
&=(\eta^z-1)d\ln\xi^z\left[\frac{\ell\eta^\ell}{z\eta^z+\ell\eta^\ell}\right]\\
&=(\eta^z-1)\frac{1}{1+\left(\eta^z/\eta^\ell\right)\left(z/\ell\right)}\,d\ln\xi^z.
\end{align*}
Evaluating at $(z^N,\ell^N)$ with $d\ln\xi^z\approx\delta$:
\begin{equation}
\label{eq:app_betaz}
\beta_z^{GPT}
\approx
\frac{\eta^z-1}{1+\left(\eta^z/\eta^\ell\right)\left(z^N/\ell^N\right)}\cdot \delta.
\end{equation}

Similarly, for leisure:
\begin{align*}
d\ln\ell
&=-\eta^\ell\cdot\frac{z(\eta^z-1)}{z\eta^z+\ell\eta^\ell}d\ln\xi^z\\
&=(1-\eta^z)\cdot\frac{\eta^\ell/\eta^z}{1+\left(\eta^\ell/\eta^z\right)\left(\ell/z\right)}\,d\ln\xi^z,
\end{align*}
yielding
\begin{equation}
\label{eq:app_betal}
\beta_\ell^{GPT}
\approx
\frac{(1-\eta^z)\cdot\left(\eta^\ell/\eta^z\right)}{1+\left(\eta^\ell/\eta^z\right)\left(\ell^N/z^N\right)}\cdot \delta.
\end{equation}

\paragraph{Relative impact.}
Dividing the log-differentials by their respective elasticities and subtracting eliminates $d\ln\omega$:
\begin{equation}
\frac{\beta_z^{GPT}}{\eta^z}-\frac{\beta_\ell^{GPT}}{\eta^\ell}
\approx
\frac{\eta^z-1}{\eta^z}\cdot \delta.
\end{equation}
or, without the first-order approximation:
\begin{equation}
\label{eq:app_difference}
\frac{\beta_z^{GPT}}{\eta^z}-\frac{\beta_\ell^{GPT}}{\eta^\ell}
\approx
\frac{\eta^z-1}{\eta^z}\cdot \ln(1+ \delta).
\end{equation}




\subsection{Digital-time Engel Curves and the Implied GenAI Productivity Shock}
\label{app:engel_to_delta}

This appendix summarizes how we use an Engel-curve (``time demand system'') approach---following \citet{aguiar2021}---to (i) estimate relative curvature parameters for different digital activities and (ii) translate observed adoption-induced reallocations of digital time into an implied productivity shock from generative AI.

\subsection{A digital-time demand system}
\label{app:engel_setup}

Let the household allocate \emph{total} digital time $H_{it}$ across three mutually exclusive categories: digital leisure $\ell_{it}$, digital productive home-production $z_{it}$, and other digital time $o_{it}$:
\begin{equation}
\label{eq:app_digital_time_budget}
\ell_{it} + z_{it} + o_{it} \;=\; H_{it}.
\end{equation}
(In the stylized model in the main text we normalize $H_{it}=1$; here we keep $H_{it}$ explicit because it is useful for Engel-curve estimation and welfare calculations.)

Preferences over the composition of digital time are summarized by the additively separable ``digital activity aggregator''
\begin{equation}
\label{eq:app_activity_aggregator}
v(\ell_{it},z_{it},o_{it};\theta_{it},\xi_{it})
=
\sum_{a\in\{\ell,z,o\}}
\frac{\big(\theta_{it}^{a}\,\xi_{it}^{a}\,h_{it}^{a}\big)^{1-\frac{1}{\eta^{a}}}}{1-\frac{1}{\eta^{a}}}
\qquad\text{where}\qquad
(h_{it}^{\ell},h_{it}^{z},h_{it}^{o})\equiv(\ell_{it},z_{it},o_{it}).
\end{equation}
The parameter $\eta^{a}>0$ governs diminishing returns in activity $a$: for $\eta^{a}>1$, $a$ is relatively ``luxurious'' in that higher efficiency can raise time allocated to $a$; for $\eta^{a}<1$, $a$ behaves like a ``necessity'' in that efficiency improvements tend to reduce time allocated to $a$ (holding total digital time fixed), pushing time into other activities.

Given total digital time $H_{it}$, the household solves the within-digital-time allocation subproblem
\begin{equation}
\label{eq:app_subproblem}
\max_{\{\ell_{it},z_{it},o_{it}\}} \; v(\ell_{it},z_{it},o_{it};\theta_{it},\xi_{it})
\quad \text{s.t.}\quad
\ell_{it}+z_{it}+o_{it}=H_{it}.
\end{equation}
Let $\hat{\omega}_{it}$ denote the \emph{normalized} shadow price of digital time (the Lagrange multiplier on \eqref{eq:app_digital_time_budget} divided by the marginal utility of the aggregator if $v$ is embedded in a higher-level utility function).\footnote{This normalization is convenient because it drops out of relative demands across activities. See \citet{aguiar2021} for details.} The first-order condition for activity $a$ implies a constant-elasticity ``demand'' for time in activity $a$:
\begin{equation}
\label{eq:app_activity_demand}
h_{it}^{a}
=
\big(\theta_{it}^{a}\,\xi_{it}^{a}\big)^{\eta^{a}-1}\;
\hat{\omega}_{it}^{-\eta^{a}}.
\end{equation}
Taking logs yields
\begin{equation}
\label{eq:app_log_foc}
\ln h_{it}^{a}
=
(\eta^{a}-1)\big(\ln\theta_{it}^{a}+\ln\xi_{it}^{a}\big)
-\eta^{a}\ln\hat{\omega}_{it}.
\end{equation}
Equation \eqref{eq:app_log_foc} is the key object: it shows how (i) activity-specific shifters $(\theta^{a},\xi^{a})$ and (ii) the common shadow price of digital time $\hat{\omega}$ jointly determine the allocation of time across activities.

\subsection{Engel curves in time}
\label{app:engel_curves}

Define the share of digital time devoted to activity $a$ as
\begin{equation}
s_{it}^{a} \equiv \frac{h_{it}^{a}}{H_{it}},\qquad \sum_{a}s_{it}^{a}=1.
\end{equation}
Let $\bar{\eta}_{it}$ denote the share-weighted average curvature parameter:
\begin{equation}
\label{eq:app_eta_bar_def}
\bar{\eta}_{it} \equiv \sum_{a\in\{\ell,z,o\}} s_{it}^{a}\,\eta^{a}.
\end{equation}
The model implies an Engel elasticity (``leisure Engel curve'' in \citet{aguiar2021}) for each activity:
\begin{equation}
\label{eq:app_engel_elasticity}
\beta^{a}
\;\equiv\;
\frac{\partial \ln h_{it}^{a}}{\partial \ln H_{it}}
\;=\;
\frac{\eta^{a}}{\bar{\eta}_{it}}.
\end{equation}
Thus $\beta^{a}>1$ means that activity $a$ increases \emph{more than proportionally} with total digital time $H$ (a ``time luxury''), while $\beta^{a}<1$ means it increases \emph{less than proportionally} (a ``time necessity'').

The empirical advantage of \eqref{eq:app_engel_elasticity} is that it identifies \emph{relative} curvature parameters. In particular,
\begin{equation}
\label{eq:app_eta_ratio}
\frac{\eta^{a}}{\eta^{b}}=\frac{\beta^{a}}{\beta^{b}}
\qquad\text{for any activities }a,b.
\end{equation}
This log-log relationship can be used directly to estimate the Engel curve elasticities from 
$$
\ln h_{it}^{a}
=
\delta_{t}^{a}
+
X_{it}'\alpha^{a}
+
\beta^{a}\,\ln H_{it}
+
u_{it}^{a},
$$
where we use an IV approach based on rainfall shocks to identify exogenous changes in $\ln H_{it}$.

\paragraph{Estimating Engel curves with an AIDS-style share equation.} \label{app:aids_share}
Following \citet{aguiar2021}, a convenient empirical specification uses shares $s_{it}^{a}$ and regresses them on $\ln H_{it}$ (plus controls and fixed effects). A first-order approximation of the almost ideal demand system (AIDS) implies an equation of the form
\begin{equation}
\label{eq:app_share_reg}
s_{it}^{a}
=
\delta_{t}^{a}
+
X_{it}'\alpha^{a}
+
\gamma^{a}\,\ln H_{it}
+
u_{it}^{a},
\end{equation}
where $\delta_t^{a}$ are time fixed effects and $X_{it}$ collects additional controls (demographics, region/industry cells, etc.). The coefficient $\gamma^{a}$ maps into the Engel elasticity $\beta^{a}$ via
\begin{equation}
\label{eq:app_gamma_to_beta}
\gamma^{a}=\bar{s}^{a}\,(\beta^{a}-1)
\qquad\Longrightarrow\qquad
\widehat{\beta}^{a}=1+\frac{\widehat{\gamma}^{a}}{\bar{s}^{a}},
\end{equation}
where $\bar{s}^{a}$ is the sample mean share for activity $a$ in the estimation sample.

With three activities, the add-up restriction $\sum_{a}s_{it}^{a}=1$ implies $\sum_{a}\gamma^{a}=0$ (and likewise for the fixed effects). In practice, one can (i) estimate \eqref{eq:app_share_reg} for two activities and recover the third by add-up, or (ii) estimate all three equations and check that the estimated coefficients approximately satisfy add-up.

\paragraph{From Engel elasticities to the curvature parameters $\eta^{a}$} \label{app:beta_to_eta}
The Engel elasticities $\{\beta^{a}\}$ identify $\{\eta^{a}\}$ only up to a common scale, because $\beta^{a}=\eta^{a}/\bar{\eta}$. To obtain \emph{levels} of $\eta^{a}$, we would have to impose (or calibrate) a value for the average curvature $\bar{\eta}$:
\begin{equation}
\label{eq:app_calibration_eta}
\eta^{a}=\beta^{a}\,\bar{\eta}.
\end{equation}
Operationally, this means that once we choose $\bar{\eta}$ from the literature (or from an auxiliary calibration), the Engel-curve estimates deliver $\eta^{\ell}$, $\eta^{z}$, and $\eta^{o}$.

\subsection{Backing out the GenAI productivity shock $\delta$ from adoption-induced time changes.} \label{app:delta_inversion} 

Suppose adopting generative AI increases the efficiency of productive digital tasks only:
\begin{equation}
\label{eq:app_gpt_shock}
\xi_{it}^{z,\text{Adopt}}=(1+\delta)\,\xi_{it}^{z,\text{No}},
\qquad
\xi_{it}^{\ell,\text{Adopt}}=\xi_{it}^{\ell,\text{No}},
\qquad
\xi_{it}^{o,\text{Adopt}}=\xi_{it}^{o,\text{No}},
\end{equation}
and that $\theta^{a}$ does not change at adoption. Let
\begin{equation}
\beta^{GPT}_{a}\equiv
\ln h_{it}^{a,\text{Adopt}}-\ln h_{it}^{a,\text{No}}
\qquad (a\in\{\ell,z,o\})
\end{equation}
denote the causal (adoption-induced) change in log time allocated to activity $a$ in the post period.

Take the log first-order condition \eqref{eq:app_log_foc} for productive time $z$ and subtract the log FOC for a \emph{reference} activity $r\in\{\ell,o\}$ whose technology is unchanged by adoption. Differencing across the Adopt and No-Adopt regimes eliminates the common shadow price $\hat{\omega}$ and isolates the technology shock, after using Engel-curve estimates to substitute $\eta^{z}/\eta^{r}=\beta^{z}/\beta^{r}$:
\begin{equation}
\label{eq:app_delta_identity_eta}
(\eta^{z}-1)\ln(1+\delta)
=
\beta^{GPT}_{z}
-\beta^{z}/\beta^{r}\,\beta^{GPT}_{r}.
\end{equation}
Equation \eqref{eq:app_delta_identity_eta} has a direct interpretation: the adoption-induced increase in productive efficiency shows up as a residual increase in $z$ \emph{relative} to the change in the reference activity $r$, after accounting for the fact that the shadow price of time affects activities with different elasticities $\eta^{a}$ differently.


\end{document}